\documentclass[prb,twocolumn,nofootinbib,longbibliography,superscriptaddress]{revtex4-2}
\usepackage{graphicx}
\usepackage{dcolumn}
\usepackage{bm}
\usepackage{xcolor}
\usepackage[normalem]{ulem}

\usepackage{amsmath,amsfonts,amssymb}

\usepackage[mathlines]{lineno}

\let\oldequation\equation
\let\oldendequation\endequation
\renewenvironment{equation}
  {\linenomathNonumbers\oldequation}
  {\oldendequation\endlinenomath}
  
\let\oldalign\align
\let\oldendalign\endalign

\renewenvironment{align}
  {\linenomathNonumbers\oldalign}
  {\oldendalign\endlinenomath}

\newcommand{\state}[1]{|#1\rangle}

\begin{document}

\title{Classification of spin-$1/2$ fermionic quantum spin liquids on the trillium lattice }
\author{Ming-Hao Li}
	\affiliation{Rudolf Peierls Centre for Theoretical Physics, Parks Road, Oxford, OX1 3PU, UK}

\author{Sounak Biswas}
	\affiliation{Rudolf Peierls Centre for Theoretical Physics, Parks Road, Oxford, OX1 3PU, UK}
	\affiliation{Institut f\"ur Theoretische Physik und Astrophysik, Universit\"at W\"urzburg, 97074 W\"urzburg, Germany}
\author{S.A. Parameswaran}
	\affiliation{Rudolf Peierls Centre for Theoretical Physics, Parks Road, Oxford, OX1 3PU, UK}

\date{\today}

\begin{abstract}

We study fermionic quantum spin liquids (QSLs) on the three-dimensonal trillium lattice of corner-sharing triangles. We are motivated by recent experimental and theoretical investigations that have explored various classical and quantum spin liquid states on similar networks of triangular motifs with strong geometric frustration. Using the framework of Projective Symmetry Groups (PSG), we obtain a classification of all symmetric $\mathsf{Z}_2$ and $\mathsf{U}(1)$ QSLs on the trillium lattice.  
We find 2 $\mathsf{Z}_2$ spin-liquids, and a single $\mathsf{U}(1)$ spin-liquid which is proximate to one of the $\mathsf{Z}_2$ states. The small number of solutions reflects the constraints imposed by the two non-symmorphic symmetries in the space group of trillium.  Using self-consistency conditions of 
the mean-field equations, we obtain the spinon band-structure and spin structure factors corresponding to these states. All three of our spin liquids are gapless at their saddle points: the $\mathsf{Z}_2$ QSLs are both nodal, while the  $\mathsf{U}(1)$ case hosting a spinon Fermi surface. One of our $\mathsf{Z}_2$ spin liquids hosts a stable gapless nodal star, that is protected by projective symmetries against additions of further neighbour terms in the mean field ansatz.
 We comment on directions for further work.

\end{abstract}

\maketitle

\section{Introduction} \label{sec:intro}

Spin liquids are magnetic systems that fail to order at the temperatures expected on the basis of their exchange energy scale, while  exhibiting cooperative behaviour that  distinguish them from  high-temperature paramagnet~\cite{Science_qslrev,SavaryBalents}. A natural ingredient
leading to such lack of ordering  is geometric frustration, wherein the lattice structure eliminates  simple ground state configurations
which minimise the exchange interaction energy between magnetic
moments~\cite{MoessnerRev,Chalker_review},  and which would typically lead to symmetry-breaking in the thermodynamic limit. This is  manifest, for instance, in a
system of three classical spins with pairwise antiferromagnetic Heisenberg interactions: the lowest-energy state of the triangle cannot be described in terms of minimal-energy configurations of each of the individual bonds. Frustrated lattices can be assembled by tiling space with such elementary
units --- typically triangles or tetrahedra --- in order to form
edge-sharing or corner-sharing structures: common examples are the triangular and
kagome lattices in two dimensions (2D), and the pyrochlore and hyperkagome lattices in 3D.
Classical ground states of antiferromagnets on such lattices are macroscopically
degenerate~\cite{MoessnerChalker_PRB}. These degeneracies can often be
understood in the exactly solvable large-$N$ limit: frustration is signaled by  a
macroscopically degenerate manifold of continuously connected ordering
wavevectors~\cite{Canals_Kagome,IsakovEA_pyrochlore,Canals_HHK,
IsakovEA_trilliumHHK, HopkinsonEA_hk}. Such ``classical spin liquids'' often
order at very low temperatures $T$ (much lower than the scale set by exchange
couplings), in accord with the third law of thermodynamics, that forbids the finite $T\to 0$ entropy associated with an extensive ground-state degeneracy. Typically, thermal or quantum fluctuations select an ordering
wave vector out of this manifold, in a phenomenon termed ``order by
disorder"~\cite{Villain, ShenderHoldsworthChalker, IsakovEA_trilliumHHK}. 
However, in some cases the system is sufficiently frustrated that the quantum mechanical ground state selected  as $T\to 0$ continues to exhibit no broken symmetries, and instead is a quantum spin characterized by an emergent deconfined gauge structure. The resulting quantum spin liquid (QSL) is often strikingly characterized by the appearance of fractionalized excitations, whereas its gauge structure is more subtly encoded in certain long-range entanglement properties of the ground-state wavefunction.

A powerful framework to understand  QSL
ground states of quantum spin systems is provided by the projective symmetry group~\cite{Wen2001}. This framework, which  makes the emergent gauge structure especially transparent, builds on the so-called ``parton
construction"~\cite{BaskaranAnderson, BaskaranZouAnderson, AffleckEA,
DagottoEA, WenLee, WenMFT}, and proceeds by representing  each spin in terms of auxiliary fermionic `spinons',
$\vec{S}=\frac{1}{2}f^{\dagger}_i \vec{\sigma}_{ij} f_j$. The physical
Hilbert space of quantum spins is recovered via the 
projection i.e. by imposing the constraint that there is exactly one fermion per site. The resulting  Hamiltonian of these
auxiliary (or Abrikosov) fermions is generically quartic and can then be studied within a mean field decoupling wherein 
operators corresponding to fermion hopping, fermion-pair creation, and
fermion-pair annihilation are self-consistently determined, leading to a quadratic mean-field ``ansatz". By construction, ground states of such
ansatzes correspond to symmetric, disordered wavefunctions, \textit{i.e.},
candidate QSL states. 

This parton (or ``projective'') construction suggests low-energy effective descriptions for these phases in terms
of spinons coupled to gauge fields. Of course, the resulting strongly-coupled problem can be challenging to treat in a controlled fashion, particularly in cases where the spinon degrees of freedom are gapless. Nevertheless, a key feature of the parton approach is that it provides a systematic framework to enumerate and classify candidate variational  wavefunctions in terms of their topological structure, in much the same way that the Landau-Ginzburg formalism provides a useful starting point to investigate broken symmetries as captured by conventional  mean-field wavefunctions. Such a classification is facilitated by the projection of the mean-field Hamiltonian from the large Hilbert space
of auxiliary fermions back to the physical spin Hilbert space --- essential in order to obtain a sensible spin wave function ---
which confers a 
``gauge structure" to the fermion Hilbert space. Specifically, mean field
ansatzes which correspond to the same physical wave function after projection
are related by a gauge transformation. Consequently, the mean field fermion
ansatzes are
only  required to be invariant under physical symmetries up to an associated gauge transformation. In other words, the mean-field ansatz is invariant under a  \emph{projective symmetry
group} (PSG) which is usually larger
than the physical symmetry group of the QSL wave function. 
However, there exists a group of pure gauge transformations ---  typically $\mathsf{Z}_2$, $\mathsf{U}(1)$ or
$\mathsf{SU}(2)$) ---  termed the invariant gauge group (IGG),  which leaves the mean field
anstaz invariant. The IGG and PSG together  characterize the low-energy, long-wavelength fluctuations around the mean field ansatz: these involve
fermions coupled to gauge fields, with the gauge group specified precisely by
the IGG, and  fermions in a mean-field dispersion  classified by representations of the PSG.

In other words, different PSGs capture distinct
``quantum orders" of QSL phases with a specified IGG, in much the same way that the physical symmetry groups
characterise broken symmetries. 

Notably, there can be distinct PSGs corresponding to the same physical symmetry manifest in the spin wavefunction, underscoring the fact that these ``quantum orders'' can be richer than their classical counterparts.

Experimental evidence for QSLs and the resulting need to characterize their emergent low-energy properties has driven a systematic program of applying the projective construction to a variety of frustrated lattices~\cite{Wen2001,wen_trian_psg, Lu_psg_trian, LuYan_psg_hcmb, YouEA_psg_hcmb,
LuEA_psg_kagome, BieriKagome,
Bieri_chiral,Lawler_Paramekanti_YB_Balents,HuanEA_PSG_HK,ChernEA_HHKPSG,
IqbalReuther_bccfcc,
IqbalEA_squactagon,ChauhanEA_diamond,Chern_YBK_Castelnovo_breathingPyrochlore,
LiuHalasozBalents_Pyrochlore, YB_psg_hyperhcmb}. The resulting  mean-field
ansatzes provide starting points for more refined calculations where the
projected fermion wavefunctions can be calculated variationally using Monte
Carlo approaches~\cite{variational1,variational2, variational3,
variational4,variational5,variational5,variational6,variational7}. (Alternative parton constructions that split
the spins into bosons~\cite{WangVishwanath_psg_bosonic, Wang_bosonic_hcmb, hhk_psg_bosonic} offer a complementary set of insights into the phenomenology of possible QSLs and  their possible proximate phases.)

\begin{figure}[t!]
    \centering
    \includegraphics[width=\columnwidth]{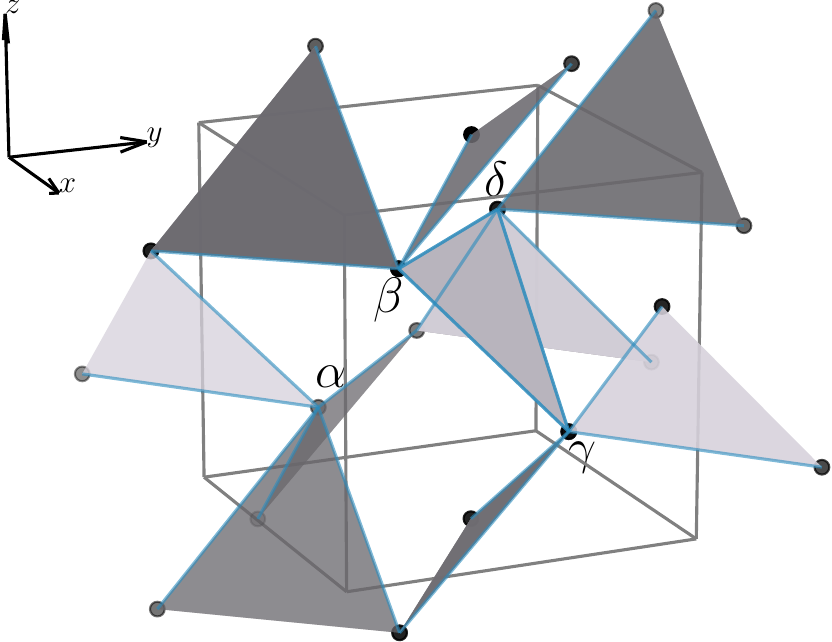}
    \caption{The three-dimensional trillium lattice of corner-sharing triangles. Each site is shared by three triangular plaquettes. The Bravais lattice is cubic, with a basis of 4 sublattice sites labelled $\alpha, \beta, \delta$ and $\gamma$.}
    \label{fig:trillium_latt}
\end{figure}

In this work, we continue this program by classifying  symmetric spin liquid states on
the trillium lattice~\cite{HopkinsonKee_Trillium}, a three-dimensional network of corner sharing
triangles displayed in Fig.~\ref{fig:trillium_latt}. A natural theoretical motivation of this problem is that the motif of corner-sharing triangles is expected to seed significant magnetic frustration,  like the better-known kagome and hyperkagome lattices. At a more experimentally-grounded level, trillium is  the magnetic lattice of MnSi, or that of the Ce moments in
CeIrSi, which has been considered before in the context of frustrated
magnetism~\cite{IsakovEA_trilliumHHK}. Recent characterisations of the quantum
spin liquid material
K\textsubscript{2}Ni\textsubscript{2}(SO\textsubscript{4})\textsubscript{3}~\cite{langbeinite_QSL,
langbeinite_dynamics, langbeinite_neutrondiff} show that the magnetic
$\mathrm{Ni}^{2+}$ ions, with $S=1$, lie on two interconnected trillium lattices, having
exactly the same set of symmetries as single trillium lattice--- implying that
these structures share the classification of symmetric spin liquid states in
terms of projective symmetry groups. Another compound KSrFe\textsubscript{2}(PO\textsubscript{4})\textsubscript{3} with structures similar to K\textsubscript{2}Ni\textsubscript{2}(SO\textsubscript{4})\textsubscript{3}, with $S=5/2$, has been shown to exhibit no long range order down to $T=0.19K$ \cite{boya2022signatures}. Our interest in trillium is also seeded
by its remarkable similarity to the hyperhyperkagome (HHK) lattice which
describes the network of coupled Cu\textsuperscript{2+} ions in
PbCuTe\textsubscript{2}O\textsubscript{6}, which was  shown to host a QSL
ground state in a series of recent
experiments~\cite{KoteswararaoEA,ChillalEA,KhuntiaEA}, leading to theoretical
work on spin liquid states on the underlying HHK
structure~\cite{ChernEA_HHKPSG,hhk_psg_bosonic}. Both trillium and HHK are
three-dimensional networks of corner-sharing triangles with a cubic Bravais
lattice where each site belongs to three corner-sharing triangles. Classical
frustrated magnets on these lattices share similar
phenomenology~\cite{IsakovEA_trilliumHHK}: a large regime with classical spin
liquid behaviour,  eventually yielding to co-planar ordering at very low
temperatures. For both lattices, large-$N$ approaches yield ``partial
ordering"~\cite{Canals_HHK,
IsakovEA_trilliumHHK,WangVishwanathVolborthite,BiswasDamle}, characterized by a
macroscopic but sub-extensive set of ordering wave vectors, whose manifold forms a line (HHK) or surface (trillium) in three
dimensional reciprocal space. This is distinct from the large-$N$ signatures of a classical spin liquid, where this manifold would be extensive~\cite{Canals_Kagome,IsakovEA_pyrochlore,HopkinsonEA_hk}, as obtained, e.g. for the pyrochlore, kagome and hyperkagome lattices. [Note that a recent study of classical Ising models with  three-spin interactions on the trillium and HHK showed that both  host very similar  classical fractal spin liquids, with ``fractonic''  glassy behaviour arising out of kinetic constraints~\cite{BiswasKwanParameswaran}; however this is unlikely to be directly relevant to the QSL problem studied here.]

 The  HHK lattice has the same space group and hence the same classification of PSGs as the three-dimensional hyperkagome lattice~\cite{ChernEA_HHKPSG}, in which each site is shared by two (rather than three)
  corner sharing triangles. The latter has been the subject of  numerous theoretical
  investigations of its ordered and spin-liquid
  states~\cite{HopkinsonEA_hk, Lawler_Paramekanti_YB_Balents,
  LawlerEA_bosonicHK, ZhouLeeEA_HK, HuanEA_PSG_HK, hhk_psg_bosonic} motivated by its relevance to the candidate QSL material
Na\textsubscript{4}Ir\textsubscript{3}O\textsubscript{8}~\cite{NaIr_exp1,
  NaIr_exp2, NaIr_exp3, NaIr_exp4}. The corresponding $\mathsf{P4_1 32}$ space group  has both 3-fold rotations and a 4-fold
  non-symmorphic screw rotation, with the latter  known to
  cause a drastic reduction of total number of QSL states~\cite{HuanEA_PSG_HK}. In contrast, 
  the $\mathsf{P2_1 3}$ space group of trillium  has a three-fold rotation, along with
  {\it two} non-symmorphic screw rotations~\cite{IsakovEA_trilliumHHK}. In the light
  of the preceding discussion, it is natural to ask what QSL phases are
  consistent with symmetries of the trillium lattice. To this end, in this
  paper we undertake a classification of PSGs for spin-$1/2$ QSLs on this lattice. Although  experiments~\cite{langbeinite_QSL,
langbeinite_dynamics, langbeinite_neutrondiff} indicate that on K\textsubscript{2}Ni\textsubscript{2}SO\textsubscript{4} is best understood as an effective spin-$1$ system, understanding the simpler spin-$1/2$ case is an important first step towards a more comprehensive study of the higher-spin problem. Accordingly, we
  hope that the present work will guide the
  interpretation of results of future experiments, and add to our theoretical
  understanding of QSL phases in three dimensions.

  The rest of this paper is organised as follows. In Sec.~\ref{subsec:trillium} we introduce the crystal structure of the trillium lattice and the symmetry 
generators of its space group. In Sec.~\ref{subsec:psg_trillium} we present the symmetry group relations of
trillium, and outline the classification of its PSGs using them. We also present the gauge transformations accompanying
physical symmetries for all of the PSGs. The details are relegated to Appendices ~\ref{sec: z2psg} and ~\ref{sec: u1psg}.

\section{Background: Projective symmetry group Formalism} \label{sec:psg_review}
We briefly review the projective symmetry group classification of parton mean-field theories, as applied to spin models with Heisenberg exchange interactions. Readers familiar with the parton approach can jump ahead to the next section, but may wish to quickly skim this section to orient themselves with our notation and conventions. We begin with the Heisenberg model on a given spatial lattice,
\begin{equation}
    H=\sum_{\{i, j\}}J_{ij}\vec{S}_i\cdot\vec{S}_j.
\end{equation}

In order to implement the PSG, we first enlarge the Hilbert space by decomposing spins into Abrikosov fermions as follows:
\begin{equation}
    \vec{S}_i=\sum_{\alpha,\beta}\frac{1}{2}f^\dag_{i\alpha}\vec{\sigma}_{\alpha\beta}f_{i\beta}. \label{eq:abrikosov}
\end{equation}
The above equation maps the spin Hilbert space to the subspace of the Abrikosov fermion Hilbert space in which the fermion occupation number on each site is $1$. This means that, on the operator level, we strictly have $\sum_{\alpha}f^\dag_{i\alpha}f_{i\alpha}=\mathrm{Id}$. Indeed, by using the identity, we can verify that $[S^m,S^n]=i\epsilon_{lmn}S^l$. In fact, a second constraint, is also introduced as a consequence of the first: $\sum_{\alpha, \beta} f_{i\alpha}f_{i\beta}\epsilon_{\alpha\beta} = 0$.  (One can verify by considering $\sum_{\alpha, \beta} f_{i\alpha}f_{i\beta}\epsilon_{\alpha\beta} \sum_{\gamma}f^\dag_{i\gamma}f_{i\gamma}\state{\psi}$, where $\sum_{\gamma}f^\dag_{i\gamma}f_{i\gamma}\state{\psi} = \state{\psi}$ by virtue of single-occupancy.)

\par
In terms of the Abrikosov fermions, the Heisenberg Hamiltonian reads (up to some constants)
\begin{align}
    H&=\sum_{\{i, j\}}\sum_{\alpha\beta\mu\nu}J_{ij}\frac{1}{4}(f^\dag_{i\alpha}\vec{\sigma}_{\alpha\beta}f_{i\beta})\cdot(f^\dag_{j\mu}\vec{\sigma}_{\mu\nu}f_{j\nu})\label{eq:abrikosov_hamiltonian} \\
    &=\sum_{\{i, j\}}\sum_{\alpha\beta}-\frac{1}{2}J_{ij}(f^\dag_{i\alpha}f_{j\alpha}f^\dag_{j\beta}f_{i\beta}+\frac{1}{2}f^\dag_{i\alpha}f_{i\alpha}f^\dag_{j\beta}f_{j\beta}),
    \nonumber
\end{align}
 We now study $H$ within a mean-field approximation, by introducing parameters for expectation values of operators
\begin{equation}
    \eta_{ij}\epsilon_{\alpha\beta}=-2\langle f_{i\alpha}f_{j\beta}\rangle, \quad \chi_{ij}\delta_{\alpha\beta}=2\langle f^\dag_{i\alpha}f_{j\beta}\rangle;
    \label{eq:mf_parameters}
\end{equation}
where $\eta_{ij}=\eta_{ji}$ and $\text{ }\chi_{ij}=\chi^\dag_{ji}$.

As is usual, we expand operators in  Eq.~\ref{eq:abrikosov_hamiltonian} in terms of fluctuations about their expectation values and ignore terms which are quadratic in fluctuations,  leading to 
\begin{align}
   H_{\mathrm{MFT}} =& - \sum_{\langle i, j\rangle} \frac{3}{8}J_{ij}(\chi_{ji} f^\dag_{i\mu}f_{j\mu} 
   + \eta_{ij} f^\dag_{i\mu}f^\dag_{j\mu} + \text{h.c.} \nonumber\\
   &- |\chi_{ij}|^2 - |\eta_{ij}|^2)  
   +\sum_{i} (\mu^3_{i}(f^\dag_{i\uparrow}f_{i\uparrow} - f_{i\downarrow}f^\dag_{i\downarrow}) \nonumber \\
   &+ \frac{1}{2}(\mu^1_{i}+i \mu^2_{i})f_{i\mu}f_{i\nu}\epsilon_{\mu\nu}+ \text{h.c.}).
   \label{eq:mf_hamiltonian1} ,
\end{align}
where we have introduced the Lagrange multipliers $\mu^{1,2,3}_i$ to impose the  
one-fermion-per-site   constraint at a mean-field level. These, as well as the parameters $\chi_{ij}$ and $\eta_{ij}$, are determined self consistently.

To discuss the $\mathsf{SU}(2)$ gauge structure of the mean-field Hamiltonian, it is convenient to introduce a spinor representation
\begin{equation}
    \psi \equiv \begin{bmatrix}
        \psi_1\\
        \psi_2
    \end{bmatrix}\equiv \begin{bmatrix}
        f_\uparrow\\
        f^\dag_\downarrow
    \end{bmatrix}.
    \label{eq:spinor}
\end{equation}

In terms of these spinors, the  $H_{\text{MFT}}$ can be be compactly rewritten as:
\begin{align}
   H_{\mathrm{MFT}} &= \sum_{\langle i, j\rangle}\frac{3}{8} J_{ij}\left[\frac{1}{2} \mathrm{Tr}(U^\dag_{ij}U_{ij}) - (\psi^\dag_i U_{ij} \psi_j + \text{h.c.})\right] \nonumber \\
   & \,\,\,\,\,\,\,\,\,\,\,\,+ \sum_{i,l} \mu^l_i\psi^\dag_i\tau^l \psi_i ,
   \label{eq:mf_hamiltonian2}
\end{align}
where the matrix $U_{ij}$ captures both mean-field parameters via 
\begin{equation}
    U_{ij} \equiv \begin{bmatrix}
        \chi^\dag_{ij} & \eta_{ij}\\
        \eta^\dag_{ij} & -\chi_{ij}
    \end{bmatrix} .
\end{equation}
The constraint implementing projection into the spin Hilbert space at the mean-field level now has the form:
\begin{equation}
    \langle \psi^\dag_i \tau^l \psi_i \rangle = 0, \,\,\,\, l=1,2,3.
\end{equation}
The $\{U_{ij}\}$ and $\{\mu^{m}_i\ \}$ together constitute variational parameters that specify the mean-field ``ansatz''for the  Hamiltonian and the  corresponding ground state wavefunction. Variationally optimizing the parameters to obtain the lowest energy ground state is equivalent to determining the parameters self-consistently.

It is crucial to realize that the ground state of the mean field spinon Hamiltonian is not a valid spin wave function, since the on-site constraints are only enforced {\it on average}. The final  wavefunction in terms of the physical spin degrees of freedom is constructed from the mean-field spinon state by Gutzwiller projection, i.e. $\state{\Psi_{\mathrm{spin}}} = P_G\state{\Psi_{\mathrm{MFT}}}$.

The spinor representation makes the $\mathsf{SU}(2)$ gauge redundancy  of the
mean-field Hamiltonian manifest. The Hamiltonian is invariant, trivially, under
the site-dependent gauge transformation $\psi_{i} \mapsto W_{i}\psi_{i}$, 
$U_{ij} \mapsto W_i U_{ij} W_{j}^{\dagger}$ and $\mu^m_i \mapsto W_i \mu^m_{i} W_{i}^{\dagger}$, with $W_i\in \mathsf{SU}(2)$ since this leaves physical spin operator invariant [cf. ~\ref{eq:abrikosov}]. Therefore, the mean-field anstaze
parametrised by $U_{ij},\mu^m_i$ and  $W_i U_{ij}
W_j^{\dagger}, W_i \mu^m_i W_{i}^{\dagger}$ share the same physical spin wavefunctions, \textit{i.e.} after projection into the
 spin Hilbert space. This has significant consequences for what we
require of symmetric mean-field ansatzes. Consider the action of a symmetry $g:
U_{ij}\mapsto U_{g(i)g(j)}$. For a symmetric ansatz we no longer
require $U_{g(i)g(j)}=U_{ij}$; instead, we only require  that   there exist transformations $G_g(n) \in
\mathsf{SU}(2)$ for all sites $n$, such that $G_g(g(i)) U_{g(i)g(j)}G^\dag_g(g(j)) = U_{ij}$. The gauge redundancy then implies that physical properties of the state represented
by the ansatz has not changed. The physical transformations $g$ together with the gauge transformation, $(G_g(i),g)$, which leaves
the ansatz invariant, constitute the \emph{projective symmetry group} (PSG).
The PSG characterises the symmetries of the ansatz, and serves to classify and
characterise different mean field spin liquid states. \par
The PSG also determines the low-energy  description of fluctuations about the
mean field states.  From the preceding discussion on the gauge structure it is
clear that not all fluctuations of the mean-field parameters $\{ U_{ij}\}$ are
physical: the unphysical fluctuations between gauge inequivalent states must be
described by gauge fields in the effective theory. The effective theories,
then, are likely to be fermions coupled to gauge fields. The gauge structure of
the low energy theory is in general not   given by the high energy
gauge group $\mathsf{SU}(2)$, but is instead determined by the
``invariant gauge group'' (IGG)~\cite{Wen2001}. The IGG is a subgroup of the PSG
comprised of pure gauge transformations which leave the ansatz invariant,
\textit{i.e.}, $\mathcal{G} = \{ W_i|W_i U_{ij}W^\dag_j=u_{ij}, W\in
\mathsf{SU}(2)\}$.  Given the central importance of the IGG, one usually labels
QSLs by the IGG, leading to the terminology of ``$\mathsf{Z}_2$,
$\mathsf{U}(1)$, or $\mathsf{SU}(2)$" QSLs. The PSGs, therefore, play a role
for mean-field QSL phases akin to that of ordinary symmetry groups for broken-symmetry phases, distinguishing quantum
disordered states with the same physical symmetries but different emergent properties.\par
This is a good place to flag one final complication: namely, that that certain PSGs do not correspond to non-zero mean field ansatzes. Therefore, simply tabulating the list of PSGs does not conclude the classification of PSGs; it is essential to investigate the physical constraints that each imposes on the mean field ansatzes. Despite this complication, it is nevertheless useful to organize the investigation of symmetric spin liquid ground states on a
given spatial lattice in terms of an enumeration of all PSGs, for a given set of physical
symmetries (typically, the full lattice space group as well as time reversal symmetry) and the IGG. This
allows the construction of the corresponding mean-field ansatzes and  spin liquid wavefunctions.  In the balance of this paper, we implement such a program
 for the trillium lattice.

\section{PSGs of the Trillium lattice} 
\label{sec:latt_details}
\begin{table} 
\begin{ruledtabular}
\begin{tabular}{ccc}
 $\zeta$ & $\vec{u}_j - \vec{u}_i$ & $(s_i, s_j)$\\  \hline
$1$ & $(0,0,0)$ & $(\beta, \gamma)$ \\
$2$ & $(0,0,1)$ & $(\beta, \gamma)$ \\
$3$ & $(0,1,1)$ & $(\delta, \alpha)$ \\
$4$ & $(0,1,0)$ & $(\delta, \alpha)$ \\
$5$ & $(0,0,0)$ & $(\gamma, \delta)$ \\
$6$ & $(1,0,0)$ & $(\gamma, \delta)$ \\
$7$ & $(1,0,1)$ & $(\beta, \alpha)$ \\
$8$ & $(0,0,1)$ & $(\beta, \alpha)$ \\
$9$ & $(0,0,0)$ & $(\delta, \beta)$ \\
$10$ & $(0,1,0)$ & $(\delta, \beta)$ \\
$11$ & $(1,1,0)$ & $(\gamma, \alpha)$ \\
$12$ & $(1,0,0)$ & $(\gamma, \alpha)$ 
\end{tabular}
\end{ruledtabular}
\caption{\label{table:bond_list} The labelling of the $12$ translationally inequivalent nearest neighboring links for a unit cell, indexed by $\zeta$. Each link is specified by the unit-cell positions and sublattice indices of the two lattice sites making up the link. For a given label $\zeta$, the head of the bond is labeled $i$ and the end is labeled $j$. $\vec{u}_{i/j}$ is the position of the unit cell, whereas $s_{i/j}$ is the sub-lattice index.}
\end{table}

\begin{table*} 
\begin{ruledtabular}
\begin{tabular}{ccccccc|c}
 & $G_x$ & $G_y$ & $G_z$ & $G_a$ & $G_b$ & $G_c$ & $G_{\mathcal{T}}$ \\  \hline
 $(\vec{u}, \alpha)$ & $\tau_0$ & $\tau_0$ & $\tau_0$ & $\tau_0$ & $\mathcal{A}^{\dag}$ & $\mathcal{A}$ & $\mathcal{E}$  \\
 $(\vec{u}, \beta)$ & $\tau_0$&$\tau_0$ &$\tau_0$ &$\mathcal{A}$ &$ \mathcal{A}^{\dag}$ & $\tau_0$&$\mathcal{E}$ \\
$(\vec{u}, \gamma)$ &$\tau_0$ &$\tau_0$ &$\tau_0$ &$ \mathcal{A}^{\dag}$ &$ \mathcal{A}$ &$\tau_0$ & $\mathcal{E}$\\
 $(\vec{u}, \delta)$ &$\tau_0$ &$\tau_0$ &$\tau_0$ &$ \tau_0$ &$\mathcal{A}$ &$\tau_0$ & $\mathcal{E}$\\ \hline
\end{tabular}
\end{ruledtabular}
\caption{\label{table:psg_sol} The  $\mathsf{Z}_2$ PSG solutions for the trillium lattice with the symmetry group $\mathsf{P}2_13\times \mathsf{Z}^{\mathcal{T}}_2$. Here $\mathcal{A}=\tau_0, e^{i\frac{2\pi}{3}\tau_z}$, and $\mathcal{E}=\tau_0, i\tau_z$. Thus in total we have $4$ $\mathsf{Z}_2$ PSG. We will, however, note that the $\mathcal{E}=\tau_0$ cases do not produce physical mean-field ansatzes. If TRS is not included, we have $2$ PSG solutions. }
\end{table*}

\begin{table*} 
\begin{ruledtabular}
\begin{tabular}{ccccccc|cc}
 & $G_x$ & $G_y$ & $G_z$ & $G_a$ & $G_b$ & $G_c$ & $G_{\mathcal{T}}$ $(n_{\mathcal{T}}=1)$ & $G_{\mathcal{T}}$ $(n_{\mathcal{T}}=0)$ \\  \hline
 $(\vec{u}, \alpha)$ & $\tau_0$ & $\tau_0$ & $\tau_0$ & $\tau_0$ & $e^{-iA\tau_z}$ & $e^{iA\tau_z}$ & $i\tau_x e^{iA\tau_z}$ & $i\tau_z$ \\
 $(\vec{u}, \beta)$ & $\tau_0$&$\tau_0$ &$\tau_0$ &$e^{iA\tau_z}$ &$e^{-iA\tau_z}$ & $\tau_0$& $i\tau_x $ & $i\tau_z$\\
$(\vec{u}, \gamma)$ &$\tau_0$ &$\tau_0$ &$\tau_0$ &$e^{-iA\tau_z}$ & $e^{iA\tau_z}$ &$\tau_0$ & $i\tau_x e^{-iA\tau_z}$ & $i\tau_z$\\
 $(\vec{u}, \delta)$ &$\tau_0$ &$\tau_0$ &$\tau_0$ &$\tau_0$ &$e^{iA\tau_z}$ &$\tau_0$ & $i\tau_x e^{iA\tau_z}$ & $i\tau_z$\\ \hline
\end{tabular}
\end{ruledtabular}
\caption{\label{table:psg_sol_u1} The $\mathsf{U}(1)$ PSG solutions for the trillium lattice with the symmetry group $\mathsf{P}2_13\times \mathsf{Z}^{\mathcal{T}}_2$. When TRS is not included, the PSG solutions are characterised by $A$, where we have $A=0,\frac{2\pi}{3}$. When TRS is included, there are two classes of PSG solutions. 1.) $n_{\mathcal{T}}=1$: in this class, no new constraint is introduced; 2.) $n_{\mathcal{T}}=0$: in this class, we have $G_{\mathcal{T}}=i\tau_z$ uniformly. Later we will see that only the case with $A=0$ and $n_{\mathcal{T}}=1$ leads to physical nearest neighbor mean field ansatz invariant under the PSG actions. Thus in total we have $4$ $\mathsf{U}(1)$ PSG. If TRS is not included, we have $2$ PSG solutions.}
\end{table*}

\subsection{The trillium lattice}
\label{subsec:trillium}

We begin by describing the trillium lattice and its spatial symmetries. These, along with time reversal, will constitute the physical symmetries that our QSL ground states (after projection to the correct Hilbert space) must respect, and are hence central to the PSG construction.

The trillium lattice, shown in Fig.~\ref{fig:trillium_latt},  has a simple cubic Bravais lattice with four sub-lattices: $\alpha, \beta, \gamma$ and  $\delta$. The positions of the sublattice sites relative to the unit cell center are given by:
\begin{align}
\vec{r}^0_\alpha &= (\kappa , \kappa , \kappa ), \vec{r}^0_\beta = \left(\frac{1}{2}+\kappa , \frac{1}{2}-\kappa , 1-\kappa \right), \\
\vec{r}^0_\gamma &= \left(1-\kappa , \frac{1}{2}+\kappa , \frac{1}{2}-\kappa \right), \vec{r}^0_\delta = \left(\frac{1}{2}-\kappa , 1-\kappa , \frac{1}{2}+\kappa \right), \nonumber
\end{align}
where $\kappa$ is a free parameter. As mentioned before, the nearest neighbor bonds on the lattice form a network of corner sharing triangles, with each site participating in three triangles, which are the elementary frustrated motifs.

We  denote the  position of a unit cell $i$ by the vector $\vec{u_i}=(x,y,z)$, where $x,y,z$ are integers.  
A generic lattice site $i$  is referred to by specifying its unit cell position and sublattice as $i\equiv (x,y,z;s)$; such a site lies at position $\vec{u}_i+\vec{r}^0_s$. Since the mean-field parameters $\{U_{ij}\}$ specifying the ansatz are associated with the links, it is  convenient to uniquely label all links for the purpose of further discussion. We do so by exploiting lattice translation invariance: there are $12$ links per unit cell, all of which are translationally inequivalent. We introduce the labels $\zeta=(1,2\ldots 12)$ for these links, and specify each of these links in Tab.~\ref{table:bond_list}.

Trillium has  space group $\mathsf{P2_1 3}$, with the symmetry generators $\{T_x, T_y, T_z, g_a, g_b, g_c \}$. Here, $T_i$s are the three translational generators. $g_c$ is a threefold rotation about the $(1,1,1)$ axis passing through the origin of an unit cell. $g_a$ and $g_b$ are the generators of the $2$-fold non-symmorphic screw rotations. $g_a$ involves a $\pi$ rotation about an axis in the $(0,0,1)$ direction passing through the point $(1/2,0,0)$, followed by a translation by $1/2$ of the unit-cell distance along the rotation axis. $g_b$ involves a $\pi$ rotation about an axis in the direction $(0,1,0)$ 
passing through the point $(0,0,1/2)$ followed by a translation of $1/2$ of the untit-cell distance along the rotation axis.  It has been noted previously~\cite{HuanEA_PSG_HK} that non-symmorphic symmetries generally lead to strong constraints on possible PSGs, and a consequent reduction of their number.

The generators $g_a,g_b$ and $g_c$ act on a lattice site $i\equiv(x,y,z,s)$ via
\begin{align}
\nonumber g_a: & (x,y,z;\alpha)\mapsto (-x, -y-1, z;\delta), \\
& (x,y,z;\beta)\mapsto (-x-1, -y-1, z+1;\gamma),\nonumber \\
& (x,y,z;\gamma)\mapsto (-x-1, -y-1, z;\beta),\nonumber \\
& (x,y,z;\delta)\mapsto (-x, -y-1, z+1;\alpha),\nonumber \\
g_b: & (x,y,z;\alpha)\mapsto (-x-1, y, -z;\gamma),\nonumber \\
& (x,y,z;\beta)\mapsto (-x-1, y, -z-1;\delta),\nonumber \\
& (x,y,z;\gamma)\mapsto (-x-1, y+1, -z;\alpha),\nonumber \\
& (x,y,z;\delta)\mapsto (-x-1, y+1, -z-1;\beta),\nonumber \\
g_c: & (x,y,z;\alpha)\mapsto (z, x, y;\alpha),\nonumber \\
& (x,y,z;\beta)\mapsto (z, x, y;\gamma),\nonumber \\
& (x,y,z;\gamma)\mapsto (z, x, y;\delta),\nonumber \\
& (x,y,z;\delta)\mapsto (z, x, y;\beta).
\label{eq:symmetry_action}
\end{align}

\subsection{PSG classification on the trillium lattice}
\label{subsec:psg_trillium}
The PSG involves the group of the transformations ${(G_g(n),g)}$ that leaves
the mean-field ansatz invariant. Here $g$ is a physical symmetry
transformation, and $G_g(n) \in \mathsf{SU(2)}$ is the associated
site-dependent gauge transformation, with $n$ denoting the physical site.  $(G_g(n),g)$ acts on a mean-field
parameter  $U_{ij}$ as 
\begin{equation}
  (G_g(n),g): U_{ij}\mapsto G_g(g(i)) U_{g(i)g(j)}G^\dag_g(g(j)). 
  \label{eq:action_on_ansatz}
\end{equation}
It follows from consecutive action on the ansatz that the product of two PSG elements is given by the group compatibility condition
\begin{align}
  (G_{g_1}(n),g_1) \circ (G_{g_2}(n),g_2) = (G_{g_1}(n)G_{g_2}(g^{-1}_1 n),g_1 g_2),
\label{eq:psg_prod}
\end{align}
and the group inverse by
\begin{align}
  (G_{g_1}(n), g_1 )^{-1}  = (G^{\dag}_{g_1}(g_1(n)),g^{-1}_1 ).
\label{eq:psg_inv}
\end{align}
Since the
elements of the IGG $\mathcal{G}$ are pure gauge transformations which leave
the ansatz invariant, it is clear that whenever $(G_g(i),g)$ is an element of
the PSG, $(W G_g(i),g)$, for all $W \in \mathcal{G}$, is also an element of the
PSG. If one considers a gauge-equivalent ansatz, $W_i U_{ij} W^{\dag}_j$, the
PSG element $(G_g(i),g)$ changes to $(W_i G_g(i) W^{\dag}_{g(i)})$. PSGs related
  by such gauge transformations are equivalent; they are associated with
  gauge-equivalent ansatzes and represent the same QSL phase. Our task is to
  find all such  equivalence classes; in other words, to single out one 
  representative from each class by fixing the gauge freedom.

  It is convenient to carry out this task purely ``algebraically'',
  \textit{i.e.}, by making no reference to the ansatz. To do this, we note that given a
  physical symmetry group and the IGG $\mathcal{G}$, the PSG can be viewed as a
  group equipped with a projection $\mathcal{P}$ to the physical symmetry
  group, such that $\mathcal{P}: (G_g(i),g) \mapsto g$. From the discussion in
  the previous paragraph, $\mathcal{P}: (W G_g(i),g) \mapsto g$ for $W \in
  \mathcal{G}$ . As a corollary, $\mathcal{P}$ projects pure gauge
  transformations in the IGG back to the identity element, $\mathcal{P}: (W ,e)
  \mapsto e$ for $W \in \mathcal{G}$. 

The projection map between the PSG and the physical symmetry group implies that
the gauge transformation $G_g(i)$ associated with the symmetry transformation
$g$ is constrained by the relations between symmetry group elements $g$.  
These constraints on
$G_{g}$ can be used to enumerate all gauge-inequivalent choices of $G_{g}$ for
all symmetry transformations $g$, and hence enumerate all PSGs.

To see this, one begins with the relations between the symmetry generators $\{T_x,
T_y, T_z, g_a, g_b, g_c, \mathcal{T} \}$ which completely specify the group.
Each such relation will lead to an equation constraining the associated PSG
elements. The minimal set of such relations that specify the group is called
the ``presentation'' of the group. Using the GAP computer algebra package~\cite{GAP4}, we obtain the
finite presentation of trillium space group of : \begin{subequations}
\begin{align}
& g^3_c=e,\label{sg:1}\\
& T^{-1}_zg^2_a=e,\label{sg:2}\\
& T^{-1}_yg^2_b=e,\label{sg:3}\\
& T^{-1}_xT^{-1}_yT_xT_y=e,\label{sg:4}\\
& T^{-1}_yT^{-1}_zT_yT_z=e,\label{sg:5}\\
& T^{-1}_zT^{-1}_xT_zT_x=e,\label{sg:6}\\
& g^{-1}_aT_xg_aT_x=e,\label{sg:7}\\
& g^{-1}_aT_yg_aT_y=e,\label{sg:8}\\
& g^{-1}_aT^{-1}_zg_aT_z=e,\label{sg:9}\\
& g^{-1}_bT_xg_bT_x=e,\label{sg:10}\\
& g^{-1}_bT^{-1}_yg_bT_y=e,\label{sg:11}\\
& g^{-1}_bT_zg_bT_z=e,\label{sg:12}\\
& g^{-1}_cT^{-1}_yg_cT_x=e,\label{sg:13}\\
& g^{-1}_cT^{-1}_zg_cT_y=e,\label{sg:14}\\
& g^{-1}_cT^{-1}_xg_cT_z=e,\label{sg:15}\\
& g^{-1}_ag^{-1}_cg^{-1}_bT^{-1}_xT_yg_ag_c=e,\label{sg:16}\\
& g^{-1}_bg^{-1}_aT_xT^{-1}_yT_zg_bg_a=e,\label{sg:17}\\
& g^{-1}_cg^{-1}_bT^{-1}_xT_yg_ag_bg_cg_b=e,\label{sg:18}
\end{align}
\end{subequations}
where $e$ denotes the identity of the symmetry group.\par
We focus further on QSLs on the trillium lattice which respect time-reversal symmetry (TRS).  
The TRS operator $\mathcal{T}$ acts on the mean-field ansatz by complex conjugating  the mean-field parameters
$U_{ij}$ and $\mu_{i}$. It is convenient to include a global gauge transformation $i \tau_2$ in the definition of
$G_{\mathcal{T}}$, such that we have 
\begin{align}
\nonumber  (G_{\mathcal{T}},\mathcal{T}):  U_{ij}\mapsto &G_{\mathcal{T}}(i) i \tau_2 U^{*}_{ij} (-i \tau_2) G^{\dag}_{\mathcal{T}}(j)\\ 
 =& -G_{\mathcal{T}}(i) U_{ij} G^{\dag}_{\mathcal{T}}(j).
\end{align}
Including  TRS introduces the following additional relations, which express the fact that $\mathcal{T}$ commutes with generators in the space group:
\begin{subequations}
\begin{align}
& \mathcal{T}^2=e, \label{sg:19}\\
& \mathcal{T}^{-1}T^{-1}_x\mathcal{T}T_x=e,\label{sg:20}\\
& \mathcal{T}^{-1}T^{-1}_y\mathcal{T}T_y=e,\label{sg:21}\\
& \mathcal{T}^{-1}T^{-1}_z\mathcal{T}T_z=e,\label{sg:22}\\
& \mathcal{T}^{-1}g^{-1}_a\mathcal{T}g_a=e,\label{sg:23}\\
& \mathcal{T}^{-1}g^{-1}_b\mathcal{T}g_b=e,\label{sg:24}\\
& \mathcal{T}^{-1}g^{-1}_c\mathcal{T}g_c=e.\label{sg:25}
\end{align}
\end{subequations}
Chiral spin liquids, which break TRS and some lattice symmetries separately
while preserving their combinations,  have also been considered in the
literature~\cite{Bieri_chiral, MessioEA_chiralbosonic, SchneiderEA_pyrochlore}. For chiral PSGs, one considers the symmetry
group generated by ${g \mathcal{T}^{\epsilon_g}}$ instead of the usual symmetry group generated by $\{g\}$~\cite{MessioEA_chiralbosonic,Bieri_chiral}. 
 $\epsilon_g=\{0,1\}$ specifies whether the lattice symmetry $g$ is preserved on its own ($\epsilon_g=0$), or 
preserved only up to TRS ($\epsilon_g=1$). 
The trillium SG relations given by Eqs.~\ref{sg:1}-\ref{sg:18} 
impose the constraint $\epsilon_g=0$ for all generators. This can be easily seen from the fact that
for each generator $g$, there exists one SG relation which has only an odd number of appearances of that generator, which forces $\epsilon_g=0$. One could still consider spin liquids which respect all lattice symmetries but not TRS. In all our PSG calculations, we first derive the PSG classification without TRS, and then impose TRS at the end.  While this immediately gives us  the PSGs without TRS, we forego a consideration of mean-field ansatzes corresponding to such PSGs, restricting ourselves to the study of physical fully symmetric spin liquids.
Ground states for classical spins on the trillium lattice~\cite{IsakovEA_trilliumHHK}
are also known to be non-chiral (which, for classical spin configurations, is equivalent to co-planarity).

The projective
relation between the symmetry group elements and the corresponding PSG elements
allow us to translate the above symmetry relations
(Eqs~\ref{sg:1}-~\ref{sg:25}) into constraint equations for the  PSG elements.
Consider a general symmetry group relation among a set of elements, $\prod_{\nu}g_{\nu}= e$. The product of the corresponding 
PSG elements are given by $(\tilde{G}, \prod_{\nu}g_{\nu}=e)$, where 
  $\tilde{G}$ can be constructed from the matrices $G_{g_{\nu}}(i)$ using
  Eq.~\ref{eq:psg_prod}.
Under the projection $\mathcal{P}$ to the symmetry group elements $(\tilde{G},e)\mapsto e)$; this
  immediately implies a constraint equation expressing that $\tilde{G}$ must be a member of the IGG, $\tilde{G}\in \mathcal{G}$.

The unknowns in these
equations are of two kinds: first, the site-dependent 
gauge transformation matrices $\{ G_x , G_y, G_z, G_a, G_b, G_c, G_{\mathcal{T}} \}$ 
accompanying each symmetry transformation in $\{ T_x, T_y,
T_z, g_a, g_b, g_z, \mathcal{T} \}$; 
and second, an
element of the IGG $W\in \mathcal{G}$ corresponding to each symmetry group relation in 
Eqs~\ref{sg:1}-\ref{sg:25}. Solving
these equations, along with choice of gauge described earlier, leads to
the different inequivalent PSGs. 

Explicit   procedures for solving these equations in a fixed gauge are detailed for specific lattices in
Ref.~\cite{Wen2001}, as well as several later works that classifying PSGs in
different spatial lattices~\cite{HuanEA_PSG_HK,Bieri_chiral}. We have
undertaken this procedure to enumerate and classify all symmetric spin liquids
with the IGG set to both $\mathsf{Z}_2$ and $\mathsf{U}(1)$. The calculations
are tedious, and  hence we have relegated their details to the
Appendices~\ref{sec: z2psg} and ~\ref{sec: u1psg} for conciseness. Each inequivalent PSG is uniquely specified by  the 
expressions for site-dependent
gauge transformations $\{ G_x , G_y, G_z, G_a, G_b, G_c, G_{\mathcal{T}} \}$
which accompany the symmetry transformation. We now summarize our results by specifying these gauge transformations for all the PSGs that we identify.

When the IGG is fixed to $\mathsf{Z}_2$, we find 4 inequivalent PSGs.  Once the global
gauge freedoms are fixed, the gauge transformation matrices associated with
lattice translations are uniform, with no position or sublattice
dependence for all PSGs, \textit{i.e.}, $G_{x}=G_{y}=G_{z}=1$. The PSGs can be
uniquely indexed by constraints on the gauge transformation matrices obtained
from the PSG equations. First, the transformation corresponding to
time-reversal $G_{\mathcal{T}}$ takes the values $\tau_0$ or $i \tau_z$, though
the PSGs corresponding to $G_{\mathcal{T}}=\tau_0$ do not lead to any non-zero
mean-field ansatzes. Gauge transformation matrices associated with other
symmetry generators are also unit-cell independent, although they retain a
sublattice dependence. Second, the gauge transformation associated with the
rotation $g_c$ acting on sites of sublattice $\alpha$,
$G_c(\alpha)=\mathcal{A}$,  takes the 2 values $\exp(i k(2\pi/3)  \tau_z)$
for $k=\{0,1\}$. All other gauge transformations can be specified in
terms of these three, as detailed in Tab.~\ref{table:psg_sol}. The 2
possible values of $G_c(\alpha)$ and
the 2 possible values of $G_{\mathcal{T}}$ lead to 4 inequivalent PSGs, out
of which only 2 (corresponding to $G_{\mathcal{T}}=i \tau_z $) lead to
non-zero mean field ansatzes.

Next, we fix the IGG to $\mathsf{U}(1)$. As in the case
of $\mathsf{Z}_2$, the gauge transformations corresponding to the three translations
are uniform, $G_{x}=G_{y}=G_{z}=1$. The other gauge transformations, however, acquire 
both a unit-cell and a sublattice dependence. The PSGs can again be indexed by the parameters
specifying certain gauge transformations. The gauge transformation associated with the
rotation $g_c$ acting on sites of sublattice $\alpha$,
$G_c(\alpha)=\mathcal{A}$,  takes the 2 values $\exp(i k(2\pi/3)  \tau_z)$
for $k=\{0,1\}$. On including time reversal,
we find two possibilities for the associated gauge-transformation $G_{\mathcal{T}}$. First, $G_{\mathcal{T}}$ can be $i\tau_z$ uniformly, and this case does not lead to any physical spin-liquid ansatz with non-zero mean-field parameters, and so we do not consider these PSGs further. Second, $G_{\mathcal{T}}$ can acquire a space-dependent form depending on $A$, which leads to physical spin liquids. Following the second possibility, therefore, we have 
1 $\mathsf{U}(1)$ spin liquid, as we will later show that only the $k=0$ case leads to nearest neighbor mean field spin liquid states. We specify the PSGs by expressing all gauge transformations in terms of the parameters $A$ in Tab.~\ref{table:psg_sol_u1}. In the next section, we will  construct mean-field ansatzes for spin liquids corresponding to these PSGs and proceed to investigate them.

\section{Mean-field spin liquid phases} 
\label{sec:mft}
\begin{figure*}[t!] 
\centering
\includegraphics[width=0.16\textwidth]{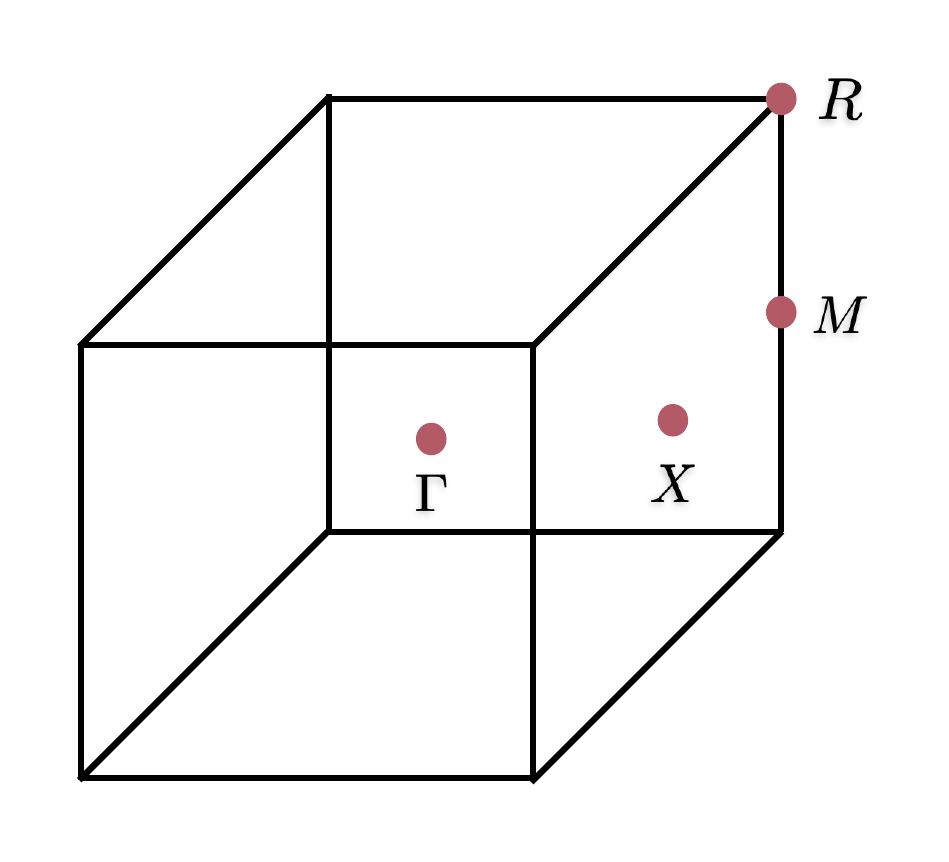}%
\includegraphics[width=0.74\textwidth]{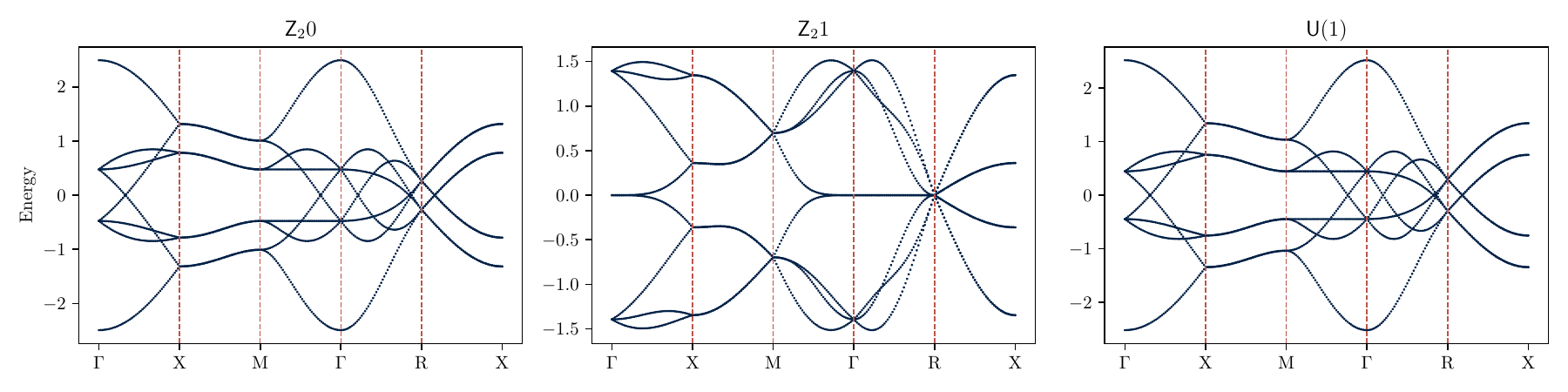}
\caption{\label{fig:bands} Left: The high symmetry points of the Brillouin zone for the simple cubic lattice. Right: Mean-field spinon band structures of the various quantum spin liquids along high symmetry lines in the Brillouin zone. Note that the saddle point (mean-field) parameters of the nearest-neighbor ansatz of the $\mathsf{Z}_20$ QSL looks identical to that of the $\mathsf{U}(1)$ QSL, as explained at in Sec.~\ref{sec:mft}  }
\end{figure*}    
\begin{table*} 
\begin{ruledtabular}
\begin{tabular}{cll}
QSL label & \multicolumn{1}{c}{$U_\zeta$} & \multicolumn{1}{c}{$\mu_s$}\\  \hline
$\mathsf{Z}_20$ & $U_\zeta = U^x\tau_x + U^y\tau_y, \quad \zeta \in \{1, \dots, 12\}.$ &  $\mu_s = \mu^x\tau_x + \mu^y\tau_y, \quad   s\in \{\alpha,\dots,\delta \}.$ \\ \hline

 $\mathsf{Z}_21$ & $U_\zeta = U^x\tau_x + U^y\tau_y, \quad \zeta \in \{1, 2, 3, 4, 5, 6, 9, 10 \}, $ & $\mu_s =0, \quad   s\in \{\alpha,\dots,\delta \}.$ \\
 & $U_\zeta = U^{x'}\tau_x + U^{y'}\tau_y, \quad \zeta \in \{11, 12 \},$ & \\
 & $U_\zeta = U^{x''}\tau_x + U^{y''}\tau_y, \quad \zeta \in \{7, 8 \}.$& \\ \hline

 $\mathsf{U}(1)$ & $U_\zeta = U^z \tau_z, \quad  \zeta \in \{1,\dots,12 \}.$& $\mu_s  = \mu^z \tau_z, \quad s\in \{\alpha,\dots,\delta \}.$ \\ 
\end{tabular}
\end{ruledtabular}
\caption{\label{table:mft_sol_form}Forms of the nearest-neighbor mean field ansatz corresponding to the various algebraic PSG solutions. Given a generic label $\mathsf{Z}_2x$, we can read off the phases in our PSG solutions: $\mathcal{A} = \exp(ix\frac{2\pi}{3} \tau_z)$. The parameters denoted $U^{x'}$ etc. are related to $U^x$ and $U^y$ via Eq.~\ref{eq:rotate_ux_uy}.}
\end{table*}

\begin{table*} 
\begin{ruledtabular}
\begin{tabular}{clc}
 QSL label & Numerical values of the mean field parameters & Energy density \\  \hline
$\mathsf{Z}_20$ & $U^x = -0.161242, U^y = -0.333897, \mu^x = 0.115602, \mu^y = 0.239386$. & -1.647184\\
  $\mathsf{Z}_21$ & $U^x = -0.110700, \quad U^y = 0.330299$. & -1.458114\\
$\mathsf{U}(1)$ & $U^z = 0.370536, \quad\mu^z = -0.295063$. & -1.646913\\
\end{tabular}
\end{ruledtabular}
\caption{\label{table:mft_sol} Numerical values of the mean field parameters and their energetics for nearest-neighbor ansatzes. As explained in the main text, the nearest neighbor ansatzes for the $\mathsf{Z}_20$ and $\mathsf{U}(1)$ QSLs are related by a global gauge transformation at the saddle point, and therefore have the same energy. The $\mathsf{Z}_21$ QSL state has a higher energy than this.}
\end{table*}
In this section, we construct the mean field QSL solutions to the Heisenberg Hamiltonian on the trillium lattice. For this purpose, we will consider only  nearest-neighbor interactions and set $J=8/3$  henceforth.  The form of the ansatzes are constrained by the PSG. 
Concretely, a given PSG   $(G_g,g),g\in \mathsf{P}2_13\times \mathsf{Z}^{\mathcal{T}}_2 $ requires that,
\begin{align}
    \forall g:\,\, & G_g(g(i)) U_{g(i)g(j)}G^\dag_g(g(j)) = U_{ij},\nonumber \\
    & G_g(g(i)) \mu_{g(i)}G^\dag_g(g(i)) = \mu_{i}.
    \label{eq:ansatz_constraint}
\end{align}

The  ansatz for each PSG is derived by systematically imposing Eq.~\ref{eq:ansatz_constraint} using the 
gauge transformations detailed in Tab.~\ref{table:psg_sol} and Tab.~\ref{table:psg_sol_u1}; the results of this procdure, detailed in Appendix~\ref{app:ansatzes}, are  tabulated in Table~\ref{table:mft_sol_form}. The labeling scheme in the table is such that, Given a generic label $\mathsf{Z}_2x$, we can read off the phases in our PSG solutions: $\mathcal{A} = \exp(ix\frac{2\pi}{3} \tau_z)$. There is only $1$ $\mathsf{U}(1)$ QSL, which is labeled as such. The primed parameters, i.e. quantities like $U^{x'}$, are defined as follows:
\begin{align}
    \begin{bmatrix}
        U^{x'}\\
        U^{y'}
    \end{bmatrix} &= \frac{1}{2}\begin{bmatrix}
        -1 & \sqrt{3}\\
        -\sqrt{3} & -1
    \end{bmatrix} \begin{bmatrix}
        U^x\\
        U^y
    \end{bmatrix},\nonumber \\
    \begin{bmatrix}
        U^{x''}\\
        U^{y''}
    \end{bmatrix} &= \frac{1}{2}\begin{bmatrix}
        -1 & -\sqrt{3}\\
        \sqrt{3} & -1
    \end{bmatrix} \begin{bmatrix}
        U^x\\
        U^y
    \end{bmatrix}. \label{eq:rotate_ux_uy}
\end{align}

Concretely, this means we need to find $\{U_{ij}, \mu_i\}$ such that the following self-consistency equations and on-site constraints are satisfied:
\begin{align}
    \chi_{ij}&=\langle f^\dag_{i\uparrow}f_{j\uparrow} \rangle+ \langle f^\dag_{i\downarrow}f_{j\downarrow} \rangle, \nonumber \\
    \eta_{ij}&=\langle f_{i\downarrow}f_{j\uparrow} \rangle- \langle f_{i\uparrow}f_{j\downarrow} \rangle, \nonumber \\
    1 & = \langle f^\dag_{i\uparrow}f_{i\uparrow} \rangle+ \langle f^\dag_{i\downarrow}f_{i\downarrow} \rangle, \nonumber \\
    0 & = \langle f_{i\uparrow}f_{i\downarrow} \rangle. \label{eq:mftconstraint}
\end{align}
We can impose symmetry conditions on the mean-field solutions to reduce the number of parameters, and the PSG determines these symmetry conditions by requiring that Eq.~\ref{eq:action_on_ansatz} is respected. 
We discuss how the symmetry conditions constrain the mean-field ansatzes in detail in
Appendix~\ref{sec: z2psg} and Appendix~\ref{sec: u1psg}, and the results are tabulated in Tab.~\ref{table:mft_sol_form}.
Once we obtain the form of the mean-field Hamiltonians, we assemble the
Hamiltonian using our variational parameters, and solve the non-linear equation set Eq.~\ref{eq:mftconstraint} using the \textsc{NLSolve} package \cite{julia_nlsolve} available in
\textsc{julia} \cite{Julia-2017}. We set up our system with periodic boundary conditions (PBC),
with $L=99$ in the three directions. 

The numerical values of mean-field parameters and the corresponding energies obtained from the self-consistent solutions are 
summarized in Table~\ref{table:mft_sol}. Note that the  mean-field energies of $\mathsf{Z}_20$ and $\mathsf{U}(1)$ state are the same, while $\mathsf{Z}_21$ has a higher energy. A comment on the close energies of the $\mathsf{Z}_20$ and $\mathsf{U}(1)$ state is in order. The ansatzes corresponding to these states are uniform, \textit{i.e}, $U_{ij}=U^x \tau_x +U^y \tau_y $ ($\mu^i= \mu^x \tau_x + \mu^y \tau_y$) for all links (sites) of the $\mathsf{Z}_20$ state while $U_{ij}=U^z \tau_z $ ($\mu_i = \mu^z \tau_z$) uniformly on all links (sites) of the $\mathsf{U}(1)$ state. The saddle-point  solutions for the mean-field parameters of the $\mathsf{Z}_20$ state have the property $|\mu^x/\mu^y| = |U^x/U^y|$. Thus we can use a global $\mathsf{SU}(2)$ transformation to change this ansatz --- only at the saddle point --- to the $\mathsf{U}(1)$ ansatz displayed in Table ~\ref{table:mft_sol_form}. Similar phenomenon has been noted in previous works such as \cite{HuanEA_PSG_HK}.
Therefore, while the nearest-neighbor ansatzes for both QSLs are the same at the saddle point, the general ansatzes displayed in Table ~\ref{table:mft_sol_form} refer to different QSL states with different IGGs.

\begin{figure*}[ht] 

\includegraphics[width=\textwidth]{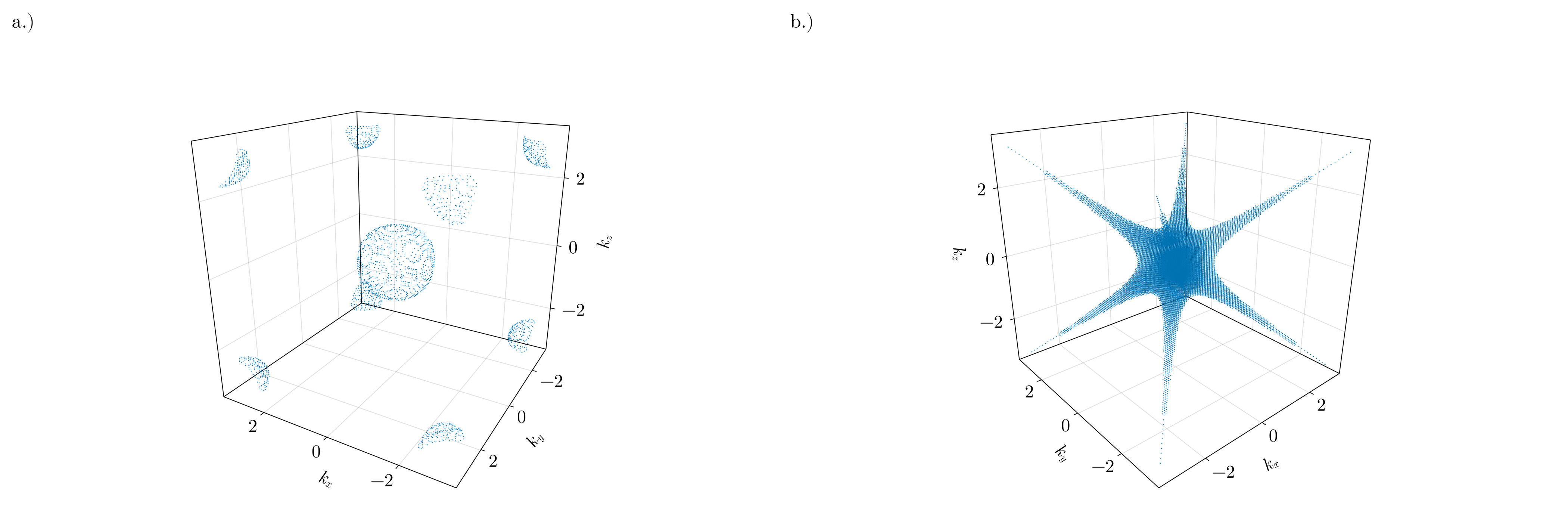}
\caption{In this figure, we plot the collection of gapless points for certain mean field QSL states. a.) $\mathsf{U}(1)$: this state features the two sheeted spinon fermi surfaces, with one located at the center of BZ, and another one at the corners; b.) $\mathsf{Z}_21$: this state has a star-shaped gapless manifold, and features a dispersion-less band along the diagonals of the BZ.\label{fig:sfs}}

\end{figure*}    

\subsection{Relations between the QSLs }
We can also infer connections between the 3 distinct QSLs. First, we note that the $\mathsf{U}(1)$ QSL is the parent state of the $\mathsf{Z}_20$ QSL.  This can be seen by first performing a gauge transformation on $\mathsf{U}(1)$:
\begin{equation}
    W(\alpha)= W(\beta)= W(\gamma)= W(\delta)= e^{-i\frac{\pi}{4}\tau_y},
\end{equation}
so that for this QSL we now have
\begin{align}
&U_\zeta = U^z \tau_x, \quad \zeta \in \{1,\dots,12 \},\nonumber \\
&\mu_s  = \mu^z \tau_x, \quad s\in \{\alpha,\dots,\delta \}.
\end{align}
From this, we see that introducing the perturbations $\Delta U_\zeta \sim \tau_y$ and $\Delta \mu_s \sim \tau_y$ breaks the $\mathsf{U}(1)$ symmetry down in a manner that results in the $\mathsf{Z}_20$ state. The $\mathsf{Z}_21$ QSL state cannot be obtained by perturbing around the $\mathsf{U}(1)$ QSL state.

\subsection{Spinon spectra and the nodal star}
Our mean-field ansatz allow us to determine the structure of excitations on the mean-field ground state. We compute the eigenvalues of the mean-field Hamiltonian for different wave-vectors to obtain the spinon dispersion spectra in the Brillouin zone.

Fig.~\ref{fig:bands} shows the spinon band structures for the different mean-field QSL states. As explained earlier, the nearest-neighbour ansatzes for the $\mathsf{Z}_20$ and $\mathsf{U}(1)$ states are related by a global $\mathsf{SU}(2)$ at the saddle point, leading to identical band structures.
To illustrate the structure of possible gapless modes, we plot the set of gapless points in the Brillouin zone (BZ) for $\mathsf{U}(1)$ and $\mathsf{Z}_21$ states in Fig.~\ref{fig:sfs}. The collection of gapless points at the saddle point for the $\mathsf{Z}_20$ QSL is identical to that of the $\mathsf{U}(1)$ QSL.

We note that all the mean field states we obtained are gapless at saddle point. The $\mathsf{U}(1)$ possesses a spinon fermi surface at the center of the BZ, with another sheet of spinon fermi surface at the corners.
The $\mathsf{Z}_2$ mean-field states are also gapless. The $\mathsf{Z}_20$ state has similar spectrum to those of the $\mathsf{U}(1)$.  We note that the gaplessness of the $\mathsf{Z}_20$ state seems to not be protected by symmetries and one might generically expect a gap to open up when our nearest-neighbor ansatzes are extended to include further neighbor terms.

The $\mathsf{Z}_21$ state hosts a spectrum with a ``nodal star" of gapless points, with dispersion-less bands running from the center of the BZ to its 8 corners. This can be seen from Fig.~\ref{fig:sfs}.  This nodal star is not a specific property of the short-range ansatz we use to display the bands in Fig.~\ref{fig:sfs}; rather it is robust to the addition of arbitrary links in the ansatz. In App.~\ref{app:ansatzes} we prove that the gapless nodal star is protected by projective symmetries of the $\mathsf{Z}_21$ phase. 
Such gapless nodal stars have received significant attention in the pyrochlore lattice~\cite{Burnell_SondhiEA,LiuHalasozBalents_Pyrochlore}, where two gapless bands along the nodal star were recently proven to be protected by the projective symmetries~\cite{LiuHalasozBalents_Pyrochlore}. Such lines were also observed in FCC structures in Ref.~\cite{IqbalReuther_bccfcc}, where the whole mean field Hamiltonian vanishes along the nodal star. Gapless nodal loops were observed in diamond lattice ~\cite{ChauhanEA_diamond} where strong evidence of symmetry-protection was provided by showing that the  gapless nodal loops  persist despite longer range bond amplitudes being  included in the ansatz.

Our proof of the protected nodal star is algebraic and close in spirit to that of Ref.~\cite{LiuHalasozBalents_Pyrochlore}. We  look at the symmetries of the mean-field Hamiltonian directly in momentum-space
\begin{align}
  \nonumber H_{\rm MFT}=\sum_{\vec{k}} \psi^{\dag}(\vec{k}) H_{\rm MFT} (\vec{k}) \psi (\vec{k}),
  \label{eq:mf_hamiltoniank}
\end{align}
When the spinors (Eq.~\ref{eq:spinor}) are arranged as $\psi(\vec{k})=\big(\psi^{\alpha}_1(\vec{k}),\psi^{\beta}_1(\vec{k}),\psi^{\gamma}_1(\vec{k}),\psi^{\delta}_1(\vec{k}),\psi^{\alpha}_2(\vec{k}),\psi^{\beta}_2(\vec{k}),\psi^{\gamma}_2(\vec{k}),\psi^{\delta}_2(\vec{k})\big)$, time-reversal already implies that $H_{\rm MFT}(\vec{k})$ takes the form 
  \begin{align}
  H_{\rm MFT}(\vec{k})= 
  \begin{pmatrix}
    0_{4\times4} & h_{4 \times 4}(\vec{k}) \\
    h^{\dag}_{4\times4}(\vec{k}) & 0_{4\times4} 
\end{pmatrix}.
  \end{align}
This block off-diagonal hermitian structure implies that the eigenvalues come in symmetric pairs of $\pm E(\vec{k})$ everywhere in the BZ. Next, we work out the most general form of $h_{4 \times 4}(\vec{k})$ allowed by the projective representations of
the symmetries $(G_a,g_a), (G_b,g_b)$ and $(G_c,g_c)$. 
Restricting the general form of $h_{4\times4}(\vec{k})$ to the ``nodal star'' wavevectors  $\vec{k}=(\pm k, \pm k, \pm k)$, we show, using elementary linear algebraic techniques,  that it has a maximum rank of $3$. This implies that $H_{\rm MFT}$ has a maximum rank of $6$ along the nodal star, proving the existence of two gapless bands. The complete proof  involves explicit expressions for the most general $h_{4 \times 4}(\vec{k})$ allowed by projective symmetries, and is fleshed out in Appendix~\ref{app:ansatzes}.\par
\begin{figure}[t] 
\includegraphics[width=0.5\textwidth]{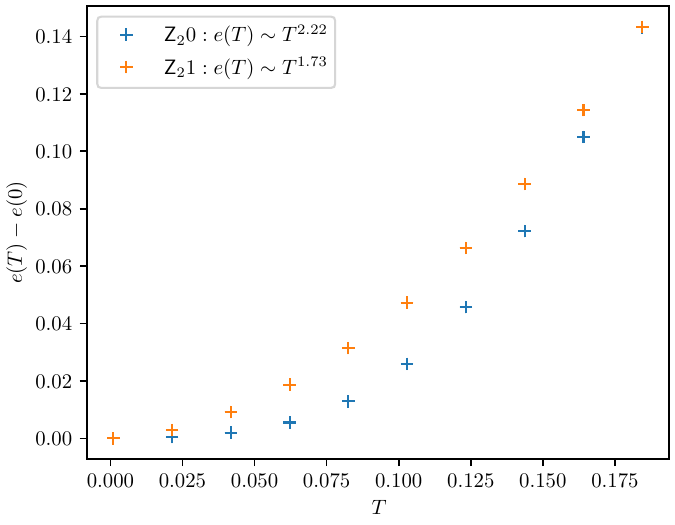}

\caption{\label{fig:energy_temp} Temperature-dependence of the mean-field energies $e(T)$ for the $\mathsf{Z}_2$ spin liquid ansatzes, which can be used to extract the respective specific heat scaling via $C_V(T) \sim \partial e/\partial T$. We note that the $\mathsf{Z}_20$ and $\mathsf{Z}_21$ spin liquids exhibit different scaling behaviors.}
\end{figure}

By computing the equilibrium state energy of the mean field spinon models, we estimate the temperature dependence of the specific heat for the nodal star spin liquid state. Specifically we see that for $\mathsf{Z}_20$, $C_v\sim T^{1.22}$, where as for $\mathsf{Z}_21$, $C_v\sim T^{0.73}$. The numerical results are given in Fig.~\ref{fig:energy_temp}. The above analysis is not performed for the $\mathsf{U}(1)$ spin liquid, since we expect the gauge field excitations at low energies to modify the results from the calculations of the non-interacting model.

\subsection{Spin structure factors}

\begin{figure*}[ht] 
\includegraphics[width=\textwidth]{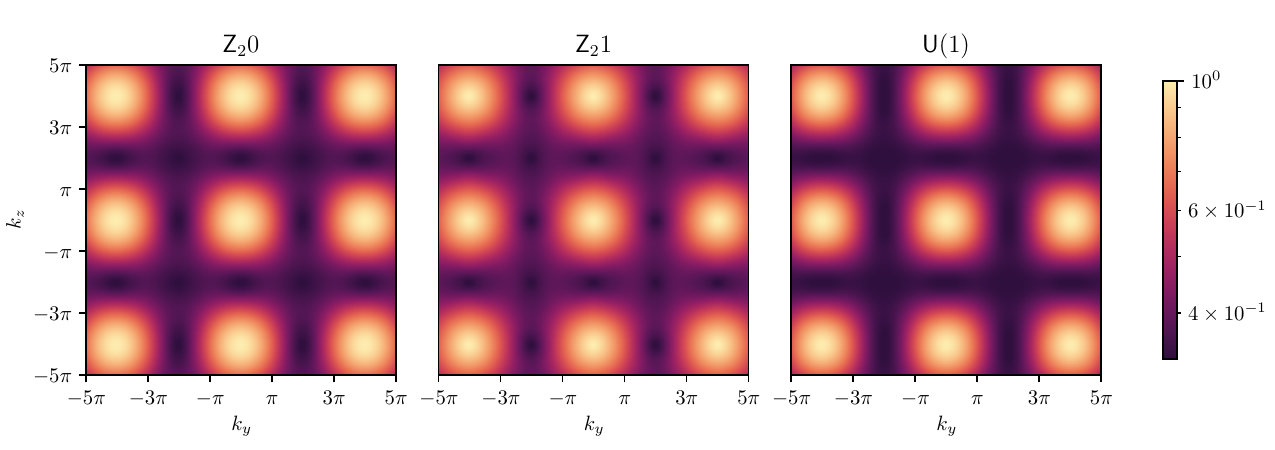}

\caption{\label{fig:ssf} Static structure factors for $\mathsf{Z}_20$, $\mathsf{Z}_21$ and $\mathsf{U}(1)$ mean field states, plotted in the $k_y$-$k_z$ plane. We note that the $\mathsf{U}(1)$ QSL exhibits remarkably broadened static structure factors, whereas those of the $\mathsf{Z}_21$ QSL are the most featureful. The seemingly four-fold rotation symmetry in the $k_y$-$k_z$ plane is due to the two screw symmetries on the lattice.}
\end{figure*}

In Fig.~\ref{fig:ssf}, we plot the static structure factor of $\mathsf{Z}_20$, $\mathsf{Z}_21$ and $\mathsf{U}(1)$ mean field states in the $k_y$-$k_z$ plane. \par
The definition of the static structure factor is:
\begin{equation}
    \mathfrak{S}^{s_i,s_j}(\vec{q}) \equiv \frac{1}{N} \sum_{\vec{R}} e^{-i\vec{q}\cdot (\vec{R} + \vec{d}^0_{ij})} \langle \vec{S}_{(0;s_i)} \cdot \vec{S}_{(\vec{R};s_j)}\rangle,
\end{equation}
where $s_i$ and $s_j$ are the sub-lattice indices of site $i$ and $j$, $\vec{R}$ is the distance between the two unit cells, and $\vec{d}^0_{ij}$ is the distance between the two sub-lattice sites within in the unit cell. And we compute the sum of all these components:
\begin{equation}
    \mathfrak{S}(\vec{q}) \equiv \sum_{s_i, s_j} \mathfrak{S}^{s_i,s_j}(\vec{q}),
\end{equation}
then plot the normalized results.\par
The structure factor plots indicate that we have obtained remarkable quantum spin liquid states, especially the $\mathsf{U}(1)$ QSLs, which are visibly featureless, implying a sharp departure from the ordered states. We also note that the $\mathsf{Z}_21$ state is the most featureful among the three. 

\section{Conclusions and outlook} \label{sec:conclusions}
In this work, we have computed the PSGs for the trillium lattice both with and without time reversal symmetry. In the former case we implement the full  construction of the nearest neighbor mean field (fermionic) parton Hamiltonian of the corresponding quantum spin liquid states.  We find two distinct such QSLs with a  $\mathsf{Z}_2$ gauge group, and a single example of a QSL with a  $\mathsf{U}(1)$ gauge group. We also obtained the corresponding mean-field spinon  band structures and static structure factors, providing some basic thermodynamic and spectral information on these states.  Our main results  are reported in Tab.~\ref{table:psg_sol} and Tab.~\ref{table:psg_sol_u1}.

\par
As noted in the introduction, one of our principal motivations is the recent report of QSL-type behaviour in 
K\textsubscript{2}Ni\textsubscript{2}SO\textsubscript{4}; our work represents a stepping stone towards a parton mean-field analysis of this system, which hosts a double trillium lattice with spin-$1$ moments. Accordingly, a natural next step in this program is to modify the PSG analysis to account for these differences, and 
perform variational Monte Carlo studies of the Gutzwiller projected mean field QSL wavefunctions to compare with the available experimental data. These tasks are currently underway, and we hope to report on them in the near future. 

\begin{acknowledgments}
SAP acknowledges useful conversations with Yasir Iqbal, and the hospitality of the Max Planck Institute for Complex Systems (MPI-PKS) Dresden, where part of this work was completed. SB acknowledges useful discussions with Atanu Maity.
This work was supported by the European Research Council under the European Union Horizon 2020 Research and Innovation Programme, Starting Grant [Agreement No. 804213-TMCS],  a Buckee Scholarship at Merton College (ML), the Deutsche Forschungsgemeinschaft via Grant No. AS120/16-1,  Project No. 493886309 (SB) and the Gutzwiller Fellowship of the MPI-PKS (SAP)
\end{acknowledgments}

\appendix

\section{$\text{IGG}=\mathsf{Z}_2$} \label{sec: z2psg}
One can translate the SG relations to the PSG relations:
\begingroup
\allowdisplaybreaks
\begin{subequations}
\begin{align}
& G_c(g_c^3(i))G_c(g_c^2(i))G_c(g_c(i))=\eta_c,\label{psg:1}\\
& G^\dag_z(g_a^2(i))G_a(g_a^2(i))G_a(g_a(i))=\eta_a,\label{psg:2}\\
& G^\dag_y(g_b^2(i))G_b(g_b^2(i))G_b(g_b(i))=\eta_b,\label{psg:3}\\
& G^\dag_x(T^{-1}_yT_xT_y(i))G^\dag_y(T_xT_y(i))G_x(T_xT_y(i))G_y(T_y(i))=\eta_{xy},\label{psg:4}\\
& G^\dag_y(T^{-1}_zT_yT_z(i))G^\dag_z(T_yT_z(i))G_y(T_yT_z(i))G_z(T_z(i))=\eta_{yz},\label{psg:5}\\
& G^\dag_z(T^{-1}_xT_zT_x(i))G^\dag_x(T_zT_x(i))G_z(T_zT_x(i))G_x(T_x(i))=\eta_{zx},\label{psg:6}\\
& G^\dag_a(T_xg_aT_x(i))G_x(T_xg_aT_x(i))G_a(g_aT_x(i))G_x(T_x(i))=\eta_{ax},\label{psg:7}\\
& G^\dag_a(T_yg_aT_y(i))G_y(T_yg_aT_y(i))G_a(g_aT_y(i))G_y(T_y(i))=\eta_{ay},\label{psg:8}\\
& G^\dag_a(T^{-1}_zg_aT_z(i))G^\dag_z(g_aT_z(i))G_a(g_aT_z(i))G_z(T_z(i))=\eta_{az},\label{psg:9}\\
& G^\dag_b(T_xg_bT_x(i))G_x(T_xg_bT_x(i))G_b(g_bT_x(i))G_x(T_x(i))=\eta_{bx},\label{psg:10}\\
& G^\dag_b(T^{-1}_yg_bT_y(i))G^\dag_y(g_bT_y(i))G_b(g_bT_y(i))G_y(T_y(i))=\eta_{by},\label{psg:11}\\
& G^\dag_b(T_zg_bT_z(i))G_z(T_zg_bT_z(i))G_b(g_bT_z(i))G_z(T_z(i))=\eta_{bz},\label{psg:12}\\
& G^\dag_c(T^{-1}_yg_cT_x(i))G^\dag_y(g_cT_x(i))G_c(g_cT_x(i))G_x(T_x(i))=\eta_{cyx},\label{psg:13}\\
& G^\dag_c(T^{-1}_zg_cT_y(i))G^\dag_z(g_cT_y(i))G_c(g_cT_y(i))G_y(T_y(i))=\eta_{czy},\label{psg:14}\\
& G^\dag_c(T^{-1}_xg_cT_z(i))G^\dag_x(g_cT_z(i))G_c(g_cT_z(i))G_z(T_z(i))=\eta_{cxz},\label{psg:15}\\
& G^\dag_a(g^{-1}_cg^{-1}_bT^{-1}_xT_yg_ag_c(i))G_c^\dag(g^{-1}_bT^{-1}_xT_yg_ag_c(i))\nonumber \\
&\times G_b^\dag(T^{-1}_xT_yg_ag_c(i))G^\dag_x(T_yg_ag_c(i))\nonumber \\
&\times G_y(T_yg_ag_c(i))G_a(g_ag_c(i))G_c(g_c(i))=\eta_{acb},\label{psg:16}\\
& G^\dag_b(g^{-1}_aT_xT^{-1}_yT_zg_bg_a(i))G^\dag_a(T_xT^{-1}_yT_zg_bg_a(i)) \nonumber \\
&\times G_x(T_xT^{-1}_yT_zg_bg_a(i))G^\dag_y(T_zg_bg_a(i))\nonumber \\
&\times G_z(T_zg_bg_a(i))G_b(g_bg_a(i))G_a(g_a(i))=\eta_{ab},\label{psg:17}\\
& G^\dag_c(g^{-1}_bT^{-1}_xT_yg_ag_bg_cg_b(i))G^\dag_b(T^{-1}_xT_yg_ag_bg_cg_b(i))\nonumber \\
&\times G^\dag_x(T_yg_ag_bg_cg_b(i))G_y(T_yg_ag_bg_cg_b(i))G_a(g_ag_bg_cg_b(i)) \nonumber \\
&\times G_b(g_bg_cg_b(i))G_c(g_cg_b(i))G_b(g_b(i))=\eta_{cba}.\label{psg:18}
\end{align}
\end{subequations}
\endgroup
The $G$s in the above relations are $\mathsf{SU}(2)$ matrices, and are associated with the $\mathsf{SU}(2)$ gauge symmetry, which transforms the $G$ in the following way:
\begin{equation}
G_g(i)\mapsto W(g^{-1}(i))G_g(i)W^\dag(i), \quad W\in \mathsf{SU}(2);
\end{equation}
This equation can be understood as follows: under gauge transformation, we have:
\begin{equation}
    U_{ij}\mapsto \tilde{U}_{ij}\equiv W(i)U_{ij}W^\dag(j),
\end{equation}
and the requirement for the gauge transformed PSG is:
\begin{equation}
    \tilde{G}_g(g(i))\tilde{U}_{g(i)g(j)}\tilde{G}^\dag_g(g(j)) = \tilde{U}_{ij}.
\end{equation}
From the above relations we derived the gauge transformation of $G$s.\par
Aside from the gauge symmetry, we note that we can replace a generic element $G$ with $\mathfrak{g}G$, where $\mathfrak{g}\in \text{IGG}=\{\tau_0, -\tau_0 \}$. Wisely making use of this fact is going to help us reduce the number of phases on the right hand side of the PSG equations.\par
By performing:
\begin{align}
    &G_x\mapsto \eta_{cyx} G_x, \quad G_z\mapsto \eta_{czy} G_z, \nonumber \\
    &G_a\mapsto \eta_{acb}\eta_{cba}G_a, \quad G_b\mapsto \eta_{acb} \eta_{cyx} G_b, \nonumber \\
    &G_c\mapsto \eta_c G_c,
\end{align}
we eliminate the phases on the right hand side (RHS) of Eq.~\ref{psg:1}, Eq.~\ref{psg:13}, Eq.~\ref{psg:14}, Eq.~\ref{psg:16} and Eq.~\ref{psg:18}.

\subsection{Solving for the translational Elements}
Let us start by considering the following equations Eq.\ref{psg:4}, Eq.\ref{psg:5} and Eq.\ref{psg:6} that arise because of the commutation of translational generators. Canonically, this gives us the following expressions of $G_x, G_y, G_z$:
\begin{align}
G_x(x,y,z;s)&=\tau_0, G_y(x,y,z;s)=\eta^x_{xy}\tau_0, \nonumber \\ G_z(x,y,z;s)&=\eta^x_{zx}\eta^y_{yz}\tau_0.
\end{align}

\subsection{Solving for $G_c$}
Using the IGG $\mathsf{Z}_2$ gauge symmetry, we had eliminated the phases on the RHS of Eq.\ref{psg:13}, Eq.\ref{psg:14}. To solve for $G_c$, one then plug the canonical expressions of the translational PSG elements into Eq.\ref{psg:13}, Eq.\ref{psg:14} and Eq.\ref{psg:15}. One arrives at the following expressions:
\begin{subequations}
\begin{align}
    G^\dag_c(T^{-1}_y(i))\eta^{-x}_{xy}G_c(i)&=\tau_0,\\
    G^\dag_c(T^{-1}_z(i))\eta^{-x}_{zx}\eta^{-y}_{yz}G_c(i)G_y(g^{-1}_c(i))&=\tau_0,\\
    G^\dag_c(T^{-1}_x(i))G_c(i)G_z(g^{-1}_c(i))&=\eta_{cxz}.
\end{align}
\end{subequations}
Further simplifying the expressions, one arrives at:
\begin{subequations}
\begin{align}
    G_c(x,y,z)&=\eta^{x}_{xy}G_c(x,y-1,z),\\
    G_c(x,y,z)&=\eta^{x}_{zx}\eta^{y}_{yz}\eta^{-y}_{xy}G_c(x,y,z-1),\\
    G_c(x,y,z)&=\eta^{-y}_{zx}\eta^{-z}_{yz}\eta_{cxz}G_c(x-1,y,z).
\end{align}
\end{subequations}
The above expressions are valid for all sub-lattice indices, and we have suppressed the $s$ indices. One then assumes that the following form is valid for $G_c$: $G_c\equiv f_c(x,y,z;s)\mathfrak{M}_c(s)$. Because of the mentioned reason, we have $f_c(x,y,z;s)=f_c(x,y,z)$. Then the separation of variables allows one to arrive at:
\begin{subequations}
\begin{align}
    f_c(x,y,z)&=\eta^{x}_{xy}f_c(x,y-1,z),\\
    f_c(x,y,z)&=\eta^{x}_{zx}\eta^{y}_{yz}\eta^{-y}_{xy}f_c(x,y,z-1),\\
    f_c(x,y,z)&=\eta^{-y}_{zx}\eta^{-z}_{yz}\eta_{cxz}f_c(x-1,y,z).
\end{align}
\end{subequations}
For $f_c$ to be a path-independent function, there are certain constraints that the phases have to satisfy. For example, one considers two paths to arrive at $f_c(x+1, y+1, z)$: 1.) $f_c(x, y, z)\mapsto f_c(x+1, y, z)\mapsto f_c(x+1, y+1, z)$; 2.) $f_c(x, y, z)\mapsto f_c(x, y+1, z)\mapsto f_c(x+1, y+1, z)$. One then compares the phases resulting from the two paths, and enforces them to be identical. Such a process produces the relevant constraints on the phases. We check the path independence on the $xy$, $yz$ and $zx$ planes respectively, and arrive at the following constraint:
\begin{equation}
    \eta_{xy}=\eta^{-1}_{zx};\eta_{yz}=\eta_{xy}; \eta_{zx}=\eta^{-1}_{yz}.
\end{equation}
It follows then $\eta_{xy}=\eta_{yz}=\eta_{zx}=\eta_1$. The previous equations on $f_c$ become:
\begin{subequations}
\begin{align}
    f_c(x,y,z)&=\eta^{x}_{1}f_c(x,y-1,z),\\
    f_c(x,y,z)&=\eta^{x}_{1}f_c(x,y,z-1),\\
    f_c(x,y,z)&=\eta^{-(y+z)}_{1}\eta_{cxz}f_c(x-1,y,z).
\end{align}
\end{subequations}
Therefore, at this point we claim that $G_c=\eta^{xy+xz}_{1}\eta^x_{cxz}\mathfrak{M}_c(s)$.\par
Let us take a look at Eq.\ref{psg:1}. One can eliminate the phase on the RHS by making use of the IGG gauge symmetry. Plugging the above expression into Eq.\ref{psg:1}, we arrive at:
\begin{subequations}
\begin{align}
    \mathfrak{M}^3_c(\alpha)&=\tau_0; \\
    \mathfrak{M}_c(\delta)\mathfrak{M}_c(\gamma)\mathfrak{M}_c(\beta)&=\tau_0; \\
 \eta_{cxz} & =1.
\end{align}
\end{subequations}
It is useful to make a summary before we close this subsection:\\
1.) $\eta_{c}=\eta_{cyx}=\eta_{czy}=\eta_{cxz}=1$;\\
2.) $\eta_{xy}=\eta_{yz}=\eta_{zx}=\eta_1$;\\ 3.)$G_c=\eta^{xy+xz}_{1}\mathfrak{M}_c(s)$, for which the following relations are satisfied:
\begin{align}
    \mathfrak{M}^3_c(\alpha)&=\tau_0; \\
    \mathfrak{M}_c(\delta)\mathfrak{M}_c(\gamma)\mathfrak{M}_c(\beta)&=\tau_0; \label{eq: cyclic} 
\end{align}

\subsection{Solving for $G_a$}
To solve for $G_a$, one plugs the simplified expressions of the translational PSG elements into Eq.\ref{psg:7}, Eq.\ref{psg:8} and Eq.\ref{psg:9}. One arrives at the following expressions:
\begin{subequations}
\begin{align}
    G^\dag_a(T_x(i))G_a(i)&=\eta_{ax},\\
    G^\dag_a(T_y(i))G_y(T_y(i))G_a(i)G_y(g^{-1}_a(i))&=\eta_{ay},\\
    G^\dag_a(T^{-1}_z(i))G^\dag_z(i)G_a(i)G_z(g^{-1}_a(i))&=\eta_{az}.
\end{align}
\end{subequations}
One makes the usual ansatz $G_a(i)\equiv f_a(x,y,z;s)\mathfrak{M}_a(s)$, only this time one does not have $f_a(x,y,z;s)=f_a(x,y,z)$, for the evaluation of $G_{y/z}(g^{-1}_a(i))$ is not $s$-independent.\\
We have, for $s=\alpha/\delta$, the following conditions for $f_a$:
\begin{subequations}
\begin{align}
    \eta^{-1}_{ax}f_a(x,y,z;\alpha/\delta)&=f_a(x+1,y,z;\alpha/\delta),\\
    \eta^{-1}_{ay}f_a(x,y,z;\alpha/\delta)&=f_a(x,y+1,z;\alpha/\delta),\\
    \eta_{az}\eta_1f_a(x,y,z;\alpha/\delta)&=f_a(x,y,z+1;\alpha/\delta);
\end{align}
\end{subequations}
and for $s=\beta/\gamma$, the following conditions for $f_a$:
\begin{subequations}
\begin{align}
    \eta^{-1}_{ax}f_a(x,y,z;\beta/\gamma)&=f_a(x+1,y,z;\beta/\gamma),\\
    \eta^{-1}_{ay}\eta^{-1}_1f_a(x,y,z;\beta/\gamma)&=f_a(x,y+1,z;\beta/\gamma),\\
    \eta_{az}f_a(x,y,z;\beta/\gamma)&=f_a(x,y,z+1;\beta/\gamma).
\end{align}
\end{subequations}
Note that this time we do not have to check the path independence of $f_a$, as the phases appearing in the above equations are constants. We then arrive at the following expressions:
\begin{align}
    &f_a(x,y,z;\alpha/\delta)=\eta^{-x}_{ax}\eta^{-y}_{ay}\eta^z_{az}\eta^z_1,\nonumber \\
    &f_a(x,y,z;\beta/\gamma)=\eta^{-x}_{ax}\eta^{-y}_{ay}\eta^{-y}_1\eta^z_{az},
\end{align}
from which we write:
\begin{align}
    &G_a(x,y,z;\alpha/\delta)=\eta^{-x}_{ax}\eta^{-y}_{ay}\eta^z_{az}\eta^z_1\mathfrak{M}_a(\alpha/\delta),\nonumber \\
    &G_a(x,y,z;\beta/\gamma)=\eta^{-x}_{ax}\eta^{-y}_{ay}\eta^{-y}_1\eta^z_{az}\mathfrak{M}_a(\beta/\gamma).
\end{align}
In plugging these expressions into Eq.\ref{psg:2}, we first consider $i\equiv (x,y,z;\alpha)$. The condition we arrive at is:
\begin{equation}
    G^\dag_z(-x, -y-1, z;\delta)G_a(-x, -y-1, z;\delta)G_a(x,y,z;\alpha)=\eta_a,
\end{equation}
which further simplifies to:
\begin{equation}
    \eta^{x+y+1}_1\eta_{ay}\mathfrak{M}_a(\delta)\mathfrak{M}_a(\alpha)=\eta_a.
\end{equation}
The above equation dictates that $\eta_1=1$. Consequently, one has:
\begin{equation}
    \mathfrak{M}_a(\delta)\mathfrak{M}_a(\alpha)=\eta_a\eta^{-1}_{ay}.
\end{equation}
We then consider other sublattice sites, and they give us:
\begin{subequations}
\begin{align}
G_a(-x-1, -y-1,z+1;\gamma)G_a(x,y,z;\beta)&=\eta_a,\\
G_a(-x-1, -y-1,z;\beta)G_a(x,y,z;\gamma)&=\eta_a,\\
G_a(-x, -y-1,z+1;\alpha)G_a(x,y,z;\delta)&=\eta_a.
\end{align}
\end{subequations}
Plugging the explicit forms into the above equations, and we arrive at:
\begin{subequations}
\begin{align}
    \mathfrak{M}_a(\gamma)\mathfrak{M}_a(\beta)&=\eta_a\eta^{-1}_{ax}\eta^{-1}_{ay}\eta^{-1}_{az},\\
    \mathfrak{M}_a(\beta)\mathfrak{M}_a(\gamma)&=\eta_a\eta^{-1}_{ax}\eta^{-1}_{ay},\\
    \mathfrak{M}_a(\alpha)\mathfrak{M}_a(\delta)&=\eta_a\eta^{-1}_{ay}\eta^{-1}_{az}.
\end{align}
\end{subequations}
It is useful to make a summary again before we close this subsection:\\
1.) $\eta_{xy}=\eta_{yz}=\eta_{zx}=\eta_1=1$, and as a result, $G_x=G_y=G_z=\tau_0$;\\ 2.) We have $G_a(x,y,z;s)=\eta^{-x}_{ax}\eta^{-y}_{ay}\eta^z_{az}\mathfrak{M}_a(s)$; for which the following relations are satisfied:
\begin{subequations}
\begin{align}
\mathfrak{M}_a(\delta)\mathfrak{M}_a(\alpha)&=\eta_a\eta^{-1}_{ay}, \label{eq:eta_4_be_killed_1}\\
    \mathfrak{M}_a(\gamma)\mathfrak{M}_a(\beta)&=\eta_a\eta^{-1}_{ax}\eta^{-1}_{ay}\eta^{-1}_{az},\\
    \mathfrak{M}_a(\beta)\mathfrak{M}_a(\gamma)&=\eta_a\eta^{-1}_{ax}\eta^{-1}_{ay},\\
    \mathfrak{M}_a(\alpha)\mathfrak{M}_a(\delta)&=\eta_a\eta^{-1}_{ay}\eta^{-1}_{az}. \label{eq:eta_4_be_killed_2}
\end{align}
\end{subequations}

\subsection{Solving for $G_b$}
Now we attack Eq.\ref{psg:10}, Eq.\ref{psg:11} and Eq.\ref{psg:12}. Since $G_x$, $G_y$ and $G_z$ are trivial now, the equations are reduced to the following form:
\begin{subequations}
\begin{align}
    G^\dag_b(T_x(i))G_b(i)&=\eta_{bx},\\
    G^\dag_b(T^{-1}_y(i))G_b(i)&=\eta_{by},\\
    G^\dag_b(T_z(i))G_b(i)&=\eta_{bz}.
\end{align}
\end{subequations}
We make the ansatz $G_b\equiv f_b(x,y,z)\mathfrak{M}_b(s)$, where the separation of variables is possible because the above conditions are $s$-independent. One can quickly arrive at the condition that $G_b=\eta^{-x}_{bx}\eta^{y}_{by}\eta^{-z}_{bz}\mathfrak{M}_b(s)$.\par
Plugging the expression into Eq.\ref{psg:3}, one ends up with:
\begin{equation}
    G_b(g_b(i))G_b(i)=\eta_b.
\end{equation}
Considering the individual sublattice sites respectively, one arrives at:
\begin{subequations}
\begin{align}
\mathfrak{M}_b(\gamma)\mathfrak{M}_b(\alpha)&=\eta_b\eta^{-1}_{bx},\\
    \mathfrak{M}_b(\delta)\mathfrak{M}_b(\beta)&=\eta_b\eta^{-1}_{bx}\eta^{-1}_{bz},\\
    \mathfrak{M}_b(\alpha)\mathfrak{M}_b(\gamma)&=\eta_b\eta^{-1}_{bx}\eta^{-1}_{by},\\
    \mathfrak{M}_b(\beta)\mathfrak{M}_b(\delta)&=\eta_b\eta^{-1}_{bx}\eta^{-1}_{by}\eta^{-1}_{bz}.
\end{align}
\end{subequations}
A quick summary:\\
We have $G_b(x,y,z;s)=\eta^{-x}_{bx}\eta^{y}_{by}\eta^{-z}_{bz}\mathfrak{M}_b(s)$; for which the following relations are satisfied:
\begin{subequations}
\begin{align}
\mathfrak{M}_b(\gamma)\mathfrak{M}_b(\alpha)&=\eta_b\eta^{-1}_{bx},\\
    \mathfrak{M}_b(\delta)\mathfrak{M}_b(\beta)&=\eta_b\eta^{-1}_{bx}\eta^{-1}_{bz},\\
    \mathfrak{M}_b(\alpha)\mathfrak{M}_b(\gamma)&=\eta_b\eta^{-1}_{bx}\eta^{-1}_{by},\\
    \mathfrak{M}_b(\beta)\mathfrak{M}_b(\delta)&=\eta_b\eta^{-1}_{bx}\eta^{-1}_{by}\eta^{-1}_{bz}.
\end{align}
\end{subequations}
\subsection{Solving Eq.\ref{psg:16}, Eq.\ref{psg:17} and Eq.\ref{psg:18}}
To attack the remaining three equations, note that we can use the IGG gauge symmetry of $G_a$ and $G_b$ to eliminate $\eta_{acb}$ and $\eta_{cba}$. The equations are reduced to:
\begin{subequations}
\begin{align}
&G^\dag_a(g^{-1}_cg^{-1}_b(i))G^\dag_c(g^{-1}_b(i))G^\dag_b(i) \nonumber \\
&\times G_a(T^{-1}_yT_x(i))G_c(g^{-1}_aT^{-1}_yT_x(i))=\tau_0, \label{eq:acb}\\
&G^\dag_b(g^{-1}_a(i))G^\dag_a(i)G_b(T^{-1}_zT_yT^{-1}_x(i)) \nonumber \\
&\times G_a(g^{-1}_bT^{-1}_zT_yT^{-1}_x(i))=\eta_{ab}\tau_0, \\
&G^\dag_c(g^{-1}_bT^{-1}_xT_y(i))G^\dag_b(T^{-1}_xT_y(i))G_a(i)\nonumber \\
&\times G_b(g^{-1}_a(i))G_c(g^{-1}_bg^{-1}_a(i))G_b(g^{-1}_cg^{-1}_bg^{-1}_a(i))=\tau_0. \label{eq:cba}
\end{align}
\end{subequations}
We consider first $i=(x,y,z;\alpha)$. Plugging this into Eq.\ref{eq:acb} gives us the following:
\begin{align}
&G^\dag_a(y-1,-z,-x-1;\beta)\mathfrak{M}_c(\gamma)G^\dag_b(x,y,z;\alpha)\nonumber \\
&\times G_a(x+1,y-1,z;\alpha)\mathfrak{M}_c(\delta)=\tau_0.
\end{align}
Recalling $G_a=\eta^{-x}_{ax}\eta^{-y}_{ay}\eta^z_{az}\mathfrak{M}_a(s)$ and $G_b=\eta^{-x}_{bx}\eta^{y}_{by}\eta^{-z}_{bz}\mathfrak{M}_b(s)$, the LHS of the above expression is evaluated as:
\begin{align}
&\eta^{y-1}_{ax}\eta^{-z}_{ay}\eta^{x+1}_{az}\mathfrak{M}^\dag_a(\beta)\mathfrak{M}^\dag_c(\gamma)\nonumber \\
&\times \eta^{x}_{bx}\eta^{-y}_{by}\eta^{z}_{bz}\mathfrak{M}^\dag_b(\alpha)\eta^{-x-1}_{ax}\eta^{-y+1}_{ay}\eta^z_{az}\mathfrak{M}_a(\alpha)\mathfrak{M}_c(\delta)\nonumber \\
=&(\eta_{az}\eta_{bx}\eta^{-1}_{ax})^x(\eta_{ax}\eta^{-1}_{ay}\eta^{-1}_{by})^y(\eta_{az}\eta_{bz}\eta^{-1}_{ay})^z\eta_{az}\eta_{ay}\nonumber \\
& \times \mathfrak{M}^\dag_a(\beta)\mathfrak{M}^\dag_c(\gamma)\mathfrak{M}^\dag_b(\alpha)\mathfrak{M}_a(\alpha)\mathfrak{M}_c(\delta).
\end{align}
Since the RHS of the previous expression is unit cell independent, we would have the following equations:
\begin{align}
\eta_{az}\eta_{bx}&=\eta_{ax}, \nonumber \\
\eta_{by}\eta_{ay}&=\eta_{ax}, \nonumber \\
\eta_{az}\eta_{bz}&=\eta_{ay}.
\end{align}
Naming $\eta_2\equiv \eta_{ax}$, $\eta_3\equiv \eta_{ay}$ and $\eta_4\equiv \eta_{az}$, we would have $\eta_{bx}=\eta_2\eta^{-1}_4$, $\eta_{by}=\eta_2\eta^{-1}_3$ and $\eta_{bz}=\eta_3\eta^{-1}_4$. Also, we now have $G_a=\eta^{-x}_{2}\eta^{-y}_{3}\eta^z_{4}\mathfrak{M}_a(s)$ and $G_b=\eta^{-x+y}_{2}\eta^{-y-z}_{3}\eta^{x+z}_{4}\mathfrak{M}_b(s)$. The previous constraint becomes:
\begin{equation}
\mathfrak{M}^\dag_a(\beta)\mathfrak{M}^\dag_c(\gamma)\mathfrak{M}^\dag_b(\alpha)\mathfrak{M}_a(\alpha)\mathfrak{M}_c(\delta)=\eta^{-1}_3\eta^{-1}_4.
\end{equation}
What happens now for Eq.\ref{eq:cba}? Plugging $i=(x,y,z;\alpha)$ into the equation gives us:
\begin{align}
&\mathfrak{M}^\dag_c(\gamma)G^\dag_b(x-1,y+1,z;\alpha) \nonumber \\
&\times G_a(x,y,z;\alpha)G_b(-x,-y-1,z-1;\delta) \nonumber \\
&\times \mathfrak{M}_c(\beta)G_b(-y-1,-z,x-1;\delta)=\tau_0.
\end{align}
Evaluating the LHS of the above equation gives us:
\begin{align}
&\eta^{x-y}_2\eta^{y+z+1}_3\eta^{-x-z+1}_4\eta^{-x}_2\eta^{-y}_3\eta^{-z}_4\eta^{x-y-1}_2\nonumber \\
&\times \eta^{y-z}_3\eta^{-x+z-1}_4\eta^{y+1-z}_2\eta^{z-x+1}_3\eta^{-y+x}_4\nonumber \\
\times &\mathfrak{M}^\dag_c(\gamma)\mathfrak{M}^\dag_b(\alpha)\mathfrak{M}_a(\alpha)\mathfrak{M}_b(\delta)\mathfrak{M}_c(\beta)\mathfrak{M}_b(\delta)\nonumber \\
=&(\eta^{-1}_2\eta^{-1}_3\eta_4)^x(\eta^{-1}_3\eta^{-1}_4\eta_2)^y(\eta_4\eta^{-1}_2\eta_3)^z\nonumber \\
&\times \mathfrak{M}^\dag_c(\gamma)\mathfrak{M}^\dag_b(\alpha)\mathfrak{M}_a(\alpha)\mathfrak{M}_b(\delta)\mathfrak{M}_c(\beta)\mathfrak{M}_b(\delta).
\end{align}
Again, the unit cell independence gives us an extra condition $\eta_2=\eta_3\eta_4$. The original equation becomes:
\begin{equation}
\mathfrak{M}^\dag_c(\gamma)\mathfrak{M}^\dag_b(\alpha)\mathfrak{M}_a(\alpha)\mathfrak{M}_b(\delta)\mathfrak{M}_c(\beta)\mathfrak{M}_b(\delta)=\tau_0.
\end{equation}
No further conditions on the phases can be derived from the three equations. We are in the position to write $G_a=\eta^{-x-y}_{3}\eta^{-x+z}_{4}\mathfrak{M}_a(s)$ and $G_b=\eta^{-x-z}_{3}\eta^{y+z}_{4}\mathfrak{M}_b(s)$. Iterating scenarios with different $s$ for $i\equiv (x,y,z;s)$, we arrive at the following constraints:
\begin{subequations}
\begin{align}
&\mathfrak{M}^\dag_a(\beta)\mathfrak{M}^\dag_c(\gamma)\mathfrak{M}^\dag_b(\alpha)\mathfrak{M}_a(\alpha)\mathfrak{M}_c(\delta)=\eta^{-1}_3\eta^{-1}_4, \\
&\mathfrak{M}^\dag_a(\gamma)\mathfrak{M}^\dag_c(\delta)\mathfrak{M}^\dag_b(\beta)\mathfrak{M}_a(\beta)\mathfrak{M}_c(\gamma)=\eta_4, \\
&\mathfrak{M}^\dag_a(\alpha)\mathfrak{M}^\dag_c(\alpha)\mathfrak{M}^\dag_b(\gamma)\mathfrak{M}_a(\gamma)\mathfrak{M}_c(\beta)=\tau_0, \\
&\mathfrak{M}^\dag_a(\delta)\mathfrak{M}^\dag_c(\beta)\mathfrak{M}^\dag_b(\delta)\mathfrak{M}_a(\delta)\mathfrak{M}_c(\alpha)=\eta_3, \\
&\mathfrak{M}^\dag_b(\delta)\mathfrak{M}^\dag_a(\alpha)\mathfrak{M}_b(\alpha)\mathfrak{M}_a(\gamma)=\eta_{ab}\eta_3\eta_4, \\
&\mathfrak{M}^\dag_b(\gamma)\mathfrak{M}^\dag_a(\beta)\mathfrak{M}_b(\beta)\mathfrak{M}_a(\delta)=\eta_{ab}\eta_3\eta^{-1}_4, \\
&\mathfrak{M}^\dag_b(\beta)\mathfrak{M}^\dag_a(\gamma)\mathfrak{M}_b(\gamma)\mathfrak{M}_a(\alpha)=\eta_{ab}\eta_3\eta^{-1}_4, \\
&\mathfrak{M}^\dag_b(\alpha)\mathfrak{M}^\dag_a(\delta)\mathfrak{M}_b(\delta)\mathfrak{M}_a(\beta)=\eta_{ab}\eta_3\eta^{-1}_4, \\
&\mathfrak{M}^\dag_c(\gamma)\mathfrak{M}^\dag_b(\alpha)\mathfrak{M}_a(\alpha)\mathfrak{M}_b(\delta)\mathfrak{M}_c(\beta)\mathfrak{M}_b(\delta)=\tau_0,\\
&\mathfrak{M}^\dag_c(\delta)\mathfrak{M}^\dag_b(\beta)\mathfrak{M}_a(\beta)\mathfrak{M}_b(\gamma)\mathfrak{M}_c(\alpha)\mathfrak{M}_b(\alpha)=\eta^{-1}_{3},\\
&\mathfrak{M}^\dag_c(\alpha)\mathfrak{M}^\dag_b(\gamma)\mathfrak{M}_a(\gamma)\mathfrak{M}_b(\beta)\mathfrak{M}_c(\delta)\mathfrak{M}_b(\gamma)=\eta_{3}\eta_{4},\\
&\mathfrak{M}^\dag_c(\beta)\mathfrak{M}^\dag_b(\delta)\mathfrak{M}_a(\delta)\mathfrak{M}_b(\alpha)\mathfrak{M}_c(\gamma)\mathfrak{M}_b(\beta)=\eta_{4}.
\end{align}
\end{subequations}

\subsection{Solving for the $\mathfrak{M}$s}
We start by performing some $\mathsf{SU}(2)$ gauge transformations. Let us consider a type of gauge transformation $W(i)\equiv W(s)$. Had we started at a generic gauge, we could always make the following gauge transformation:
\begin{align}
&W(\alpha) = \mathfrak{M}_a(\alpha), W(\beta)=\mathfrak{M}_c(\beta), \nonumber \\
&W(\gamma)=\mathfrak{M}^\dag_c(\delta), W(\delta) = \tau_0.\label{eq:z2su2gauge1}
\end{align}
so that:
\begin{subequations}
\begin{align}
g.t.: \quad &\mathfrak{M}_c(\delta)\mapsto W(\beta)\mathfrak{M}_c(\beta)W^\dag(\beta)=\tau_0, \\
&\mathfrak{M}_c(\gamma)\mapsto W(\delta)\mathfrak{M}_c(\delta)W^\dag(\delta)=\tau_0, \\
& \mathfrak{M}_a(\delta)\mapsto W(\alpha)\mathfrak{M}_a(\alpha)W^\dag(\alpha)=\tau_0. \label{eq:z2maalphagauge}
\end{align}
\end{subequations}
Now we make use of Eq.\ref{eq: cyclic}, and arrive at $\mathfrak{M}_c(\beta)=\mathfrak{M}_c(\gamma)=\mathfrak{M}_c(\delta)=\tau_0$. It should be noted that we are silent on $\mathfrak{M}_c(\alpha)$. Indeed, it is not possible to use gauge symmetry alone to trivialise $\mathfrak{M}_c(\alpha)$. However, as we shall see, other equations will bring enough restrictions on the form of $\mathfrak{M}_c(\alpha)$.\par
Before we take a step further, let us note that taking traces over Eq.\ref{eq:eta_4_be_killed_1} and Eq.\ref{eq:eta_4_be_killed_2} dictates that $\eta_4=1$. After the simplification, we have:
\begin{subequations}
\begin{align}
    \mathfrak{M}^3_c(\alpha)&=\tau_0, \label{eq: comp_1}\\
    \mathfrak{M}_a(\delta)\mathfrak{M}_a(\alpha)&=\eta_a\eta^{-1}_{3},\label{eq: comp_2}\\
    \mathfrak{M}_a(\gamma)\mathfrak{M}_a(\beta)&=\eta_a,\label{eq: comp_3}\\
    \mathfrak{M}_a(\beta)\mathfrak{M}_a(\gamma)&=\eta_a,\label{eq: comp_4}\\
    \mathfrak{M}_a(\alpha)\mathfrak{M}_a(\delta)&=\eta_a\eta^{-1}_{3}, \label{eq: comp_5}\\
    \mathfrak{M}_b(\gamma)\mathfrak{M}_b(\alpha)&=\eta_b\eta^{-1}_{3},\label{eq: comp_6}\\
    \mathfrak{M}_b(\delta)\mathfrak{M}_b(\beta)&=\eta_b,\label{eq: comp_7}\\
    \mathfrak{M}_b(\alpha)\mathfrak{M}_b(\gamma)&=\eta_b\eta^{-1}_{3},\label{eq: comp_8}\\
    \mathfrak{M}_b(\beta)\mathfrak{M}_b(\delta)&=\eta_b,\label{eq: comp_9}\\
    \mathfrak{M}^\dag_a(\beta)\mathfrak{M}^\dag_b(\alpha)\mathfrak{M}_a(\alpha)&=\eta^{-1}_3, \label{eq: comp_10}\\
\mathfrak{M}^\dag_a(\gamma)\mathfrak{M}^\dag_b(\beta)\mathfrak{M}_a(\beta)&=\tau_0, \label{eq: comp_11} \\
\mathfrak{M}^\dag_a(\alpha)\mathfrak{M}^\dag_c(\alpha)\mathfrak{M}^\dag_b(\gamma)\mathfrak{M}_a(\gamma)&=\tau_0, \label{eq: comp_12}\\
\mathfrak{M}^\dag_a(\delta)\mathfrak{M}^\dag_b(\delta)\mathfrak{M}_a(\delta)\mathfrak{M}_c(\alpha)&=\eta_3, \label{eq: comp_13}\\
\mathfrak{M}^\dag_b(\delta)\mathfrak{M}^\dag_a(\alpha)\mathfrak{M}_b(\alpha)\mathfrak{M}_a(\gamma)&=\eta_{ab}\eta_3, \label{eq: comp_14}\\
\mathfrak{M}^\dag_b(\gamma)\mathfrak{M}^\dag_a(\beta)\mathfrak{M}_b(\beta)\mathfrak{M}_a(\delta)&=\eta_{ab}\eta_3, \label{eq: comp_15}\\
\mathfrak{M}^\dag_b(\beta)\mathfrak{M}^\dag_a(\gamma)\mathfrak{M}_b(\gamma)\mathfrak{M}_a(\alpha)&=\eta_{ab}\eta_3, \label{eq: comp_16}\\
\mathfrak{M}^\dag_b(\alpha)\mathfrak{M}^\dag_a(\delta)\mathfrak{M}_b(\delta)\mathfrak{M}_a(\beta)&=\eta_{ab}\eta_3, \label{eq: comp_17}\\
\mathfrak{M}^\dag_b(\alpha)\mathfrak{M}_a(\alpha)\mathfrak{M}_b(\delta)\mathfrak{M}_b(\delta)&=\tau_0,\label{eq: comp_18}\\
\mathfrak{M}^\dag_b(\beta)\mathfrak{M}_a(\beta)\mathfrak{M}_b(\gamma)\mathfrak{M}_c(\alpha)\mathfrak{M}_b(\alpha)&=\eta^{-1}_{3},\label{eq: comp_19}\\
\mathfrak{M}^\dag_c(\alpha)\mathfrak{M}^\dag_b(\gamma)\mathfrak{M}_a(\gamma)\mathfrak{M}_b(\beta)\mathfrak{M}_b(\gamma)&=\eta_{3},\label{eq: comp_20}\\
\mathfrak{M}^\dag_b(\delta)\mathfrak{M}_a(\delta)\mathfrak{M}_b(\alpha)\mathfrak{M}_b(\beta)&=\tau_0. \label{eq: comp_21}
\end{align}
\end{subequations}
Let us first look at Eq.\ref{eq: comp_19}, which can be rewritten as:
\begin{equation}
\mathfrak{M}_b(\alpha)\mathfrak{M}^\dag_b(\beta)\mathfrak{M}_a(\beta)\mathfrak{M}_b(\gamma)\mathfrak{M}_c(\alpha)=\eta^{-1}_{3}. \nonumber
\end{equation}
The above expression, when combined with Eq.\ref{eq: comp_20} and Eq.\ref{eq: comp_4}, gives us:
\begin{equation}
\mathfrak{M}_b(\alpha)\mathfrak{M}_b(\gamma)=\eta^{-1}_a,
\end{equation}
that, when combined with Eq.\ref{eq: comp_8}, gives us:
\begin{equation}
\eta^{-1}_a=\eta_b\eta^{-1}_3.
\end{equation}
Now we look at Eq.\ref{eq: comp_15}, which can be rewritten as:
\begin{equation}
\mathfrak{M}_a(\delta)\mathfrak{M}^\dag_b(\gamma)\mathfrak{M}^\dag_a(\beta)\mathfrak{M}_b(\beta)=\eta_{ab}\eta_3. \nonumber
\end{equation}
The above expression, when combined with Eq.\ref{eq: comp_16} and Eq.\ref{eq: comp_4}, gives us:
\begin{equation}
\mathfrak{M}_a(\delta)\mathfrak{M}_a(\alpha)=\eta_a,
\end{equation}
that, when combined with Eq.\ref{eq: comp_2}, gives us:
\begin{equation}
\eta_3=1.
\end{equation}
Looking at Eq.\ref{eq: comp_17} and Eq.\ref{eq: comp_18}, with the help of Eq.\ref{eq: comp_2}, we reach:
\begin{equation}
\mathfrak{M}_a(\beta)=\eta_{ab}\eta_a\mathfrak{M}_b(\delta).\label{eq:173}
\end{equation}
Similarly, Eq.\ref{eq: comp_14} and Eq.\ref{eq: comp_21}, with the help of Eq.\ref{eq: comp_2}, give us:
\begin{equation}
\mathfrak{M}_a(\gamma)=\eta_{ab}\eta_a\mathfrak{M}_b(\beta).\label{eq:174}
\end{equation}
Consider, now, Eq.\ref{eq: comp_20} and Eq.\ref{eq: comp_12}. The coupled equations can be manoeuvred to give us:
\begin{equation}
\mathfrak{M}_a(\alpha)\mathfrak{M}_b(\beta)\mathfrak{M}_b(\gamma)=\tau_0.
\end{equation}
The above equation, when paired with Eq.\ref{eq: comp_10}, gives us (note that $\eta_a=\eta_b$ now):
\begin{equation}
\mathfrak{M}_b(\beta)\mathfrak{M}_a(\beta)=\eta^{-1}_a,
\end{equation}
which can be coupled with Eq.\ref{eq: comp_9} and Eq.\ref{eq:173} to give us:
\begin{equation}
\eta_{ab}=\eta_a.
\end{equation}
At this point, there is only one phase left in the problem: $\eta_a=\eta_{b}=\eta_{ab}$. Let us look at Eq.\ref{eq: comp_11}, which gives us:
\begin{equation}
\mathfrak{M}^3_b(\beta)=\eta^{-1}_a,
\end{equation}
which then implies that:
\begin{equation}
\mathfrak{M}^3_b(\delta)=\tau_0.
\end{equation}
We note that, at this point, there are four independent $\mathsf{SU}(2)$ matrices, denoted as follows:
\begin{equation}
\mathcal{A}\equiv \mathfrak{M}_c(\alpha), \mathcal{B}\equiv \mathfrak{M}_b(\delta), \mathcal{C}\equiv \mathfrak{M}_a(\alpha), \mathcal{D}\equiv \mathfrak{M}_b(\alpha). \label{eq:independence}
\end{equation}
More completely, the $\mathfrak{M}_a$s and $\mathfrak{M}_b$s are represented as follows:
\begin{align}
\mathfrak{M}_a(\alpha)&=\mathcal{C},\nonumber \\
\mathfrak{M}_a(\beta)&=\mathcal{B},\nonumber \\
\mathfrak{M}_a(\gamma)&=\eta_a \mathcal{B}^\dag,\nonumber \\
\mathfrak{M}_a(\delta)&= \eta_a \mathcal{C}^\dag;
\end{align}
and 
\begin{align}
\mathfrak{M}_b(\alpha)&=\mathcal{D},\nonumber \\
\mathfrak{M}_b(\beta)&=\eta_a \mathcal{B}^\dag,\nonumber \\
\mathfrak{M}_b(\gamma)&=\eta_a \mathcal{D}^\dag,\nonumber \\
\mathfrak{M}_b(\delta)&=\mathcal{B}.
\end{align}
There are, in fact, only five independent constraints for these matrices:
\begin{subequations}
\begin{align}
&\mathcal{A}^3=\tau_0,\label{eq: yi-final:1}\\
&\mathcal{B}^3=\tau_0,\label{eq: yi-final:2}\\
&\mathcal{A}=\mathcal{CBC}^\dag,\label{eq: yi-final:3}\\
&\mathcal{A}=\mathcal{DBD}^\dag,\label{eq: yi-final:4}\\
&\mathcal{A}=\mathcal{DB^\dag C^\dag}.\label{eq: yi-final:5}
\end{align}
\end{subequations}
However, note that $\mathcal{C}=\tau_0$ from Eq.~\ref{eq:z2maalphagauge}, we immediately have:
\begin{equation}
    \mathcal{A} = \mathcal{B}, \quad \mathcal{D}=\mathcal{A}^\dag.
\end{equation}
In summary:\\
1.) $G_x=G_y=G_z=\tau_0$;\\
2.) $G_{a/b/c}(x,y,z;s)=\mathfrak{M}_{a/b/c}(s)$, where the $\mathfrak{M}$s have the following forms:
\begin{align}
&\mathfrak{M}_a(\alpha)=\tau_0, \mathfrak{M}_a(\beta)=\mathcal{A}, \mathfrak{M}_a(\gamma)=\eta_a \mathcal{A}^\dag, \mathfrak{M}_a(\delta)=\eta_a \tau_0;\nonumber \\
&\mathfrak{M}_b(\alpha)=\mathcal{A}^\dag, \mathfrak{M}_b(\beta)=\eta_a \mathcal{A}^\dag, \mathfrak{M}_b(\gamma)=\eta_a \mathcal{A}, \mathfrak{M}_b(\delta)=\mathcal{A};\nonumber \\
&\mathfrak{M}_c(\alpha)=\mathcal{A}, \mathfrak{M}_c(\beta)=\tau_0, \mathfrak{M}_c(\gamma)=\tau_0, \mathfrak{M}_c(\delta)=\tau_0.
\end{align}
Since we have $\eta_a=\pm 1$, and $\mathcal{A}=\tau_0, e^{i\frac{2\pi}{3}\tau_z}, e^{i\frac{4\pi}{3}\tau_z}$, corresponding to apparently $6$ solutions. However, let us note that a further gauge transformation $W(x,y,z;s)\equiv \eta_a^{(x+y+z)}$ leads us to:\\
1.) $G_x=G_y=G_z= \eta_a\tau_0$;\\
2.) $G_{a/b/c}(x,y,z;s)=\mathfrak{M}_{a/b/c}(s)$, where the $\mathfrak{M}$s have the following forms:
\begin{align}
&\mathfrak{M}_a(\alpha)=\tau_0, \mathfrak{M}_a(\beta)=\mathcal{A}, \mathfrak{M}_a(\gamma)=\mathcal{A}^\dag, \mathfrak{M}_a(\delta)=\tau_0;\nonumber \\
&\mathfrak{M}_b(\alpha)=\mathcal{A}^\dag, \mathfrak{M}_b(\beta)=\mathcal{A}^\dag, \mathfrak{M}_b(\gamma)=\mathcal{A}, \mathfrak{M}_b(\delta)=\mathcal{A};\nonumber \\
&\mathfrak{M}_c(\alpha)=\mathcal{A}, \mathfrak{M}_c(\beta)=\tau_0, \mathfrak{M}_c(\gamma)=\tau_0, \mathfrak{M}_c(\delta)=\tau_0.
\end{align}
Due to the fact that $\eta_a$ now becomes global signs which are elements of the IGG, we conclude that $\eta_a=\pm 1$ PSGs are equivalent. We also remark that a gauge transformation $W(x,y,z;s)\equiv i\tau_x$ maps the PSG solutions in which $\mathcal{A}=e^{i\frac{2\pi}{3}\tau_z}$ to that in which $\mathcal{A}=e^{i\frac{4\pi}{3}\tau_z}$.\par
In conclusion, we have:\\
1.) $G_x=G_y=G_z= \tau_0$;\\
2.) $G_{a/b/c}(x,y,z;s)=\mathfrak{M}_{a/b/c}(s)$, where the $\mathfrak{M}$s have the following forms:
\begin{align}
&\mathfrak{M}_a(\alpha)=\tau_0, \mathfrak{M}_a(\beta)=\mathcal{A}, \mathfrak{M}_a(\gamma)=\mathcal{A}^\dag, \mathfrak{M}_a(\delta)=\tau_0;\nonumber \\
&\mathfrak{M}_b(\alpha)=\mathcal{A}^\dag, \mathfrak{M}_b(\beta)=\mathcal{A}^\dag, \mathfrak{M}_b(\gamma)=\mathcal{A}, \mathfrak{M}_b(\delta)=\mathcal{A};\nonumber \\
&\mathfrak{M}_c(\alpha)=\mathcal{A}, \mathfrak{M}_c(\beta)=\tau_0, \mathfrak{M}_c(\gamma)=\tau_0, \mathfrak{M}_c(\delta)=\tau_0,
\end{align}
where $\mathcal{A}=\tau_0, e^{i\frac{2\pi}{3}\tau_z}$, corresponding to $2$ solutions.

\subsection{Adding Time-Reversal Symmetry}
Having arrived at the PSG solutions given the space group for the trillium lattice, we are at a position to add time-reversal symmetry (TRS) to the story. 
The extra relations are translated into the corresponding PSG equations:
\begin{subequations}
\begin{align}
& G_{\mathcal{T}}(i)G_{\mathcal{T}}(i)=\eta_{\mathcal{T}}, \label{psg:19}\\
& G^\dag_{\mathcal{T}}(T^{-1}_x(i))G^\dag_x(i)G_{\mathcal{T}}(i)G_x(i)=\eta_{x\mathcal{T}}, \label{psg:20}\\
& G^\dag_{\mathcal{T}}(T^{-1}_y(i))G^\dag_y(i)G_{\mathcal{T}}(i)G_y(i)=\eta_{y\mathcal{T}}, \label{psg:21}\\
& G^\dag_{\mathcal{T}}(T^{-1}_z(i))G^\dag_z(i)G_{\mathcal{T}}(i)G_z(i)=\eta_{z\mathcal{T}}, \label{psg:22}\\
& G^\dag_{\mathcal{T}}(g^{-1}_a(i))G^\dag_a(i)G_{\mathcal{T}}(i)G_a(i)=\eta_{a\mathcal{T}}, \label{psg:23}\\
& G^\dag_{\mathcal{T}}(g^{-1}_b(i))G^\dag_b(i)G_{\mathcal{T}}(i)G_b(i)=\eta_{b\mathcal{T}}, \label{psg:24}\\
& G^\dag_{\mathcal{T}}(g^{-1}_c(i))G^\dag_c(i)G_{\mathcal{T}}(i)G_c(i)=\eta_{c\mathcal{T}}. \label{psg:25}
\end{align}
\end{subequations}
We make the ansatz that $G_{\mathcal{T}}\equiv f(x,y,z; s)\mathfrak{M}_{\mathcal{T}}(s)$, and Eq.\ref{psg:20}, Eq.\ref{psg:21} and Eq.\ref{psg:22} immediately tell us that:
\begin{equation}
G_{\mathcal{T}}(i)=\eta^x_{x\mathcal{T}}\eta^y_{y\mathcal{T}}\eta^z_{z\mathcal{T}}\mathfrak{M}_{\mathcal{T}}(s).
\end{equation}
The above form, when plugged into Eq.\ref{psg:19}, gives us the following constraint:
\begin{equation}
\mathfrak{M}^2_{\mathcal{T}}(s)=\eta_{\mathcal{T}}.
\end{equation}
We now discuss about the consequences of Eq.\ref{psg:23}, Eq.\ref{psg:24} and Eq.\ref{psg:25}.
\subsubsection{Solving Eq.\ref{psg:25}}
First, we consider $i\equiv(x,y,z;\beta)$, since in this case $G_c(i)=\tau_0$. It is straightforward to reach the following constraint on the phases:
\begin{equation}
\eta_{x\mathcal{T}}=\eta_{y\mathcal{T}}=\eta_{z\mathcal{T}}\equiv \eta_5.
\end{equation}
Also the following constraints for $\mathfrak{M}_{\mathcal{T}}(s)$ arise if we iterate the sub-lattice indices:
\begin{subequations}
\begin{align}
&\mathfrak{M}^\dag_{\mathcal{T}}(\delta)\mathfrak{M}_{\mathcal{T}}(\beta)=\eta_{c\mathcal{T}}, \label{eq: killct1}\\
&\mathfrak{M}^\dag_{\mathcal{T}}(\beta)\mathfrak{M}_{\mathcal{T}}(\gamma)=\eta_{c\mathcal{T}}, \label{eq: killct2}\\
&\mathfrak{M}^\dag_{\mathcal{T}}(\gamma)\mathfrak{M}_{\mathcal{T}}(\delta)=\eta_{c\mathcal{T}}, \label{eq: killct3}\\
&\mathfrak{M}^\dag_{\mathcal{T}}(\alpha)\mathfrak{M}^\dag_{c}(\alpha)\mathfrak{M}_{\mathcal{T}}(\alpha)\mathfrak{M}_{c}(\alpha)=\eta_{c\mathcal{T}}.
\end{align}
\end{subequations}
From Eq.\ref{eq: killct1} and Eq.\ref{eq: killct2}, we can reach $\mathfrak{M}^\dag_{\mathcal{T}}(\delta)\mathfrak{M}_{\mathcal{T}}(\gamma)=\tau_0$. This statement, when coupled with Eq.\ref{eq: killct3} gives us $\eta_{c\mathcal{T}}=1$.\\
A quick summary:\\
1.) $\eta_{x\mathcal{T}}=\eta_{y\mathcal{T}}=\eta_{z\mathcal{T}}\equiv \eta_5$, $\eta_{c\mathcal{T}}=1$;\\
2.) We have $G_{\mathcal{T}}(x,y,z;s)=\eta^{x+y+z}_5\mathfrak{M}_{\mathcal{T}}(s)$; for which the following relations are satisfied:
\begin{align}
&\mathfrak{M}^\dag_{\mathcal{T}}(\alpha)\mathfrak{M}^\dag_{c}(\alpha)\mathfrak{M}_{\mathcal{T}}(\alpha)\mathfrak{M}_{c}(\alpha)=\tau_0,\\
&\mathfrak{M}_{\mathcal{T}}(\beta)=\mathfrak{M}_{\mathcal{T}}(\gamma)=\mathfrak{M}_{\mathcal{T}}(\delta).
\end{align}
\subsubsection{Solving Eq.\ref{psg:23} and Eq.\ref{psg:24}}
Iterating the sub-lattice indices, we arrive at the following constraints:
\begin{subequations}
\begin{align}
\mathfrak{M}^\dag_{\mathcal{T}}(\delta)\mathfrak{M}^\dag_a(\alpha)\mathfrak{M}_{\mathcal{T}}(\alpha)\mathfrak{M}_a(\alpha)&=\eta_{a\mathcal{T}},\\
\mathfrak{M}^\dag_{\mathcal{T}}(\gamma)\mathfrak{M}^\dag_a(\beta)\mathfrak{M}_{\mathcal{T}}(\beta)\mathfrak{M}_a(\beta)&=\eta_{a\mathcal{T}}, \label{eq: killeta5_1}\\
\mathfrak{M}^\dag_{\mathcal{T}}(\beta)\mathfrak{M}^\dag_a(\gamma)\mathfrak{M}_{\mathcal{T}}(\gamma)\mathfrak{M}_a(\gamma)&=\eta_{a\mathcal{T}}\eta_5,\label{eq: killeta5_2}\\
\mathfrak{M}^\dag_{\mathcal{T}}(\alpha)\mathfrak{M}^\dag_a(\delta)\mathfrak{M}_{\mathcal{T}}(\delta)\mathfrak{M}_a(\delta)&=\eta_{a\mathcal{T}}\eta_5,\\
\mathfrak{M}^\dag_{\mathcal{T}}(\gamma)\mathfrak{M}^\dag_b(\alpha)\mathfrak{M}_{\mathcal{T}}(\alpha)\mathfrak{M}_b(\alpha)&=\eta_{b\mathcal{T}},\\
\mathfrak{M}^\dag_{\mathcal{T}}(\delta)\mathfrak{M}^\dag_b(\beta)\mathfrak{M}_{\mathcal{T}}(\beta)\mathfrak{M}_b(\beta)&=\eta_{b\mathcal{T}}\eta_5,\\
\mathfrak{M}^\dag_{\mathcal{T}}(\alpha)\mathfrak{M}^\dag_b(\gamma)\mathfrak{M}_{\mathcal{T}}(\gamma)\mathfrak{M}_b(\gamma)&=\eta_{b\mathcal{T}}\eta_5,\\
\mathfrak{M}^\dag_{\mathcal{T}}(\beta)\mathfrak{M}^\dag_b(\delta)\mathfrak{M}_{\mathcal{T}}(\delta)\mathfrak{M}_b(\delta)&=\eta_{b\mathcal{T}}. \label{eq: birth_of_eta_6}
\end{align}
\end{subequations}
\subsubsection{Collection of Constraints}
We further specify that $\mathfrak{M}_{\mathcal{T}}(\alpha)\equiv \mathcal{E}$ and $\mathfrak{M}_{\mathcal{T}}(\beta)=\mathfrak{M}_{\mathcal{T}}(\gamma)=\mathfrak{M}_{\mathcal{T}}(\delta)\equiv \mathcal{F}$. We note that Eq.\ref{eq: killeta5_1} and Eq.\ref{eq: killeta5_2} immediately imply that $\eta_5=1$ since $\mathfrak{M}_a(\gamma)=\eta_a\mathfrak{M}^\dag_a(\beta)$. Furthermore, comparing Eq.\ref{eq: killeta5_1} and Eq.\ref{eq: birth_of_eta_6} gives us $\eta_{a\mathcal{T}}=\eta_{b\mathcal{T}}\equiv \eta_6$, since $\mathfrak{M}_a(\beta)=\mathfrak{M}_b(\delta)$.\par
In the end, we reach the following five independent constraints:
\begin{subequations}
\begin{align}
&\mathcal{E}^2=\eta_{\mathcal{T}},\label{eq: er-final:1}\\
&\mathcal{F}^2=\eta_{\mathcal{T}},\label{eq: er-final:2}\\
&\mathcal{E^\dag A^\dag EA}=\tau_0,\label{eq: er-final:3}\\
&\mathcal{F^\dag B^\dag FB}=\eta_6,\label{eq: er-final:4}\\
&\mathcal{C^\dag E^\dag CF}=\eta_6.\label{eq: er-final:5}
\end{align}
\end{subequations}
Since $\mathcal{C}=\tau_0$, and $\mathcal{A}=\mathcal{B}$, we can determine that $\eta_6=1$ and $\mathcal{E}=\mathcal{F}$. Also, when $\eta_{\mathcal{T}}=1$, we have $\mathcal{E}=\mathcal{F}=\tau_0$; when $\eta_{\mathcal{T}}=-1$, we have $\mathcal{E}=\mathcal{F}=i\tau_z$. Since without TRS, we had $2$ solutions, now we have $4$ solutions, as collected in Tab.~\ref{table:psg_sol}.

\section{$\text{IGG}=\mathsf{U}(1)$} \label{sec: u1psg}
In this section, our target is to find the PSG solutions with $\mathsf{U}(1)$ IGG. The PSG relations listed in Section \ref{sec: z2psg} still hold, only with the signs on the RHS being replaced as $\eta_{\mathfrak{g}}\equiv \exp[i\phi_{\mathfrak{g}}]$. \par
To proceed, we mention a fact which is a blessing for us. In \cite{Wen2001}, Wen proved that for PSG solutions with $\mathsf{U}(1)$ IGG, the $G$s always have the following canonical forms:
\begin{equation}
G_g(i)\equiv (i\tau_x)^{n_g}e^{i\theta_g(i)\tau_z}, \quad n_g=0 \text{ or }1,
\end{equation}
where $\theta_g\in [0, 2\pi)$. Another thing we would like to mention before moving on is that, in this section $\theta$ always stands for \emph{a function} which depends on position $i$, whereas $\phi$ always stands for \emph{a constant phase}.\par
Let us first look at Eq.\ref{psg:2}, which can be rewritten as:
\begin{equation}
G_a(g_a(i))G_a(i)=G_z(g_a(i))e^{i\phi_a\tau_z}. \label{eq:u1demo}
\end{equation}
The above equation already dictates that $n_z=0$. Why? This is straightforward to see if $n_a=0$. Now supposing $n_a=1$, we would have:
\begin{equation}
\mathrm{LHS}_{\ref{eq:u1demo}}=e^{i(-\theta_a(g_a(i))+\theta_a(i))\tau_z},
\end{equation}
which also implies that $n_z=0$ on the $\mathrm{RHS}_{\ref{eq:u1demo}}$. Similarly, due to Eq.\ref{psg:1} and Eq.\ref{psg:3}, we can conclude that $n_y=n_c=0$.\par
There is a valuable lesson from the above operation. Given a PSG equation $G_{\mathfrak{g}}G^\dag_{\mathfrak{h}}\dots=e^{i\phi_{\mathfrak{i}}\tau_z}$, we demand that $(n_{\mathfrak{g}}-n_{\mathfrak{h}}\dots )=0 \text{ mod }2$. Using the above lesson, we see from Eq.\ref{psg:17} that $n_x=0$. And Eq.\ref{psg:16} tells us that $n_b=0$, whereas Eq.\ref{psg:18} implies that $n_a=0$. This is remarkable, for $n_x=n_y=n_z=n_a=n_b=n_c=0$! What was for us originally a set of coupled $\mathsf{SU}(2)$ matrix equations now reduces to a set of coupled $\mathsf{U}(1)$ matrix equations, which are equations of compact $\mathsf{U}(1)$ phases.
\subsection{Solving for the Translational Elements and the Simplification}
Before we take a step further, let us rewrite the remaining PSG equations in terms of $\theta$s:
\begin{subequations}
\begin{align}
&\theta_c(g^2_c(i))+\theta_c(g_c(i))+\theta_c(i)=\phi_c, \label{u1psg:1}\\
&-\theta_z(g_a(i))+\theta_a(g_a(i))+\theta_a(i)=\phi_a, \label{u1psg:2}\\
&-\theta_y(g_b(i))+\theta_b(g_b(i))+\theta_b(i)=\phi_b, \label{u1psg:3}\\
&-\theta_a(T_x(i))+\theta_x(T_x(i))+\theta_a(i)+\theta_x(g^{-1}_a(i))=\phi_{ax}, \label{u1psg:4}\\
&-\theta_a(T_y(i))+\theta_y(T_y(i))+\theta_a(i)+\theta_y(g^{-1}_a(i))=\phi_{ay}, \label{u1psg:5}\\
&-\theta_a(T^{-1}_z(i))-\theta_z(i)+\theta_a(i)+\theta_z(g^{-1}_a(i))=\phi_{az}, \label{u1psg:6}\\
&-\theta_b(T_x(i))+\theta_x(T_x(i))+\theta_b(i)+\theta_x(g^{-1}_b(i))=\phi_{bx}, \label{u1psg:7}\\
&-\theta_b(T^{-1}_y(i))-\theta_y(i)+\theta_b(i)+\theta_y(g^{-1}_b(i))=\phi_{by}, \label{u1psg:8}\\
&-\theta_b(T_z(i))+\theta_z(T_z(i))+\theta_b(i)+\theta_z(g^{-1}_b(i))=\phi_{bz}, \label{u1psg:9}\\
&-\theta_c(T^{-1}_y(i))-\theta_y(i)+\theta_c(i)+\theta_x(g^{-1}_c(i))=\phi_{cyx}, \label{u1psg:10}\\
&-\theta_c(T^{-1}_z(i))-\theta_z(i)+\theta_c(i)+\theta_y(g^{-1}_c(i))=\phi_{czy}, \label{u1psg:11}\\
&-\theta_c(T^{-1}_x(i))-\theta_x(i)+\theta_c(i)+\theta_z(g^{-1}_c(i))=\phi_{cxz}, \label{u1psg:12}\\
&-\theta_a(g^{-1}_cg^{-1}_bT^{-1}_xT_yg_ag_c(i))-\theta_c(g^{-1}_bT^{-1}_xT_yg_ag_c(i)) \nonumber \\
&-\theta_b(T^{-1}_xT_yg_ag_c(i))-\theta_x(T_yg_ag_c(i))\nonumber \\
&+\theta_y(T_yg_ag_c(i))+\theta_a(g_ag_c(i)) +\theta_c(g_c(i))=\phi_{acb},\label{u1psg:13}\\
&-\theta_b(g^{-1}_aT_xT^{-1}_yT_zg_bg_a(i))-\theta_a(T_xT^{-1}_yT_zg_bg_a(i))\nonumber \\
&+\theta_x(T_xT^{-1}_yT_zg_bg_a(i))-\theta_y(T_zg_bg_a(i))\nonumber \\
&+ \theta_z(T_zg_bg_a(i))+\theta_b(g_bg_a(i))+\theta_a(g_a(i))=\phi_{ab},\label{u1psg:14}\\
& -\theta_c(g^{-1}_bT^{-1}_xT_yg_ag_bg_cg_b(i))-\theta_b(T^{-1}_xT_yg_ag_bg_cg_b(i)) \nonumber \\
&-\theta_x(T_yg_ag_bg_cg_b(i))+ \theta_y(T_yg_ag_bg_cg_b(i))\nonumber \\
&+\theta_a(g_ag_bg_cg_b(i))+\theta_b(g_bg_cg_b(i))\nonumber \\
&+\theta_c(g_cg_b(i))+\theta_b(g_b(i))=\phi_{cba}.\label{u1psg:15}
\end{align}
\end{subequations}
These $\theta$s and $\phi$s in the above equations are compact $\mathsf{U}(1)$ phase factors, and an equation $\theta = \phi$ means $\theta = \phi \text{ mod }2\pi$. The $\theta$s are associated with the $\mathsf{SU}(2)$ gauge symmetry like before, specifically the $\mathsf{U}(1)$ subgroup of $\mathsf{SU}(2)$ transforms the $\theta$ in the following way:
\begin{equation}
\theta_U(i)\mapsto \theta_U(i)-\varphi(i)+\varphi(U^{-1}(i)), \quad \varphi\in [0,2\pi);
\end{equation}
note that since we do not want to spoil the choice of $\tau_z$, we consider only the $\mathsf{U}(1)$ subgroup of $\mathsf{SU}(2)$.\par
Similar to the $\mathsf{Z}_2$ case, we eliminate the phases on the right hand side of Eq.~\ref{u1psg:1}, Eq.~\ref{u1psg:10}, Eq.~\ref{u1psg:11}, Eq.~\ref{u1psg:13} and Eq.~\ref{u1psg:15}.

\subsection{Solving for the translational Elements}
Let us start by considering the equations that arise because of the commutation of translational generators. Canonically, this gives us the following expressions of $G_x, G_y, G_z$ after a gauge fixing:
\begin{align}
&G_x(x,y,z;s)=\tau_0, G_y(x,y,z;s)=e^{ix\phi_{xy}\tau_z}, \nonumber \\
&G_z(x,y,z;s)=e^{i(x\phi_{zx}+y\phi_{yz})\tau_z}.
\end{align}
In other words, we have the following representation:
\begin{align}
&\theta_x(x,y,z;s)=0, \theta_y(x,y,z;s)=x\phi_{xy}, \nonumber \\
&\theta_z(x,y,z;s)=x\phi_{zx}+y\phi_{yz}. 
\end{align}

\subsection{Solving for $\theta_c$}
Using the IGG gauge symmetry, one can eliminate the phases on the RHS of Eq.\ref{u1psg:10}, Eq.\ref{u1psg:11} and Eq.\ref{u1psg:12}, as each of $\theta_x$, $\theta_y$ and $\theta_z$ appears only once in these equations. To solve for $\theta_c$, one then plug the canonical expressions of the translational PSG elements into Eq.\ref{u1psg:10}, Eq.\ref{u1psg:11} and Eq.\ref{u1psg:12}. We make an ansatz analogous to the one we made in the $\mathsf{Z}_2$ case: $\theta(i)\equiv f(x,y,z;s)+\mathfrak{m}(s)$. We realise that the equations under attention are valid for all sub-lattice indices, therefore we have $f_c(x,y,z;s)\equiv f_c(x,y,z)$, and:
\begin{subequations}
\begin{align}
f_c(x,y,z)&=f_c(x,y-1,z)+x\phi_{xy}, \\
f_c(x,y,z)&=f_c(x,y,z-1) \nonumber \\
&+x\phi_{zx}+y\phi_{yz}-y\phi_{xy}, \\
f_c(x,y,z)&=f_c(x-1,y,z) \nonumber \\
&-y\phi_{zx}-z\phi_{yz} + \phi_{cxz}. 
\end{align}
\end{subequations}
Checking the path-independency of $f_c$, we arrive at the following constraint:
\begin{equation}
\phi_{xy}=\phi_{yz}=-\phi_{zx}\equiv \phi_1.
\end{equation}
Eventually we arrive at the conclusion that $f_c(x,y,z)=(xy-xz)\phi_1$.\par
Let us look at Eq.\ref{u1psg:1}. We had eliminated the phase on the RHS by making use of the IGG gauge symmetry. Plugging the above expression into Eq.\ref{u1psg:1}, we arrive at:
\begin{subequations}
\begin{align}
3\mathfrak{m}_c(\alpha)&=0,\\
\mathfrak{m}_c(\beta)+\mathfrak{m}_c(\gamma)+\mathfrak{m}_c(\delta)&=0, \label{u1:bcdeq0}  \\
\phi_{cxz} & = 0.
\end{align}
\end{subequations}
\par
Before moving on, we make a brief summary:
\begin{align}
&\theta_x(i)=0, \theta_y(i)=x\phi_1, \theta_z(i)=(y-x)\phi_1, \nonumber \\
&\theta_c(i)=(xy-xz)\phi_1+\mathfrak{m}_c(s). 
\end{align}
\subsection{Solving for $\theta_a$}
To solve for $\theta_a$, one plugs the simplified expressions of the translational PSG elements into Eq.\ref{u1psg:4}, Eq.\ref{u1psg:5} and Eq.\ref{u1psg:6}. One arrives at the following expressions:
\begin{subequations}
\begin{align}
\theta_a(T_x(i))&=\theta_a(i)-\phi_{ax},\\
\theta_a(T_y(i))&=\theta_a(i)+\theta_y(T_y(i)) \nonumber \\
&+\theta_y(g^{-1}_a(i))-\phi_{ay},\\
\theta_a(i)&=\theta_a(T^{-1}_z(i))+\phi_{az} \nonumber \\
&+\theta_z(i)-\theta_z(g^{-1}_a(i)).
\end{align}
\end{subequations}
One makes the usual ansatz $\theta_a\equiv f_a(x,y,z;s)+\mathfrak{m}_a(s)$, only this time one does not have $f_a(x,y,z;s)=f_a(x,y,z)$, for the evaluation of $\theta_{y/z}(g^{-1}_a(i))$ is not $s$-independent.\\
We have, for $s=\alpha/\delta$, the following conditions for $f_a$:
\begin{subequations}
\begin{align}
f_a(x+1,y,z;\alpha/\delta)&=f_a(x,y,z;\alpha/\delta)-\phi_{ax},\\
f_a(x,y+1,z;\alpha/\delta)&=f_a(x,y,z;\alpha/\delta)-\phi_{ay},\\
f_a(x,y,z;\alpha/\delta)&=f_a(x,y,z-1;\alpha/\delta) \nonumber \\
&+\phi_{az}+(2y-2x+1)\phi_1;
\end{align}
\end{subequations}
and for $s=\beta/\gamma$, the following conditions for $f_a$:
\begin{subequations}
\begin{align}
f_a(x+1,y,z;\beta/\gamma)&=f_a(x,y,z;\beta/\gamma)-\phi_{ax},\\
f_a(x,y+1,z;\beta/\gamma)&=f_a(x,y,z;\beta/\gamma)-\phi_{ay}-\phi_1,\\
f_a(x,y,z;\beta/\gamma)&=f_a(x,y,z-1;\beta/\gamma)+\phi_{az} \nonumber \\
&+(2y-2x)\phi_1.
\end{align}
\end{subequations}
Checking the path-independency of $f_a$, we arrive at the following constraint:
\begin{equation}
2\phi_1=0\Rightarrow  \phi_1=0 \text{ or }\pi.
\end{equation}
After the path-independency is guaranteed, we arrive at the following expressions:
\begin{subequations}
\begin{align}
f_a(x,y,z;\alpha/\delta)&=-x\phi_{ax}-y\phi_{ay}+z(\phi_{az}+\phi_1),\\
f_a(x,y,z;\beta/\gamma)&=-x\phi_{ax}-y(\phi_{ay}+\phi_1)+z\phi_{az}.
\end{align}
\end{subequations}
Plugging the forms of $\theta_a\equiv f_a(x,y,z;s)+\mathfrak{m}_a(s)$ into Eq.\ref{u1psg:2}, further constraints can be derived. Specifically, we iterate the sub-lattice index. Let us consider $i\equiv (x,y,z;\alpha)$, we have:
\begin{equation}
    -\theta_{z}(-x,-y-1,z;\delta) + \theta_a(-x,-y-1,z;\delta) + \theta_a(x,y,z;\alpha) = \phi_a,
\end{equation}
which gives us:
\begin{align}
    \phi_a & = -(-y-1+x)\phi_1 + (x\phi_{ax} + (y+1)\phi_{ay} \nonumber \\
    &+ z(\phi_{az}+\phi_1))+(-x\phi_{ax} - y\phi_{ay} \nonumber \\
    &+ z(\phi_{az}+\phi_1)) + \mathfrak{m}_a(\alpha) + \mathfrak{m}_a(\delta) \nonumber \\
    & = -(-y-1+x)\phi_1 + \phi_{ay} + 2z\phi_{az} \nonumber \\
    & + \mathfrak{m}_a(\alpha) + \mathfrak{m}_a(\delta).
\end{align}
The implication from the above equation is that:
\begin{equation}
    \phi_1 = 0, \quad 2\phi_{az}=0; \quad \mathfrak{m}_a(\alpha) + \mathfrak{m}_a(\delta) = \phi_a - \phi_{ay}.
\end{equation}
For $i\equiv (x,y,z;\beta)$, we have:
\begin{equation}
\theta_a(-x-1,-y-1,z+1;\gamma) + \theta_a(x,y,z;\beta) = \phi_a,
\end{equation}
which gives us:
\begin{align}
    \phi_a & = ((x+1)\phi_{ax} + (y+1)\phi_{ay} + (z+1)\phi_{az}) \nonumber \\
    &+(-x\phi_{ax} - y\phi_{ay} + z\phi_{az}) \nonumber \\
    &+ \mathfrak{m}_a(\beta) + \mathfrak{m}_a(\gamma) \nonumber \\
    & = \phi_{ax} + \phi_{ay} + \phi_{az} + \mathfrak{m}_a(\beta) + \mathfrak{m}_a(\gamma).
\end{align}
For $i\equiv (x,y,z;\gamma)$, we have:
\begin{equation}
\theta_a(-x-1,-y-1,z;\beta) + \theta_a(x,y,z;\gamma) = \phi_a,
\end{equation}
which gives us:
\begin{align}
    \phi_a & = ((x+1)\phi_{ax} + (y+1)\phi_{ay} + z\phi_{az}) \nonumber \\
    &+(-x\phi_{ax} - y\phi_{ay} + z\phi_{az}) \nonumber \\
    &+ \mathfrak{m}_a(\beta) + \mathfrak{m}_a(\gamma) \nonumber \\
    & = \phi_{ax} + \phi_{ay}+ \mathfrak{m}_a(\beta) + \mathfrak{m}_a(\gamma).
\end{align}
Combining with the equation for $i\equiv (x,y,z;\beta)$, we have:
\begin{equation}
    \phi_{az}=0; \quad \mathfrak{m}_a(\beta) + \mathfrak{m}_a(\gamma) = \phi_a - \phi_{ay} - \phi_{ax}.
\end{equation}
Lastly, the case for $i\equiv (x,y,z;\delta)$ does not give us new relations.\\
Before moving on, we make a brief summary:
\begin{equation}
\theta_{x/y/z}(i)=0, \theta_a(i)= -x\phi_{ax} - y\phi_{ay} +\mathfrak{m}_a(s). 
\end{equation}

\subsection{Solving for $\theta_b$}
To solve for $\theta_b$, one plugs the simplified expressions of the translational PSG elements into Eq.\ref{u1psg:7}, Eq.\ref{u1psg:8} and Eq.\ref{u1psg:9}. One arrives at the following expressions:
\begin{align}
\theta_b(T_x(i))&=\theta_b(i)-\phi_{bx},\nonumber \\
\theta_b(i)&=\theta_b(T^{-1}_y(i))+\phi_{by}, \nonumber \\
\theta_b(T_z(i))&=\theta_b(i)-\phi_{bz}.
\end{align}
One makes the usual ansatz $\theta_b\equiv f_b(x,y,z;s)+\mathfrak{m}_b(s)$, we arrive at
\begin{equation}
f_b(x,y,z;s)=-x\phi_{bx} + y \phi_{by} - z\phi_{bz}.
\end{equation}
Plugging the forms of $\theta_b\equiv f_b(x,y,z;s)+\mathfrak{m}_b(s)$ into Eq.\ref{u1psg:3}, further constraints can be derived. Specifically, we iterate the sub-lattice index. Let us consider $i\equiv (x,y,z;\alpha)$, we have:
\begin{equation}
    \theta_b(-x-1,y,-z;\gamma) + \theta_b(x,y,z;\alpha) = \phi_b,
\end{equation}
which gives us:
\begin{align}
    \phi_b & =  ((x+1)\phi_{bx} +y\phi_{by} + z\phi_{bz}) \nonumber \\
    &+ (-x\phi_{bx} + y \phi_{by} - z\phi_{bz}) \nonumber \\
    &+ \mathfrak{m}_b(\alpha) + \mathfrak{m}_b(\gamma) \nonumber \\
    & = \phi_{bx} + 2y\phi_{by} + \mathfrak{m}_b(\alpha) + \mathfrak{m}_b(\gamma).
\end{align}
The implication from the above equation is that:
\begin{equation}
    2\phi_{by}=0; \quad \mathfrak{m}_b(\alpha) + \mathfrak{m}_b(\gamma) = \phi_b - \phi_{bx}.
\end{equation}
For $i\equiv (x,y,z;\beta)$, we have:
\begin{equation}
\theta_b(-x-1,y,-z-1;\delta) + \theta_b(x,y,z;\beta) = \phi_b,
\end{equation}
which gives us:
\begin{align}
    \phi_b & =((x+1)\phi_{bx} +y\phi_{by} + (z+1)\phi_{bz}) \nonumber \\
    &+ (-x\phi_{bx} + y \phi_{by} - z\phi_{bz}) \nonumber \\
    &+ \mathfrak{m}_b(\beta) + \mathfrak{m}_b(\delta) \nonumber \\
    & = \phi_{bx} + \phi_{bz} + \mathfrak{m}_b(\beta) + \mathfrak{m}_b(\delta).
\end{align}
For $i\equiv (x,y,z;\gamma)$, we have:
\begin{equation}
\theta_b(-x-1,y+1,-z;\alpha) + \theta_b(x,y,z;\gamma) = \phi_b,
\end{equation}
which gives us:
\begin{align}
   \phi_b & =((x+1)\phi_{bx} +(y+1)\phi_{by} + z\phi_{bz}) \nonumber \\
   &+ (-x\phi_{bx} + y \phi_{by} - z\phi_{bz}) \nonumber \\
   &+ \mathfrak{m}_b(\alpha) + \mathfrak{m}_b(\gamma) \nonumber \\
    & = \phi_{bx} + \phi_{by} + \mathfrak{m}_b(\alpha) + \mathfrak{m}_b(\gamma).
\end{align}
Combining with the equation for $i\equiv (x,y,z;\alpha)$, we have:
\begin{equation}
    \phi_{by}=0; \quad \mathfrak{m}_b(\alpha) + \mathfrak{m}_b(\gamma) = \phi_b - \phi_{bx}.
\end{equation}
Lastly, the case for $i\equiv (x,y,z;\delta)$ gives us the same relation from the case for $i\equiv (x,y,z;\beta)$, which is:
\begin{equation}
    \mathfrak{m}_b(\beta) + \mathfrak{m}_b(\delta) = \phi_b - \phi_{bx} - \phi_{bz}.
\end{equation}
Before moving on, we make a brief summary:
\begin{equation}
\theta_b(i)= -x\phi_{bx}- z\phi_{bz} +\mathfrak{m}_b(s). 
\end{equation}

\subsection{Solving Eq.~\ref{u1psg:13}}
The equation Eq.~\ref{u1psg:13} is then reduced to:
\begin{align}
    &-\theta_a(g^{-1}_c g^{-1}_b T^{-1}_x T_y(i)) - \theta_c(g^{-1}_b T^{-1}_x T_y(i)) \nonumber \\
    &- \theta_b(T^{-1}_x T_y(i)) + \theta_a(i) + \theta_c(g^{-1}_a(i)) = 0.
\end{align}
For $i\equiv (x,y,z;\alpha)$, the above equation is:
\begin{align}
    0&=-\theta_a(y,-z,-x;\beta) - \theta_c(-x,y,-z;\gamma) \nonumber \\
    &- \theta_b(x-1,y+1,z;\alpha) \nonumber \\
    &+ \theta_a(x,y,z;\alpha) +\theta_c(-x,-y-1, z-1;\delta)\nonumber \\
    & = x(\phi_{bx} - \phi_{ax})+ y(\phi_{ax} - \phi_{ay})+ z(\phi_{bz} - \phi_{ay})-\phi_{bx} \nonumber \\
    &-\mathfrak{m}_{a}(\beta)-\mathfrak{m}_{c}(\gamma)-\mathfrak{m}_{b}(\alpha)+\mathfrak{m}_{a}(\alpha)+\mathfrak{m}_{c}(\delta). 
\end{align}
We can conclude from the above equation that $\phi_2\equiv \phi_{ax} = \phi_{ay} = \phi_{bx} =  \phi_{bz}$. Thus we have $f_a(x,y,z) = -(x+y)\phi_2$ and $f_b(x,y,z) = -(x+z)\phi_2$.\\
For $i\equiv (x,y,z;\beta)$, the above equation is:
\begin{align}
    0&=-\theta_a(y,-z-1,-x;\gamma) - \theta_c(-x,y,-z-1;\delta) \nonumber \\
    &- \theta_b(x-1,y+1,z;\beta) + \theta_a(x,y,z;\beta) \nonumber \\
    &+\theta_c(-x-1,-y-1, z;\gamma)\nonumber \\
    & = -2\phi_2 -\mathfrak{m}_{a}(\gamma)-\mathfrak{m}_{c}(\delta)-\mathfrak{m}_{b}(\beta)\nonumber \\
    &+\mathfrak{m}_{a}(\beta)+\mathfrak{m}_{c}(\gamma).
\end{align}
For $i\equiv (x,y,z;\gamma)$, the above equation is:
\begin{align}
    0&=-\theta_a(y+1,-z,-x;\alpha) - \theta_c(-x,y+1,-z;\alpha) \nonumber \\
    &- \theta_b(x-1,y+1,z;\gamma) + \theta_a(x,y,z;\gamma) \nonumber \\
    &+\theta_c(-x-1,-y-1, z-1;\beta)\nonumber \\
    & = -\mathfrak{m}_{a}(\alpha)-\mathfrak{m}_{c}(\alpha)-\mathfrak{m}_{b}(\gamma)+\mathfrak{m}_{a}(\gamma)+\mathfrak{m}_{c}(\beta).
\end{align}
For $i\equiv (x,y,z;\delta)$, the above equation is:
\begin{align}
    0&=-\theta_a(y+1,-z-1,-x;\delta) - \theta_c(-x,y+1,-z-1;\beta) \nonumber \\
    &- \theta_b(x-1,y+1,z;\delta) + \theta_a(x,y,z;\delta) \nonumber \\
    &+\theta_c(-x,-y-1, z;\alpha)\nonumber \\
    & = -\phi_2-\mathfrak{m}_{a}(\delta)-\mathfrak{m}_{c}(\beta)-\mathfrak{m}_{b}(\delta)\nonumber \\
    &+\mathfrak{m}_{a}(\delta)+\mathfrak{m}_{c}(\alpha).
\end{align}

\subsection{Solving Eq.~\ref{u1psg:14}}
The equation is reduced to:
\begin{align}
    &-\theta_b(g^{-1}_aT_xT^{-1}_yT_z(i))-\theta_a(T_xT^{-1}_yT_z(i))\nonumber \\
    &+\theta_b(i)+\theta_a(g^{-1}_b(i))=\phi_{ab}.
\end{align}
For $i\equiv (x,y,z;\alpha)$, the above equation is:
\begin{align}
    \phi_{ab}&= -\theta_b(-x-1,-y,z;\delta) - \theta_a(x+1,y-1,z+1;\alpha)\nonumber \\
    &+\theta_b(x,y,z;\alpha)+\theta_a(-x-1,y-1,-z;\gamma)\nonumber \\
    &=\phi_2 -\mathfrak{m}_{b}(\delta)-\mathfrak{m}_{a}(\alpha)+\mathfrak{m}_{b}(\alpha)+\mathfrak{m}_{a}(\gamma).
\end{align}
For $i\equiv (x,y,z;\beta)$, the above equation is:
\begin{align}
    \phi_{ab}&= -\theta_b(-x-2,-y,z+1;\gamma) - \theta_a(x+1,y-1,z+1;\beta)\nonumber \\
    &+\theta_b(x,y,z;\beta)+\theta_a(-x-1,y-1,-z-1;\delta)\nonumber \\
    &=\phi_2 -\mathfrak{m}_{b}(\gamma)-\mathfrak{m}_{a}(\beta)+\mathfrak{m}_{b}(\beta)+\mathfrak{m}_{a}(\delta).
\end{align}
For $i\equiv (x,y,z;\gamma)$, the above equation is:
\begin{align}
    \phi_{ab}&= -\theta_b(-x-2,-y,z;\beta) - \theta_a(x+1,y-1,z+1;\gamma)\nonumber \\
    &+\theta_b(x,y,z;\gamma)+\theta_a(-x-1,y,-z;\alpha)\nonumber \\
    &=-\phi_2 -\mathfrak{m}_{b}(\beta)-\mathfrak{m}_{a}(\gamma)+\mathfrak{m}_{b}(\gamma)+\mathfrak{m}_{a}(\alpha).
\end{align}
For $i\equiv (x,y,z;\delta)$, the above equation is:
\begin{align}
    \phi_{ab}&= -\theta_b(-x-1,-y,z+1;\alpha) - \theta_a(x+1,y-1,z+1;\delta)\nonumber \\
    &+\theta_b(x,y,z;\delta)+\theta_a(-x-1,y,-z-1;\beta)\nonumber \\
    &=\phi_2 -\mathfrak{m}_{b}(\alpha)-\mathfrak{m}_{a}(\delta)+\mathfrak{m}_{b}(\delta)+\mathfrak{m}_{a}(\beta).
\end{align}

\subsection{Solving Eq.~\ref{u1psg:15}}
The equation Eq.\ref{u1psg:15} is then reduced to:
\begin{align}
    &-\theta_c(g^{-1}_bT^{-1}_xT_y(i))-\theta_b(T^{-1}_xT_y(i))+\theta_a(i) \nonumber \\
    & +\theta_b(g^{-1}_a(i))+\theta_c(g^{-1}_bg^{-1}_a(i)) \nonumber \\
    & +\theta_b(g^{-1}_cg^{-1}_bg^{-1}_a(i)) =0.
\end{align}
For $i\equiv (x,y,z;\alpha)$, the above equation is:
\begin{align}
    0&=-\theta_c(-x,y,-z;\gamma)-\theta_b(x-1,y+1,z;\alpha)\nonumber \\
    &+\theta_a(x,y,z;\alpha)+\theta_b(-x,-y-1,z-1;\delta)\nonumber \\ 
    &+\theta_c(x-1,-y-1,-z;\beta)+\theta_b(-y-1,-z,x-1;\delta)\nonumber \\
    &= 2\phi_2 -\mathfrak{m}_{c}(\gamma)-\mathfrak{m}_{b}(\alpha)+\mathfrak{m}_{a}(\alpha)\nonumber \\
    &+\mathfrak{m}_{b}(\delta)+\mathfrak{m}_{c}(\beta)+\mathfrak{m}_{b}(\delta).
\end{align}
For $i\equiv (x,y,z;\beta)$, the above equation is:
\begin{align}
    0&=-\theta_c(-x,y,-z-1;\delta)-\theta_b(x-1,y+1,z;\beta)\nonumber \\
    &+\theta_a(x,y,z;\beta)+\theta_b(-x-1,-y-1,z;\gamma)\nonumber \\ 
    &+\theta_c(x,-y-1,-z;\alpha)+\theta_b(-y-1,-z,x;\alpha)\nonumber \\
    &= \phi_2 -\mathfrak{m}_{c}(\delta)-\mathfrak{m}_{b}(\beta)+\mathfrak{m}_{a}(\beta)\nonumber \\ 
    &+\mathfrak{m}_{b}(\gamma)+\mathfrak{m}_{c}(\alpha)+\mathfrak{m}_{b}(\alpha).
\end{align}
For $i\equiv (x,y,z;\gamma)$, the above equation is:
\begin{align}
    0&=-\theta_c(-x,y+1,-z;\alpha)-\theta_b(x-1,y+1,z;\gamma)\nonumber \\ 
    &+\theta_a(x,y,z;\gamma)+\theta_b(-x-1,-y-1,z-1;\beta\nonumber \\ 
    &)+\theta_c(x,-y-2,-z;\delta)+\theta_b(-y-2,-z,x;\gamma)\nonumber \\
    &= 3\phi_2 -\mathfrak{m}_{c}(\alpha)-\mathfrak{m}_{b}(\gamma)+\mathfrak{m}_{a}(\gamma)\nonumber \\ 
    &+\mathfrak{m}_{b}(\beta)+\mathfrak{m}_{c}(\delta)+\mathfrak{m}_{b}(\gamma).
\end{align}
For $i\equiv (x,y,z;\delta)$, the above equation is:
\begin{align}
    0&=-\theta_c(-x,y+1,-z-1;\beta)-\theta_b(x-1,y+1,z;\delta)\nonumber \\ 
    &\theta_a(x,y,z;\delta)+\theta_b(-x,-y-1,z;\alpha)\nonumber \\ 
    &+\theta_c(x-1,-y-2,-z;\gamma)+\theta_b(-y-2,-z,x-1;\beta)\nonumber \\
    &= 2\phi_2 -\mathfrak{m}_{c}(\beta)-\mathfrak{m}_{b}(\delta)+\mathfrak{m}_{a}(\delta)\nonumber \\ 
    &+\mathfrak{m}_{b}(\alpha)+\mathfrak{m}_{c}(\gamma)+\mathfrak{m}_{b}(\beta).
\end{align}
\subsection{Collected equations for $\mathfrak{m}$s}
In this subsection, we summarize the coupled equations to solve for $\mathfrak{m}$s. Before doing so, we note that we can use the $\mathsf{SU}(2)$ gauge symmetry to fix certain $\mathfrak{m}$s. Recall that the action of the gauge transformation is:
\begin{equation}
    g.t.:\theta_{U}(i)\mapsto w(i) -  \theta_{U}(i) + w(U^{-1}(i)).
\end{equation}
We start off in a generic gauge where all $\mathfrak{m}$s are non-trivial. We first perform the gauge transformation $w(\beta) = \mathfrak{m}_c(\beta)$ and $w(\gamma) = -\mathfrak{m}_c(\delta)$. The consequence is that $\mathfrak{m}_c(\beta) = \mathfrak{m}_c(\delta) = 0$. And because $\mathfrak{m}_c(\beta)+ \mathfrak{m}_c(\gamma) + \mathfrak{m}_c(\delta)=0$ from one of our constraints, we have $\mathfrak{m}_c(\gamma)=0$. We then perform the gauge transformation $w(\alpha) = \mathfrak{m}_a(\alpha)$, such that $\mathfrak{m}_a(\alpha)=0$. And because $\mathfrak{m}_a(\alpha) + \mathfrak{m}_a(\delta) = \phi_a - \phi_{2}$ from one of our constraints, we have $\mathfrak{m}_a(\delta) = \phi_a - \phi_{2}$.\par
The remaining equations after the reductions are:
\begin{subequations}
\begin{align}
    &3\mathfrak{m}_c(\alpha)=0, \label{eq: u1comp_1}\\
    &\mathfrak{m}_a(\beta) + \mathfrak{m}_a(\gamma) = \phi_a - 2\phi_{2}, \label{eq: u1comp_2}\\
    &\mathfrak{m}_b(\alpha) + \mathfrak{m}_b(\gamma) = \phi_b - \phi_{2}, \label{eq: u1comp_3}\\
    &\mathfrak{m}_b(\beta) + \mathfrak{m}_b(\delta) = \phi_b - 2\phi_{2}, \label{eq: u1comp_4}\\
    &-\mathfrak{m}_{a}(\beta)-\mathfrak{m}_{b}(\alpha) = \phi_2 , \label{eq: u1comp_5}\\
     &-\mathfrak{m}_{a}(\gamma)-\mathfrak{m}_{b}(\beta)+\mathfrak{m}_{a}(\beta) = 2\phi_2, \label{eq: u1comp_6}\\
     &-\mathfrak{m}_{c}(\alpha)-\mathfrak{m}_{b}(\gamma)+\mathfrak{m}_{a}(\gamma) =0, \label{eq: u1comp_7}\\
     &-\mathfrak{m}_{b}(\delta)+\mathfrak{m}_{c}(\alpha)  = \phi_2, \label{eq: u1comp_8}\\
     &-\mathfrak{m}_{b}(\delta)+\mathfrak{m}_{b}(\alpha)+\mathfrak{m}_{a}(\gamma)  = \phi_{ab}-\phi_2, \label{eq: u1comp_9}\\
     &-\mathfrak{m}_{b}(\gamma)-\mathfrak{m}_{a}(\beta)+\mathfrak{m}_{b}(\beta) = \phi_{ab}-\phi_a, \label{eq: u1comp_10}\\
     &-\mathfrak{m}_{b}(\beta)-\mathfrak{m}_{a}(\gamma)+\mathfrak{m}_{b}(\gamma)  = \phi_{ab}+\phi_2, \label{eq: u1comp_11}\\
     &-\mathfrak{m}_{b}(\alpha)+\mathfrak{m}_{b}(\delta)+\mathfrak{m}_{a}(\beta)  = \phi_{ab}+\phi_a-2\phi_2, \label{eq: u1comp_12}\\
     &-\mathfrak{m}_{b}(\alpha)+\mathfrak{m}_{b}(\delta)+\mathfrak{m}_{b}(\delta)  = - 2\phi_2, \label{eq: u1comp_13}\\
     &-\mathfrak{m}_{b}(\beta)+\mathfrak{m}_{a}(\beta)+\mathfrak{m}_{b}(\gamma)
     +\mathfrak{m}_{c}(\alpha)+\mathfrak{m}_{b}(\alpha) = - \phi_2, \label{eq: u1comp_14}\\
     &-\mathfrak{m}_{c}(\alpha)-\mathfrak{m}_{b}(\gamma)+\mathfrak{m}_{a}(\gamma)+\mathfrak{m}_{b}(\beta)+\mathfrak{m}_{b}(\gamma) = - 3\phi_2, \label{eq: u1comp_15}\\
     &-\mathfrak{m}_{b}(\delta)+\mathfrak{m}_{b}(\alpha)+\mathfrak{m}_{b}(\beta) = -\phi_a- \phi_2. \label{eq: u1comp_16}
\end{align}
\end{subequations}
We now set $A\equiv \mathfrak{m}_c(\alpha)$, $B\equiv \mathfrak{m}_a(\beta)$, $C\equiv \mathfrak{m}_b(\alpha)$ and $D\equiv \mathfrak{m}_b(\beta)$. Eq.~\ref{eq: u1comp_5} tells us that $-B-C=\phi_2$. Also, Eq.~\ref{eq: u1comp_6} tells us that $D = 2B - \phi_a$. Thus all the $\mathfrak{m}$s can be represented using $A$ and $B$, as deduced from Eq.~\ref{eq: u1comp_1} to Eq.~\ref{eq: u1comp_6}:
\begin{align}
    \mathfrak{m}_a(\alpha) &= 0,\nonumber \\
    \mathfrak{m}_a(\beta) &= B,\nonumber \\
    \mathfrak{m}_a(\gamma) &= \phi_a-2\phi_2 - B,\nonumber \\
    \mathfrak{m}_a(\delta) &= \phi_a - \phi_2,\nonumber \\
    \mathfrak{m}_b(\alpha) &= -\phi_2-B,\nonumber \\
    \mathfrak{m}_b(\beta) &= 2B-\phi_a,\nonumber \\
    \mathfrak{m}_b(\gamma) &= \phi_b+B,\nonumber \\
    \mathfrak{m}_b(\delta) &= \phi_b -2\phi_2+\phi_a-2B,\nonumber \\
    \mathfrak{m}_c(\alpha) &= A,\nonumber \\
    \mathfrak{m}_c(\beta) &= 0,\nonumber \\
    \mathfrak{m}_c(\gamma) &= 0,\nonumber \\
    \mathfrak{m}_c(\delta) &= 0.
\end{align}
Eq.~\ref{eq: u1comp_7} tells us that:
\begin{equation}
    A+2B = \phi_a - \phi_b -2\phi_2,
\end{equation}
whereas Eq.~\ref{eq: u1comp_8} tells us that:
\begin{equation}
    A+2B = \phi_a + \phi_b -\phi_2.
\end{equation}
From which we can see that $\phi_2=-2\phi_b$. Now if we look at Eq.~\ref{eq: u1comp_9}, we have $\phi_{ab} = -\phi_b$. In fact, Eq.~\ref{eq: u1comp_10} to Eq.~\ref{eq: u1comp_12} do not tell us more than this. Eq.~\ref{eq: u1comp_13} tells us that $3B = 4\phi_b+2\phi_a$, Eq.~\ref{eq: u1comp_14} tells us that $A-B = -\phi_a-\phi_b$, Eq.~\ref{eq: u1comp_15} tells us that $A-B = -2\phi_b$ and finally  Eq.~\ref{eq: u1comp_16} tells us that $3B = \phi_a + \phi_b-2\phi_2$. The above relations allow us to assert that:
\begin{equation}
    \phi_3\equiv -\phi_{ab}=\phi_a=\phi_b, \quad \phi_2 = -2\phi_3, \quad B = A+2\phi_3,
\end{equation}
and:
\begin{align}
    \mathfrak{m}_a(\alpha) &= 0,\nonumber \\
    \mathfrak{m}_a(\beta) &= A + 2\phi_3,\nonumber \\
    \mathfrak{m}_a(\gamma) &=3\phi_3-A,\nonumber \\
    \mathfrak{m}_a(\delta) &= 3\phi_3,\nonumber \\
    \mathfrak{m}_b(\alpha) &= -A,\nonumber \\
    \mathfrak{m}_b(\beta) &= -A+3\phi_3,\nonumber \\
    \mathfrak{m}_b(\gamma) &= 3\phi_3+A,\nonumber \\
    \mathfrak{m}_b(\delta) &= 2\phi_3+A,\nonumber \\
    \mathfrak{m}_c(\alpha) &= A,\nonumber \\
    \mathfrak{m}_c(\beta) &= 0,\nonumber \\
    \mathfrak{m}_c(\gamma) &= 0,\nonumber \\
    \mathfrak{m}_c(\delta) &= 0.
\end{align}
We also have:
\begin{equation}
    f_a(i) = 2(x+y)\phi_3, \quad f_b(i) = 2(x+z)\phi_3.
\end{equation}
Similar to the $\mathsf{Z}_2$ case, we now consider a further gauge transformation:
\begin{align}
    w(x,y,z;\alpha)&=-2\phi_3+\phi_3(x+y+z),\nonumber \\
    w(x,y,z;\beta/\gamma/\delta)&=\phi_3(x+y+z),
\end{align}
we see that:\\
1.) $G_x=G_y=G_z= e^{-i\phi_3\tau_z}$;\\
2.) $G_{a/b/c}(x,y,z;s)=e^{i\mathfrak{m}_{a/b/c}(s)\tau_z}$, where the $\mathfrak{m}$s have the following forms:
\begin{align}
&\mathfrak{m}_a(\alpha)=0, \mathfrak{m}_a(\beta)=A, \mathfrak{m}_a(\gamma)=-A, \mathfrak{m}_a(\delta)=0;\nonumber \\
&\mathfrak{m}_b(\alpha)=-A, \mathfrak{m}_b(\beta)=-A, \mathfrak{m}_b(\gamma)=A, \mathfrak{m}_b(\delta)=A;\nonumber \\
&\mathfrak{m}_c(\alpha)=A, \mathfrak{m}_c(\beta)=0, \mathfrak{m}_c(\gamma)=0, \mathfrak{m}_c(\delta)=0.
\end{align}
Due to the fact that $\phi_3$ now becomes global signs which are elements of the IGG, we conclude that $\phi_3$ is redundant. We also remark that a gauge transformation $W(x,y,z;s)\equiv i\tau_x$ maps the PSG solutions in which $\mathcal{A}=e^{i\frac{2\pi}{3}\tau_z}$ to that in which $\mathcal{A}=e^{i\frac{4\pi}{3}\tau_z}$.\par
In conclusion, we have:\\
1.) $G_x=G_y=G_z= \tau_0$;\\
2.) $G_{a/b/c}(x,y,z;s)=e^{i\mathfrak{m}_{a/b/c}(s)\tau_z}$, where the $\mathfrak{m}$s have the following forms:
\begin{align}
&\mathfrak{m}_a(\alpha)=0, \mathfrak{m}_a(\beta)=A, \mathfrak{m}_a(\gamma)=-A, \mathfrak{m}_a(\delta)=0;\nonumber \\
&\mathfrak{m}_b(\alpha)=-A, \mathfrak{m}_b(\beta)=-A, \mathfrak{m}_b(\gamma)=A, \mathfrak{m}_b(\delta)=A;\nonumber \\
&\mathfrak{m}_c(\alpha)=A, \mathfrak{m}_c(\beta)=0, \mathfrak{m}_c(\gamma)=0, \mathfrak{m}_c(\delta)=0,
\end{align}
where $\mathcal{A}=\tau_0, e^{i\frac{2\pi}{3}\tau_z}$.

\subsection{Adding Time-Reversal Symmetry}
We firstly write the algebraic relations:
\begin{subequations}
\begin{align}
& G_{\mathcal{T}}(i)G_{\mathcal{T}}(i)=e^{i\phi_{\mathcal{T}}\tau_z}, \label{u1psg:trs1}\\
& G^\dag_{\mathcal{T}}(T^{-1}_x(i))G^\dag_x(i)G_{\mathcal{T}}(i)G_x(i)=e^{i\phi_{x\mathcal{T}}\tau_z}, \label{u1psg:trs2}\\
& G^\dag_{\mathcal{T}}(T^{-1}_y(i))G^\dag_y(i)G_{\mathcal{T}}(i)G_y(i)=e^{i\phi_{y\mathcal{T}}\tau_z}, \label{u1psg:trs3}\\
& G^\dag_{\mathcal{T}}(T^{-1}_z(i))G^\dag_z(i)G_{\mathcal{T}}(i)G_z(i)=e^{i\phi_{z\mathcal{T}}\tau_z}, \label{u1psg:trs4}\\
& G^\dag_{\mathcal{T}}(g^{-1}_a(i))G^\dag_a(i)G_{\mathcal{T}}(i)G_a(i)=e^{i\phi_{a\mathcal{T}}\tau_z}, \label{u1psg:trs5}\\
& G^\dag_{\mathcal{T}}(g^{-1}_b(i))G^\dag_b(i)G_{\mathcal{T}}(i)G_b(i)=e^{i\phi_{b\mathcal{T}}\tau_z}, \label{u1psg:trs6}\\
& G^\dag_{\mathcal{T}}(g^{-1}_c(i))G^\dag_c(i)G_{\mathcal{T}}(i)G_c(i)=e^{i\phi_{c\mathcal{T}}\tau_z}. \label{u1psg:trs7}
\end{align}
\end{subequations}
As usual, the canonical form of $G_{\mathcal{T}}(i) = (i\tau_x)^{n_{\mathcal{T}}}e^{i\theta_{\mathcal{T}}(i)\tau_z}$. When $n_{\mathcal{T}}=0$, we can show that $G_{\mathcal{T}} = i\tau_z$ uniformly much like the case for $\mathsf{Z}_2$. Since the derivation is very similar to the $\mathsf{Z}_2$ case, it is not included here. This group of PSG solutions does not produce mean field $\mathsf{U}(1)$ spin liquids if we consider the constraint imposed by TRS. For the rest of the appendix, let us focus on the case when $n_{\mathcal{T}}=1$.\par
Let us first look at Eq.~\ref{u1psg:trs1}, we straightforwardly conclude that $\phi_{\mathcal{T}}=\pi$. We denote $\theta_{\mathcal{T}}\equiv f_{\mathcal{T}}(x,y,z;s) + \mathfrak{m}_{\mathcal{T}}(s)$. Then Eq.~\ref{u1psg:trs2} to Eq.~\ref{u1psg:trs4} tell us that:
\begin{equation}
    f_{\mathcal{T}}(x,y,z;s)=x\phi_{x\mathcal{T}} + y\phi_{y\mathcal{T}} + z\phi_{z\mathcal{T}}.
\end{equation}
We would like to plug the above results into Eq.~\ref{u1psg:trs7}. We arrive at the folllowing constraint:
\begin{equation}
    -\theta_{\mathcal{T}}(g^{-1}_c(i)) + 2\theta_c(i) + \theta_{\mathcal{T}}(i) = \phi_{c\mathcal{T}}.
\end{equation}
For the case with $i=(x,y,z;\alpha)$, we have:
\begin{align}
\phi_{c\mathcal{T}}&=-\theta_{\mathcal{T}}(y,z,x;\alpha) + 2A +\theta_{\mathcal{T}}(x,y,z;\alpha)\nonumber \\
&=x(\phi_{x\mathcal{T}} - \phi_{y\mathcal{T}}) + y(\phi_{y\mathcal{T}}-\phi_{z\mathcal{T}}) \nonumber \\
&+ z(\phi_{z\mathcal{T}} - \phi_{x\mathcal{T}}) + 2A,
\end{align}
we arrive at $\phi_4\equiv \phi_{x\mathcal{T}} = \phi_{y\mathcal{T}} = \phi_{z\mathcal{T}}$, and $\phi_{c\mathcal{T}}=-A$, where we used $3A=0$.\\
For the case with $i=(x,y,z;\beta)$, we have:
\begin{align}
\phi_{c\mathcal{T}}&=-\theta_{\mathcal{T}}(y,z,x;\delta) +\theta_{\mathcal{T}}(x,y,z;\beta)\nonumber \\
&= -\mathfrak{m}_{\mathcal{T}}(\delta) + \mathfrak{m}_{\mathcal{T}}(\beta).
\end{align}
For the case with $i=(x,y,z;\gamma)$, we have:
\begin{align}
\phi_{c\mathcal{T}}&=-\theta_{\mathcal{T}}(y,z,x;\beta) +\theta_{\mathcal{T}}(x,y,z;\gamma)\nonumber \\
&= -\mathfrak{m}_{\mathcal{T}}(\beta) + \mathfrak{m}_{\mathcal{T}}(\gamma).
\end{align}
For the case with $i=(x,y,z;\delta)$, we have:
\begin{align}
\phi_{c\mathcal{T}}&=-\theta_{\mathcal{T}}(y,z,x;\gamma) +\theta_{\mathcal{T}}(x,y,z;\delta)\nonumber \\
&= -\mathfrak{m}_{\mathcal{T}}(\gamma) + \mathfrak{m}_{\mathcal{T}}(\delta).
\end{align}
Thus if we denote $\mathfrak{m}_{\mathcal{T}}(\beta)\equiv E$, we have $\mathfrak{m}_{\mathcal{T}}(\gamma)= E-A$ and $\mathfrak{m}_{\mathcal{T}}(\delta)= E+A$.\par
We now look at Eq.~\ref{u1psg:trs5}. Similar to the case before:
\begin{equation}
    -\theta_{\mathcal{T}}(g^{-1}_a(i)) + 2\theta_a(i) + \theta_{\mathcal{T}}(i) = \phi_{a\mathcal{T}}.
\end{equation}
For the case with $i=(x,y,z;\alpha)$, we have:
\begin{align}
\phi_{a\mathcal{T}}&=-\theta_{\mathcal{T}}(-x,-y-1,z-1;\delta) \nonumber \\
&+2\mathfrak{m}_{a}(\alpha)+\theta_{\mathcal{T}}(x,y,z;\alpha)\nonumber \\
&= - (-x-y-1+z-1)\phi_4 -\mathfrak{m}_{\mathcal{T}}(\delta) \nonumber \\
& + (x+y+z)\phi_4 + \mathfrak{m}_{\mathcal{T}}(\alpha) \nonumber \\
& = x(2\phi_4) + y(2\phi_4) \nonumber \\
&+2\phi_4 + \mathfrak{m}_{\mathcal{T}}(\alpha) - \mathfrak{m}_{\mathcal{T}}(\delta),
\end{align}
from which we have:
\begin{equation}
    2\phi_4 = 0, \quad \mathfrak{m}_{\mathcal{T}}(\alpha) - \mathfrak{m}_{\mathcal{T}}(\delta) = \phi_{a\mathcal{T}}.
\end{equation}
For the case with $i=(x,y,z;\beta)$, we have:
\begin{align}
\phi_{a\mathcal{T}}&=-\theta_{\mathcal{T}}(-x-1,-y-1,z;\gamma)  \nonumber \\
&+2\mathfrak{m}_{a}(\beta)+\theta_{\mathcal{T}}(x,y,z;\beta)\nonumber \\
& = \mathfrak{m}_{\mathcal{T}}(\beta) - \mathfrak{m}_{\mathcal{T}}(\gamma) + 2A\nonumber \\
& = 0.
\end{align}
For the case with $i=(x,y,z;\gamma)$, we have:
\begin{align}
\phi_{a\mathcal{T}}&=-\theta_{\mathcal{T}}(-x-1,-y-1,z-1;\beta) \nonumber \\
&+2\mathfrak{m}_{a}(\gamma)+\theta_{\mathcal{T}}(x,y,z;\gamma)\nonumber \\
& = 3\phi_4 + \mathfrak{m}_{\mathcal{T}}(\gamma) - \mathfrak{m}_{\mathcal{T}}(\beta) -2A  \nonumber \\
& = 3\phi_4.
\end{align}
Note that this relation combined with the one before, gives us that $\phi_4=0$.\\
For the case with $i=(x,y,z;\delta)$, we have:
\begin{align}
\phi_{a\mathcal{T}}&=-\theta_{\mathcal{T}}(-x,-y-1,z;\alpha) \nonumber \\
&+2\mathfrak{m}_{a}(\delta)+\theta_{\mathcal{T}}(x,y,z;\delta)\nonumber \\
& = \phi_4 + \mathfrak{m}_{\mathcal{T}}(\delta) - \mathfrak{m}_{\mathcal{T}}(\alpha).
\end{align}
Combining the above constraints, we arrive at a set of relations summarised here:
\begin{align}
    &\phi_{a\mathcal{T}} = \phi_4 = 0\nonumber \\
    &\quad \mathfrak{m}_{\mathcal{T}}(\alpha) = \mathfrak{m}_{\mathcal{T}}(\delta),\quad  \mathfrak{m}_{\mathcal{T}}(\beta) = \mathfrak{m}_{\mathcal{T}}(\gamma)+A.
\end{align}
Let us now look at Eq.~\ref{u1psg:trs6}:
\begin{equation}
    -\theta_{\mathcal{T}}(g^{-1}_b(i)) + 2\theta_b(i) + \theta_{\mathcal{T}}(i) = \phi_{b\mathcal{T}}.
\end{equation}
For the case with $i=(x,y,z;\alpha)$, we have:
\begin{align}
\phi_{b\mathcal{T}}&=-\theta_{\mathcal{T}}(-x-1,y-1,-z;\gamma)  \nonumber \\
&+2\mathfrak{m}_{b}(\alpha)+\theta_{\mathcal{T}}(x,y,z;\alpha)\nonumber \\
& = - 2A + \mathfrak{m}_{\mathcal{T}}(\alpha) - \mathfrak{m}_{\mathcal{T}}(\gamma).
\end{align}
For the case with $i=(x,y,z;\beta)$, we have:
\begin{align}
\phi_{b\mathcal{T}}&=-\theta_{\mathcal{T}}(-x-1,y-1,-z-1;\delta) \nonumber \\
&+2\mathfrak{m}_{b}(\beta)+\theta_{\mathcal{T}}(x,y,z;\beta)\nonumber \\
& = - 2A  + \mathfrak{m}_{\mathcal{T}}(\beta) - \mathfrak{m}_{\mathcal{T}}(\delta).
\end{align}
For the case with $i=(x,y,z;\gamma)$, we have:
\begin{align}
\phi_{b\mathcal{T}}&=-\theta_{\mathcal{T}}(-x-1,y,-z;\alpha)  \nonumber \\
&+2\mathfrak{m}_{b}(\gamma)+\theta_{\mathcal{T}}(x,y,z;\gamma)\nonumber \\
& = 2A  + \mathfrak{m}_{\mathcal{T}}(\gamma) - \mathfrak{m}_{\mathcal{T}}(\alpha).
\end{align}
For the case with $i=(x,y,z;\delta)$, we have:
\begin{align}
\phi_{b\mathcal{T}}&=-\theta_{\mathcal{T}}(-x-1,y,-z-1;\beta) \nonumber \\
&+2\mathfrak{m}_{b}(\delta)+\theta_{\mathcal{T}}(x,y,z;\delta)\nonumber \\
& = 2A  + \mathfrak{m}_{\mathcal{T}}(\delta) - \mathfrak{m}_{\mathcal{T}}(\beta).
\end{align}
Combining the above constraints, we arrive at $\phi_{b\mathcal{T}} = 0$ and no new relations.\par
We can then summarise:
\begin{align}
    \mathfrak{m}_{\mathcal{T}}(\alpha) &= E+A ,\nonumber \\
    \mathfrak{m}_{\mathcal{T}}(\beta) &= E,\nonumber \\
    \mathfrak{m}_{\mathcal{T}}(\gamma) &= E-A,\nonumber \\
    \mathfrak{m}_{\mathcal{T}}(\delta) &= E+A .
\end{align}
In the above, $E$ is a free $\mathsf{U}(1)$ phase. However, note that we did not make use of the IGG gauge degrees of freedom associated with TRS. Recalling that $G_{\mathcal{T}}\sim G_{\mathcal{T}}W_{\mathcal{T}}$, where $W_{\mathcal{T}}\in \mathsf{U}(1)$. We choose $W_{\mathcal{T}}\equiv \exp (-iE)$, thus eliminating the free phase in our solutions. We collect the $\mathsf{U}(1)$ PSG solutions into Tab.~\ref{table:psg_sol_u1}. Thus we have:
\begin{align}
    &G_{\mathcal{T}}(\vec{r},\alpha) = i\tau_x e^{iA\tau_z},\nonumber \\
    &G_{\mathcal{T}}(\vec{r},\beta) = i\tau_x,\nonumber \\
    &G_{\mathcal{T}}(\vec{r},\gamma) = i\tau_x e^{-iA\tau_z},\nonumber \\
    &G_{\mathcal{T}}(\vec{r},\delta) = i\tau_x e^{iA\tau_z}.
\end{align}

\section{Mean-field ansatzes for the PSG solutions}
\label{app:ansatzes}
Our PSG classification  obtains a set of gauge-inequivalent transformations $G_g$ for all $g \in  \mathsf{P}2_13 \times \mathsf{Z}_2$. In this appendix, we 
derive the constraints imposed on the mean-field parameters $U_{ij}$ and $\mu_i$ by requiring that an element of the PSG leaves the ansatz invariant. We repeat
this condition for convenience:
\begin{align}
    \forall g: & G_g(g(i)) U_{g(i)g(j)}G^\dag_g(g(j)) = U_{ij},\nonumber \\
    & G_g(g(i)) \mu_{g(i)}G^\dag_g(g(i)) = \mu_{i}.
    \tag{\ref{eq:ansatz_constraint}}
\end{align}
\subsection{$\mathsf{Z}_2$}
\label{sec:z2_ansatz}

Here we specify the ansatzes for the PSGs corresponding to the IGG being $\mathsf{Z}_2$.
As derived in Appendix~\ref{sec: z2psg} and displayed in Table~\ref{table:psg_sol} in the main text, the four $\mathsf{Z}_2$ PSGs can be
indexed by , $\mathcal{A}=\exp(i 2 \pi n/3)$ for $i=0,1$, and $\mathcal{E}=\tau_0$ or $\mathcal{E}=i \tau_z$, in terms 
of which all gauge transformations are listed in Tab.~\ref{table:psg_sol}.

We note that if $G_\mathcal{T}= \tau_0$  then the invariance of the ansatz under time-reversal requires $U_{ij}=-U_{ij}$ and $\mu_i = -\mu_i$, 
leading to no non-zero mean-field ansatzes.

For $G_\mathcal{T}=i \tau_z$, the invariance under TRS requires the following
form for all links and sites:
\begin{align}
  U_{ij}= U^x_{ij} \tau_x + U^y_{ij} \tau_y \\
  \mu_{i} = \mu^x_{i} \tau_x + \mu^y_{i} \tau_y.
  \label{eq:z2_trs_noz}
\end{align}

Finally, since $G_{x}=G_{y}=G_{z}=\tau_0$ in our solutions, we must require
$U_{ij}$ and $\mu_{ij}$ to be translationally invariant. We then encode
the dependence of the parameters $U_{ij}$ on the link $(ij)$ by 
determining $U$ for each of the unique links in Tab.~\ref{table:bond_list}, and determining the functions $U^x_i, U^y_i$, where $i\in \{1,2 \cdots 12\}$ specifies the link in Table~\ref{table:bond_list}. The on-site parameters
are described as $\mu_{\alpha}, \mu_{\beta}, \mu_{\gamma}$ and $\mu_{\delta}$, where the subscripts denote the sublattice dependence.

Imposing the invariance of the ansatz under the action of $(G_c, g_c)$, we get the following relations between $4$ groups of links that are closed under the 
application of $g_c$
\begin{align}
 \nonumber  U_1 &= U_5             = U_9 \\
 \nonumber  U_2 &= U_6             = U_{10} \\
 \nonumber  U_3 &= \mathcal{A} U_7 = \mathcal{A}^2 U_{11} \\
            U_4 &= \mathcal{A} U_8 = \mathcal{A}^2 U_{12} 
            \label{eq:z2_gc_ansatz}
\end{align}

The relations between different groups of links are obtained by the invariance under $(G_b,g_b)$ and $(G_a,g_a)$. The action of $(G_a,g_a)$ gives us
\begin{align}
 \nonumber  U_1 &= U_2      \\
 \nonumber  U_3 &= U_4        \\     
 \nonumber  U_5 &= \mathcal{A}^2 U_7  \\
 \nonumber  U_6 &= \mathcal{A}^2 U_8  \\
 \nonumber  U_9 &= \mathcal{A} U_{12} \\
 U_{10} &=  \mathcal{A} U_{11} 
            \label{eq:z2_ga_ansatz}
\end{align}
Similarly, the invariance of all links under $(G_b,g_b)$ give us the conditions 
\begin{align}
 \nonumber  U_1 &=  U_3      \\
 \nonumber  U_2 &=  U_4        \\     
 \nonumber  U_5 &= \mathcal{A}^2 U_8  \\
 \nonumber  U_6 &= \mathcal{A}^2 U_7  \\
 \nonumber  U_9 &= U_{10} \\
           U_{11} &= U_{12} 
            \label{eq:z2_gb_ansatz}
\end{align}

Combining the conditions in Eqs.~\ref{eq:z2_gc_ansatz}, ~\ref{eq:z2_ga_ansatz} and ~\ref{eq:z2_gb_ansatz} we find that the $U_{ij}$ for all links 
can be specified in terms of only two parameters $U^x$ and $U^y$:
\begin{align}
  \nonumber &U_1 = U^x \tau_x + U^y \tau_y;   \\
  \nonumber &U_2 =U_1; \qquad U_3 = U_1 ; \qquad U_4 =U_1;\\
  \nonumber &U_5 = U_1; \qquad U_6 = U_1;\qquad  U_7 = \mathcal{A}^2 U_1;\\
  \nonumber &U_8 = \mathcal{A}^2 U_1;\qquad  U_9 = U_1;  \qquad U_{10}= U_1; \\
  &U_{11} = \mathcal{A}^2 U_1 ;   \qquad    U_{12} =\mathcal{A}^2 U_1 
  \label{eq:z2_uij}
\end{align}

Similarly, demanding the invariance of $\mu_i$ under $(G_c,g_c)$ gives us $\mu_{\gamma}=\mu_{\delta}=\mu_{\beta}$,
and $\mu_{\alpha}=\mathcal{A}^{2} \mu_{\alpha}$. Under $(G_a, g_a)$, we have $\mu_{\alpha}= \mu_{\delta}$ and
$\mu_{\beta}=\mathcal{A}^2 \mu_{\gamma}$. This already implies that when $\mathcal{A} \neq 1$, 
$\mu =0$ on all sites. When $\mathcal{A}=1$, site-independent on-site terms of the form $\mu^x \tau_x + \mu^y \tau_y$ are allowed 
in the ansatz. 

\subsection{PSG-protected gapless nodal star in $\mathsf{Z}_21$ spin liquid}
In this Appendix, we prove that the mean-field Hamiltonian $H_{\rm MFT}(\vec{k})$ (Eqs.~\ref{eq:mf_hamiltonian2} and Eqs.~\ref{eq:mf_hamiltoniank}) for the $\mathsf{Z}_21$ QSL has 
two zero-energy eigenvalues for $\vec{k}=(\pm k, \pm k, \pm k)$. To this end, we first work out the most general PSG-allowed $H_{\rm MFT}(\vec{k})$  for the 
$\mathsf{Z}_21$ QSL. The rest of the discussion assumes the translation invariance of the ansatzes, which is true for all our QSLs.
First, we use the basis $\big(\psi^{\alpha}_1, \psi^{\alpha}_2,\cdots \psi^{\delta}_1 \psi^{\delta}_2 \big)$ to express 
the $H_{\rm MFT}(\vec{k})$ in terms of $2\times2$ blocks as 
\begin{align}
  H_{\rm MFT}(\vec{k})= 
  \begin{pmatrix}
    h_{\alpha,\alpha}(\vec{k}) & h_{\alpha,\beta}(\vec{k}) & h_{\alpha,\gamma}(\vec{k})& h_{\alpha,\gamma}(\vec{k}) \\
    h_{\beta ,\alpha}(\vec{k}) & h_{\beta ,\beta}(\vec{k}) & h_{\beta ,\gamma}(\vec{k})& h_{\beta ,\gamma}(\vec{k}) \\
    h_{\gamma,\alpha}(\vec{k}) & h_{\gamma,\beta}(\vec{k}) & h_{\gamma,\gamma}(\vec{k})& h_{\gamma,\gamma}(\vec{k}) \\
    h_{\delta,\alpha}(\vec{k}) & h_{\delta,\beta}(\vec{k}) & h_{\delta,\gamma}(\vec{k})& h_{\delta,\gamma}(\vec{k}) 
\end{pmatrix}
\label{eq:block_intro}
\end{align}
As just demonstrated in the previous section, when $\mathcal{A} \neq 1$ and $G_{\mathcal{T}}=i\tau_z$, we have 
$h_{\alpha,\alpha}=h_{\beta,\beta}=h_{\gamma,\gamma}=h_{\delta,\delta}=0$ for all $\vec{k}$. 
The 
block matrices have the  form  $U_x \tau^x +U_y \tau^y$ (Eq.~\ref{eq:z2_trs_noz}) in real space, leading to
\begin{align}
  h_{\alpha,\beta}(\vec{r},\vec{r}')= 
  \begin{pmatrix}
    0 & U_{\alpha,\beta}(\vec{r}-\vec{r}')\\
    U^{*}_{\alpha,\beta}(\vec{r}-\vec{r}') & 0
  \end{pmatrix}
  \label{eq:hab_matform}
\end{align}
for a complex amplitude $U(\vec{r})$.
The fourier-transformed equivalent is given by 
\begin{align}
h_{\alpha,\beta}(\vec{k}) &=\frac{1}{N}\sum_{\vec{r},\vec{r}'} h_{\alpha,\beta}(\vec{r},\vec{r}')e^{i\vec{k}.(\vec{r}-\vec{r}')}\\
&=  \begin{pmatrix}
    0 & U_{\alpha,\beta}(\vec{k})\\
    U^{*}_{\alpha,\beta}(-\vec{k}) & 0
  \end{pmatrix}
\label{eq:block_fourier_def}
\end{align}
Foreseeing repeated appearances of the off-diagonal form in Eq.~\ref{eq:block_fourier_def}, we introduce the shorthand $M[u(k)]$, defined by 
\begin{equation}
M[u(\vec{k})]
=  \begin{pmatrix}
    0 & u(\vec{k})\\
    u^*(-\vec{k}) & 0
  \end{pmatrix}
\end{equation}
Also note that from Eq.~\ref{eq:mf_parameters} we know that $h_{\alpha,\beta}(\vec{r},\vec{r}') = h_{\beta,\alpha}(\vec{r}',\vec{r})$. So we have from Eq.~\ref{eq:block_fourier_def}
\begin{align}
h_{\alpha,\beta}(\vec{k}) = h_{\alpha,\beta}(-\vec{k})
\label{eq:block_minusk}
\end{align}

The action of symmetries on the block matrices in $k$-space can be worked from their real-space equivalents, given by
Eq.~\ref{eq:action_on_ansatz}. To show this explicitly for a general symmetry transformation $g$, we assume the unit-cell independence of gauge transformations 
which has been shown for all our QSLs. To reduce cumbersome expressions, we   introduce the shorthand $\bar{\alpha}$ and $g_{\alpha}({\vec{r}}) $  to denote the sublattice index and unit-cell position of the operation  
  $g(\vec{r};\alpha)$.
  We have, from Eq.~\ref{eq:action_on_ansatz},
\begin{align}
\nonumber  
&(G_g,g): \big( h_{\alpha,\beta}(\vec{k}) =\frac{1}{N}\sum_{\vec{r},\vec{r}'} h_{\alpha,\beta}(\vec{r},\vec{r}')e^{i\vec{k}.(\vec{r}-\vec{r}')}\big) \\
&\mapsto 
\nonumber \frac{1}{N}\sum_{\vec{r},\vec{r}'}G_g(\bar{\alpha}) h_{\bar{\alpha} \bar{\beta} }(g_{\alpha}(\vec{r}),g_{\beta}(\vec{r}'))G^\dag_g(\bar{\beta}) e^{i\vec{k}.(\vec{r}-\vec{r}')}\\
\nonumber &= \frac{1}{N}\sum_{\vec{r},\vec{r}'}G_g(\bar{\alpha}) h_{\bar{\alpha} \bar{\beta} }(\vec{r},\vec{r}')G^\dag_g(\bar{\beta})e^{i\vec{k}.(g^{-1}_{\bar{\alpha}}\vec{r}-g^{-1}_{\bar{\beta}}(\vec{r}'))} \\
\nonumber &= \frac{1}{N}\sum_{\vec{r},\vec{r}'}G_g(\bar{\alpha}) h_{\bar{\alpha} \bar{\beta} }(\vec{r},\vec{r}')G^\dag_g(\bar{\beta}) e^{i\vec{k}'.(\vec{r}-\vec{r}')+\phi_g(\alpha,\beta)}\\
&=\nonumber G_g(\bar{\alpha})h_{\bar{\alpha},\bar{\beta}}(\vec{k}')G^\dag_g(\bar{\beta})e^{i\phi_g(\alpha,\beta)}\\
\implies
&(G_g,g): h_{\alpha,\beta}(\vec{k} \mapsto G_g(\bar{\alpha})h_{\bar{\alpha},\bar{\beta}}(\vec{k}')G^\dag_g(\bar{\beta})e^{i\phi_g(\alpha,\beta)}
\label{eq:block_trans}
\end{align}
From the third line to the fourth, we have used the fact that one can always write $ \vec{k}.(g^{-1}_{\bar{\alpha}}\vec{r}-g^{-1}_{\bar{\beta}}(\vec{r}'))$ as
$\vec{k}'.(\vec{r}-\vec{r}')+\phi_g(\alpha,\beta)$ for some $\vec{k}'$ and a constant $\phi_g(\alpha,\beta)$ independent of $\vec{r}-\vec{r}'$--- this is always true for symmetry operations which are linearly represented on the lattice sites. 

Now, let us consider the symmetry transformation $g=g_b . g_a$ acting on $h_{\alpha,\beta}$. Using Eq.~\ref{eq:block_trans} followed by Eq.~\ref{eq:block_minusk} , we find 
\begin{align}
    \nonumber h_{\alpha,\beta}(k_x,k_y,k_z) & = h_{\beta,\alpha}(k_x,-k_y,-k_z) e^{i k_x }\\
     & = h_{\alpha,\beta}(-k_x,k_y,k_z) e^{i k_x } 
     \label{eq:habeq}
\end{align}
Eq.~\ref{eq:habeq} can only be satisfied if the real space amplitudes $u(\vec{r},\vec{r}')$ in Eq.~\ref{eq:hab_matform} satisfy
\begin{align}
U_{\alpha,\beta}(\vec{r})= u(y,z)(\delta_{x,1}+ \delta_{x,0}),
\label{eq:u_def}
\end{align}
where $\delta$ is the Kronecker delta not to be confused with the sublattice index, and $u(y,z)$ is any complex function of the coordinates $y$ and $z$. This form implies that,
\begin{align}
\label{eq:ab}
h_{\alpha,\beta} = M[u(k_y,k_z)\zeta(k_x)], \\
\nonumber \zeta(k_x)= (1+\exp(i k_x))
\end{align}
The function $u(k_x,k_y)$ is the fourier transform of $u(y,z)$ defined in Eq.~\ref{eq:u_def}. All other block matrices in Eq.~\ref{eq:block_intro} can be expressed in 
terms of $u(k_x,k_y)$ by applying symmetry transformations to $h_{\alpha,\beta}$.
Applying $g_c$ to  $h_{\alpha,\beta}$ using Eq.~\ref{eq:block_trans} gives us
\begin{align*}
h_{\alpha,\gamma}=\mathcal{A}^2 M[u(k_z,k_x)\zeta(k_y)],\\
h_{\alpha,\delta}=\mathcal{A} M[u(k_x,k_y)\zeta(k_z)], \\
\nonumber \text{where } \mathcal{A} =\exp(i (2\pi/3)\tau^z).
\end{align*}
We note that 
\begin{equation}
\mathcal{A}^n M[u]=M[\omega^n u], \\
\text{where } \omega = 2 \pi/3.
\end{equation}
This gives us
\begin{align}
\nonumber h_{\alpha,\gamma}= M[\omega^2 u(k_z,k_x)\zeta(k_y)],\\
\label{eq:ag_ad}
h_{\alpha,\delta}= M[\omega u(k_x,k_y)\zeta(k_z)], \\
\nonumber \text{where } \omega =\exp(i 2\pi/3).
\end{align}
Applying $g_b$ to $h_{\gamma,\delta}$ gives us

\begin{align}
\label{eq:gd}
h_{\gamma,\delta} =M[\omega u(k_y,-k_z) \zeta(-k_x)\exp(i k_x)]
\end{align}
Finally, applying $g_c$ and $g^2_c$ to $h_{\gamma,\delta}$ gives us 
\begin{align}
\nonumber h_{\beta,\gamma} =M[\omega u(k_x,-k_y) \zeta(-k_z)\exp(i k_y)],\\
h_{\delta,\beta} =M[\omega u(k_z,-k_x) \zeta(-k_y)\exp(i k_x)].
\label{eq:bg_db}
\end{align}

Eqs.~\ref{eq:ab}, \ref{eq:ag_ad}, \ref{eq:gd}, \ref{eq:bg_db}, along with Eq.~\ref{eq:block_minusk} specify the most general form 
of all block matrices appearing in $H_{\rm MFT}(\vec{k})$  in Eq.~\ref{eq:block_intro} that is allowed by projective symmetries of the $\mathsf{Z}_21$ state.

Now we express $H_{\rm MFT}(\vec{k})$ in the basis 
$\big(\psi^{\alpha}_1, \cdots \psi^{\delta}_1, \psi^{\alpha}_2, \cdots \psi^{\delta}_2\big)$ to  have  
\begin{align}
H_{\rm MFT}(\vec{k})&=
\begin{pmatrix}
0 & h(\vec{k}) \\
h^{\dag}(\vec{k}) & 0 \\
\end{pmatrix}. 
\end{align}
The matrix the most general  $h(\vec{k})$ allowed by the PSG is given by
\begin{widetext}
\begin{align}
h(\vec{k}) &= 
\left(
\begin{array}{cccc}
 0 & \zeta \left(k_x\right) u\left(k_y,k_z\right) & \omega ^2 \zeta
   \left(k_y\right) u\left(k_z,k_x\right) & \omega  \zeta \left(k_z\right)
   u\left(k_x,k_y\right) \\
 \zeta \left(-k_x\right) u\left(-k_y,-k_z\right) & 0 & \omega  e^{i k_y} \zeta
   \left(-k_z\right) u\left(k_x,-k_y\right) & \omega  e^{-i k_x} \zeta
   \left(k_y\right) u\left(-k_z,k_x\right) \\
 \omega ^2 \zeta \left(-k_y\right) u\left(-k_z,-k_x\right) & \omega  e^{-i k_y}
   \zeta \left(k_z\right) u\left(-k_x,k_y\right) & 0 & \omega  e^{i k_z} \zeta
   \left(-k_x\right) u\left(k_y,-k_z\right) \\
 \omega  \zeta \left(-k_z\right) u\left(-k_x,-k_y\right) & \omega  e^{i k_x}
   \zeta \left(-k_y\right) u\left(k_z,-k_x\right) & \omega  e^{-i k_z} \zeta
   \left(k_x\right) u\left(-k_y,k_z\right) & 0 \\
\end{array}
\right)
\end{align}
\end{widetext}

Now, we proceed to show that $h(\vec{k})$ has a maximum rank of $3$ on the points $(\pm k, \pm k, \pm k)$.
First consider $\vec{k}=(k,k,k)$. With the further shorthands
\begin{align}
\nonumber & u = u(k,k),& & v = u(-k,-k), & &\rho = u(k,-k)\\ 
& \lambda = u(-k,k), & &\zeta = \zeta(k), & & r = e^{i k}
\end{align}

we have
\begin{align}
\nonumber
h(k,k,k)=
\left(
\begin{array}{cccc}
 0 & \zeta  u & \zeta  u \omega ^2 & \zeta  u \omega  \\
 \zeta ^* v & 0 & \zeta  \rho  \omega  & \zeta ^* \lambda  \omega  \\
 \zeta ^* v \omega ^2 & \zeta ^* \lambda  \omega  & 0 & \zeta  \rho  \omega  \\
 \zeta ^* v \omega  & \zeta  \rho  \omega  & \zeta ^* \lambda  \omega  & 0 \\
\end{array}
\right)
\label{eq:kkk_mat}
\end{align}
$h(k,k,k)$ can now be brought to the row-echelon form by the sequence of elementary row-transformations given by
\begin{align}
\nonumber &R_2 \leftrightarrow R_1,   R_3 \rightarrow \omega R_3 -R_1,  R_4 \rightarrow \omega^2 R_4  -R_1 \\
\nonumber &R_3 \rightarrow u\omega \zeta R_3   - \lambda \zeta^* R_2 ,  R_4 \rightarrow R_4 u - R_2 \rho   \\
&\begin{aligned}
 R_4 &= -u \omega  \left(\zeta  \rho  (\omega +1)-\zeta ^* \lambda  \omega \right)R_4  \\
\nonumber     &+u\zeta \omega ^2 \left(\zeta ^* \lambda +\zeta  \rho \right) R_3   .
\end{aligned}
\end{align}
in conjunction with the using the identities $\omega^3=1$, $\zeta r* =\zeta^*$ and $r r^*=1$.The result of the row transformations is 
\begin{align*}
\left(
\begin{array}{cccc}
 \zeta ^* v & 0 & \zeta ^* \rho  r \omega  & \zeta  \lambda  r^* \omega  \\
 0 & \zeta  u & \zeta  u \omega ^2 & \zeta  u \omega  \\
 0 & 0 & \zeta  u (\omega +1) \left(\zeta ^* \lambda +\zeta  \rho \right) &
   \zeta  u \left(\zeta ^* \lambda +\zeta  \rho \right) \\
 0 & 0 & 0 & 0 \\
\end{array}
\right)
\end{align*}
The appearance of the $0$ in the last diagonal element establishes that the maximum rank of $h(k,k,k)$ can be $3$.

The analysis need not be repeated for the $H_MFT$ at other gapless points 
like $(-k,k,k), (k,-k,k)$ etc.  All of $h(\pm k, \pm k, \pm k)$ are related to 
$h(k,k,k)$ by elementary ``rank-preserving" row and column transformations. $h(-k,k,k)$ 
can be obtained from $h(k,k,k)$ by the transformations 
\begin{align}
\nonumber R_1 \leftrightarrow R_2,  R_3 \leftrightarrow R_4,\\
\nonumber C_1 \leftrightarrow C_2,  C_3 \leftrightarrow C_4, \\
R_4 \rightarrow r^* C_4, C_4 \rightarrow r C_4,
\end{align}
followed by two  re-identifications: $u \leftrightarrow v$, which have been considered independent complex numbers in the proof; and $\omega \leftrightarrow \omega^2$ which survive the important properties $\omega^3=1$ and $1+\omega+\omega^2=0$. $h(k,-k,k)$ can be obtained in turn from $h(-k,k,k)$ by the transformations 
\begin{align}
R_2 \rightarrow R_3, R_3 \rightarrow R_4, R_4 \rightarrow R_2, \\
C_2 \rightarrow C_3, C_3 \rightarrow C_4, C_4 \rightarrow C_2, \\
R_1 \rightarrow \omega^2 R_1, C_1 \rightarrow \omega^2  C_1. 
\end{align}

$h(k,k,-k)$ can be obtained from $h(-k,k,k)$ by the transformations 
\begin{align}
R_2 \rightarrow R_4, R_4 \rightarrow R_3, R_3 \rightarrow R_2, \\
C_2 \rightarrow C_4, C_4 \rightarrow C_3, C_3 \rightarrow C_2, \\
R_1 \rightarrow \omega R_1, C_1 \rightarrow \omega  C_1. 
\end{align}

This completes the proof that $h(\pm k, \pm k, \pm k)$ has a maximum rank of $3$,
and consequently, $H_MFT(\pm k, \pm k, \pm k)$ has two gapless bands which are protected by projective symmetries against the addition of arbitrarily long-ranged terms
in the mean-field ansatz.

\subsection{$\mathsf{U}(1)$}
The $\mathsf{U}(1)$ spin liquid mean field ansatz has the following form:
\begin{equation}
    U_{ij} = iU^0_{ij} \tau_0 + U^z_{ij} \tau_z,
\end{equation}
dictated by the fact that the ansatz is invariant under the $\mathsf{U}(1)$ IGG gauge transformation. We would like to investigate how the PSG solutions we obtained constrain the nearest neighbor mean field ansatz by subjecting them to the following test:
\begin{align}
    \forall g\in \mathsf{P}2_13\times \mathsf{Z}^{\mathcal{T}}_2: \hat{G}_g \hat{g}(U_{ij}) = U_{ij}.
\end{align}
Among the PSGs we have, we first study the class in which $A=0$. After enumerating all the conditions imposed by the PSG, we arrive at the following nearest neighbor mean field ansatz in the class where $A = 0$.:\\
\begin{equation}
    U_i = \lambda \tau_z, \quad \text{where } i \in \{1,\dots,12\}.
\end{equation}
Next we would like to argue that, when $A\neq 0$, there would be no nearest neighbor mean field ansatz. We write $U_{1/3}\equiv i\mathfrak{U}_{1/3}\exp [i\varphi_{1/3} \tau_z]$. The TRS conditions on these two bonds give:
\begin{equation}
    2\varphi_1 + A = \pi, \quad \varphi_3 = \pi/2.
\end{equation}
However, the condition $\hat{G}_b \hat{g}_b(U_1)=U_1$ gives us:
\begin{equation}
    \varphi_3 + 2A + 4\phi_3 = \varphi_1.
\end{equation}
We time the above equation by $2$, and combined with the last two relations, we would immediately arrive at $5A = 0$. Note that we had $3A=0$. Thus the ansatz does not vanish only when $A=0$.\par
We conclude that we obtain only one nearest neighbor $\mathsf{U}(1)$ mean field ansatz:\\
\begin{align}
    \quad U_\zeta &= \lambda \tau_z, \quad  \zeta \in \{1,\dots,12 \} \nonumber \\
    \quad a_s & = \omega \tau_z, \quad i\in \{\alpha,\dots,\delta \}.
\end{align}

\bibliography{bibliothek_trillium}

\end{document}